\documentclass[12pt,preprint]{aastex}

\shorttitle{A Joint Optical and X-ray Catalog of Cluster Candidates}
\shortauthors{Donahue et al.}

\newcommand{\flux}{{\rm erg \, s^{-1} \, cm^{-2}}}

\newcommand{\etal}{{et al. }}
\newcommand{\lum}{{\rm erg \, s^{-1}} }

\newcommand{\ltsim}{{\>\rlap{\raise2pt\hbox{$<$}}\lower3pt\hbox{$\sim$}\>}}
\newcommand{\gtsim}{{\>\rlap{\raise2pt\hbox{$>$}}\lower3pt\hbox{$\sim$}\>}}

\begin{document}

\title{Distant Cluster Hunting II: A Comparison of X-ray and Optical
Cluster Detection Techniques and Catalogs from the ROX Survey}
\author{Megan Donahue\altaffilmark{1,5}, Caleb Scharf\altaffilmark{1,2}, 
Jennifer Mack\altaffilmark{1}, 
Paul Lee\altaffilmark{1}, Marc Postman\altaffilmark{1}, 
Piero Rosati\altaffilmark{3}, 
Mark Dickinson\altaffilmark{1,5}, G. Mark Voit\altaffilmark{1}, 
John T. Stocke\altaffilmark{4,5}} 
\altaffiltext{1}{Space Telescope Science Institute, 3700 San Martin Drive, 
Baltimore, MD 21218, donahue@stsci.edu}
\altaffiltext{2}{Columbia University, Columbia Astrophysics Lab, Mail Code 5247 
Pupin Hall, 550 W. 120th St., New York, NY 10027}
\altaffiltext{3}{European Southern Observatory, Karl-Schwarzschild-Str. 2, 
Garching, D-85748}
\altaffiltext{4}{University of Colorado, CASA, CB 389, Boulder, CO 80309}
\altaffiltext{5}{Visiting Astronomer, Kitt Peak National Observatory}

\begin{abstract} 
We present and analyze the optical and X-ray catalogs of
moderate-redshift cluster candidates from the ROSAT Optical X-ray
Survey, or ROXS. The survey covers the sky area contained 
in the fields of view 
of 23 deep archival ROSAT PSPC pointings, 4.8 square degrees. The  
cross-correlated cluster catalogs 
were constructed by comparing two independent 
catalogs extracted from the  optical and X-ray bandpasses, 
using a matched-filter technique for 
the optical data and a wavelet technique for the X-ray data.  We 
cross-identified cluster candidates in each catalog. As reported
in Paper I, the matched-filter 
technique found optical counterparts for at least 60\% (26 out of 43) 
of the X-ray cluster candidates; the 
estimated redshifts from the matched filter algorithm agree   
with at least 7 of 11 
spectroscopic confirmations ($\Delta z \lesssim 0.10$). 
The matched filter technique, 
with an imaging sensitivity of $m_I \sim 23$, 
identified approximately 3 times the number of
candidates (155 candidates, 142 with a detection confidence $>3\sigma$) 
found in the X-ray survey of nearly the same area. There are 
57 X-ray candidates, 43 of which are unobscured by scattered light or bright
stars in the optical images. Twenty-six of these have fairly secure
optical counterparts. We find that the matched filter 
algorithm, when applied to images with galaxy flux sensitivies of
$m_I \sim 23$, is fairly well-matched to discovering $z\leq1$ clusters
detected by wavelets in ROSAT PSPC exposures of 8,000-60,000 seconds.  
The difference
in the spurious fractions between the optical and X-ray (30\% and 10\% 
respectively) can not account for the difference in source number.
In Paper I, we compared the optical and X-ray cluster
luminosity functions  and we found that the
luminosity functions are consistent if the relationship
between X-ray and optical luminosities is steep ($L_x \propto L_{opt}^{3-4}$). 
Here, in Paper II, 
we present the cluster catalogs and a numerical simulation of 
the ROXS. We also present color-magnitude plots for several of
the cluster candidates, and examine the prominence of the red sequence
in each. We find that the X-ray clusters in our survey do
not all have a prominent red sequence. We conclude that while the red sequence
may be a distinct feature in the color magnitude plots for virialized
massive clusters, it may be less distinct in lower-mass clusters of
galaxies at even moderate redshifts.
Multiple, complementary methods
of selecting and defining clusters may be essential, 
particularly at high redshift where all methods start to 
run into completeness limits, incomplete understanding of physical
evolution, and projection effects.

\end{abstract}

\section{Introduction: Why Conduct an Optical - X-ray Survey for
Clusters of Galaxies?}

Clusters of galaxies are the most massive gravitationally-bound systems in the 
universe. Because they sample the high mass end of the mass function 
of collapsed systems, they can
be used to determine cosmological parameters such as $\Omega_{m}$
(e.g. Donahue \& Voit 1999). 
The largest clusters  ($\sim 10^{15} ~M_\odot$),
are the products of the collapse of matter from a very large volume 
of space ($r \sim 16 h^{-1}(\Omega/0.2)^{-1/3} ~\rm{Mpc}$). Therefore 
they are thought to be ``fair samples'' 
of the universe -- that the mass to light
ratio or the baryonic mass fraction defined within the domain of a cluster
of galaxies is representative of that ratio in the universe as a whole. They
are purported to be ``closed boxes'' to star formation and evolutionary  
processes that occur within their domain. In this paper we present and
analyze the catalogs from our joint optical-X-ray search for clusters
of galaxies. We conducted this survey in order to provide a sample
of clusters to test such assumptions about clusters of galaxies and
to investigate the impact of sample selection on studies of cluster
evolution and the evolution of their member galaxies. 

Understanding cosmological or galaxy
evolution studies of clusters critically requires 
an understanding of the biases in any sample of clusters. 
For example, the evolution of the 
number density of systems with cluster-sized masses 
as a function of mass and redshift is a fundamental prediction of 
cosmological structure formation models. To know the number density
of clusters, we must know the biases inherent in how 
we find them, preferably as a function of cluster 
mass.
Furthermore, testing the ``fair sample'' hypothesis requires reliable
and unbiased selection of the most massive clusters. 

Ever since Abell (1958) and Zwicky (Herzog, Wild \& Zwicky 1957; Zwicky 1961)
began publishing catalogues of optically selected clusters of galaxies, 
the definition of a cluster and the definition of biases inherent in 
the cluster detection process have been lively topics of debate. 
In 1978, the launch of the first X-ray imaging telescope, 
the Einstein observatory,  
began a new era of cluster discovery, as clusters proved to be 
luminous ($>10^{42-45} \lum$), extended ($r\sim1-5$ Mpc) 
X-ray sources, readily identified in the X-ray sky.
The intracluster gas, in nearly 
hydrostatic equilibrium with the gravitational potential of the cluster,
radiates optically thin thermal bremstrahlung and line radiation. X-ray 
selection of clusters is more robust against contamination along the line of
sight than traditional optical methods since the richest clusters are
relatively rare and since X-ray emissivity, which is proportional to the
gas density squared, is far more sensitive to physical overdensities than
is the projected number density of galaxies on the sky.

One cluster sample differs from another depending on how the
clusters were detected. 
Optical selection of clusters using traditional
methods looking for overdensities of galaxy counts (e.g. Abell 1958) was 
rife with contamination problems. 
However, modern methods such as the ``matched filter'' algorithm 
(Postman et al.\ 1996, P96 hereafter) provide automated, uniform detection 
of galaxy overdensities in deep optical images.  The matched filter technique 
searches for local density enhancements in which galaxies follow a magnitude 
distribution characteristic of that expected for a cluster of galaxies.  
The results include statistically quantifiable estimates of cluster 
richness, redshift, and significance. 
The first X-ray selection methods using sliding boxes were
optimal for point sources. Thus, the detection
 method used to construct the Extended Medium Sensitivity
Survey (EMSS; Gioia et al. 1990b) was biased somewhat towards
selecting clusters with high central surface brightnesses. Now there
are several algorithms optimized for detecting extended sources, 
including wavelets (Rosati et al. 1995) and 
Voronoi-Tesselation Percolation methods (Scharf et al. 1997).

A decade ago, optical and X-ray surveys apparently disagreed about how 
much clusters have evolved since $z\sim0.5-1.0$. Optical
surveys indicated very little evolution since
$z\sim0.5-1$ (Gunn, Hoessel \& Oke 1986), but the accurate 
measurements of survey volumes and cluster properties required 
for quantitative assessment of 
this evolution were difficult to quantify in these first high-redshift
cluster surveys and the volumes were small so uncertainties were 
large. X-ray studies suggested
modest evolution (Gioia et al. 1990a). 
The most recently compiled 
X-ray samples of clusters over a range of redshifts out to
$z\sim0.8-1.2$ agree that the X-ray luminosity function for moderate
luminosity clusters has
not evolved significantly since $z\sim0.8$ (Borgani et al. 1999; Nichol et al. 1999; 
Rosati et al. 1998, 2000; Jones et al. 1998), while the most
luminous systems, contained in the EMSS, 
might have evolved somewhat (Henry et al.
1992; Nichol et al. 1997, Vikhlinin et al 1998, 2000; Gioia et al. 2001) or
very little (Lewis et al. 2002). 
More recent optical surveys for distant clusters continue to find very little
evidence for cluster number density 
evolution at moderate redshifts (Couch et al. 1991; P96). The explanation of  
what may seem like a persisting discrepancy is that if any evolution exists in
the X-ray cluster population, it is only occuring in the highest luminosity
systems which are also the most rare systems. The optical surveys of 
Couch et al. (1991) and P96 were too small and shallow to detect the
putative evolution of the rarest systems.

While the most recent optical and X-ray results are now at least in 
statistical agreement on
the question of evolution since $z<0.8$,
the question remains whether both techniques are selecting the same clusters. 
The fundamental quantity, from the viewpoint of comparison to cosmological
simulations, is the cluster's mass. We do not know {\em a priori}
 whether optical
luminosity or X-ray luminosity should be better correlated with a cluster's mass.  
The fundamental question, from the viewpoint of ``fair sample'' techniques
of measuring universal ratios, is whether clusters are truly a ``fair 
sample''. For example, M/L ratios depend on the bandpass of the light
and the star formation history of the constituent galaxies. 
If the gas fractions or the M/L ratios vary significantly from cluster to 
cluster they are obviously not representative of the universe as a whole.

X-ray selection is generally thought to be superior to optical selection.
Observationally, the hot gas is a larger fraction of the cluster mass than
the stellar mass, and the X-ray luminosity of a cluster is far easier to measure
than its optical luminosity. For X-ray selected clusters, 
studies of gas fractions and cluster M/L ratios show that these quantitites
are statistically constant 
(Evrard 1997; Arnaud \& Evrard 1999; Carlberg et al. 1996). However, 
if X-ray selection biases the selection of the clusters,  
high-mass clusters of
galaxies  with low hot gas fractions (if they exist)  
would be omitted from such studies. Massive clusters,
under the ``fair sample'' hypothesis, should have nearly identical
baryon fractions, however they are discovered.

With the ROSAT Optical X-ray Survey (ROXS) for clusters of
galaxies, we have endeavored 
to address such issues by obtaining optical images of complete 30' by
30' fields centered on positions of deep ROSAT PSPC pointings. In contrast
to previous ROSAT PSPC serendipitous surveys such as those conducted
by Rosati et al. (1995, 1998), 
Jones et al. (1998), Romer et al. (2000), and 
Vikhlinin et al. (1998), the ROXS 
includes optical imaging for the entire field of view of each X-ray 
pointing. The X-ray
selection and optical selection of cluster candidates was then done
independently of each other. We observed 23 ROSAT pointings for a total of nearly
5 square degrees in I band. For five of these fields we also obtained
V-band imaging. In this paper (Paper II) we present 
the catalogs, survey windowing functions, data reduction
and observation details, an analysis of detection likelihoods, 
as well as an expanded discussion and 
further analysis, including numerical simulations of the survey.
In \S2, we describe the X-ray field selection criteria and the
optical observations. In \S3, we present the optical cluster candidates
catalog, cross-identification of clusters in the V and I bands. In \S4, 
we present the X-ray cluster
candidate catalogs and the X-ray/optical cross-identification procedure. In
\S5 we describe properties of the cluster candidates, 
including the distribution of observed properties of objects in
the sample, the estimated richness vs. $L_x$, 
the $V-I$ vs $I$ color magnitude diagrams for the clusters
identified in both the $V$ and $I$ bands. We discuss and summarize
our results in \S6 and \S7 respectively.

For all derived quantities, we have 
used $H_0=75 h_{75} $ km s$^{-1}$ Mpc$^{-1}$,
and $q_0=0.5$.

\section{Observing Strategy, Observations and Data Reduction}
\subsection{Field Selection and Observations}

The sample of 23 target fields were 
chosen from a sample of archival X-ray observations. All ROSAT PSPC
pointings with exposure times of more than 8,000 seconds and Galactic
latitude of $|b_{II}| >20$ degrees and declination $\delta >-20$ degrees 
were considered for optical imaging from Kitt Peak. 
We avoided fields with bright stars $(m_V
< 9)$. Even so, bright stars outside the field occasionally scattered light into the
field of view and stars with $m_I<16$ had diffraction 
spikes. Pixels affected by
bright stars and scattered light were masked in our subsequent 
analysis, and thus the total survey area is 4.8 square degrees, 
less than the nominal 5.75 square degrees that would have been
covered if optical detection success were unaffected by stars. 
Table~\ref{Fields} summarizes the fields observed as well as the
associated data.

We obtained the optical data during two observing runs at the Kitt Peak
National Observatory 4-meter telescope during March 1996, and May 1997. We
were awarded 4 observing runs for this project, but one (November 1995)
was cancelled because the dome mechanism was broken and another had
extremely poor weather (November 1996). Therefore, the entire survey was
conducted in the northern spring sky. 
The KPNO 4-meter prime focus CCD camera T2KB has a 16' field of view 
(0.47" pixel$^{-1}$).  We mosaiced $2\times2$ of these fields to
created full 30' by 30' fields,
each centered on a deep ROSAT pointing, overlapping each quadrant by $\sim1'$.

A total of 900 seconds of exposure
were obtained through
the I-band filter for each ROSAT field quadrant. Our 5$\sigma$ detection limit 
of $I=23$ (Vega magnitude) 
was sufficient to detect cluster galaxies 2 magnitudes fainter
than the typical unevolved first-ranked elliptical at $z=1$ (Postman 
et al. 1998a). 
For five of the ROX fields, we also obtained
V-band data, for a total exposure of 600 seconds for each
quadrant.  The data were obtained under near photometric conditions and
low air masses ($<1.2$, typically). When the sky was photometric, we
obtained short exposures of the center of each field for calibration
purposes, and thus all of our data were flux-calibrated (see \S3.1.1). 
The filters used were the V and I filters from the Harris set.

\subsection{Data Reduction}

The CCD images were first reduced with standard IRAF tools for bias level
subtraction, frame trimming, and flat-field division. A cosmic ray
rejection algorithm based on median filtering techniques was
used to remove cosmic rays from the images. In the first observing
run, 
each pointing was divided  
into two 450 second exposures
because the I-band sky was so bright that the dynamic range of the A-to-D
converter was swamped. We used the two images to reject cosmic rays.
For our second run, the A-to-D converter had been updated to unsigned
integers, and the cosmic-ray rejection made possible with 
two images was determined to be
of little advantage compared to the overhead time and additional data-reduction
complexity. Data obtained during subsequent telescope visits thus consisted of
single 900-second observations. The two 450 second exposures for each quadrant
obtained in earlier runs were coadded. We achieved a detection limit of
$23$ mags in I and $24$ mags in V (Vega magnitudes). 

To prepare for the galaxy catalog construction, the following processing 
steps as discussed in P96 were employed.
A sky model for each (coadded) frame was created from the data to
remove the remaining  CCD signatures. 
We fit the median sky level as a function of row, and subtracted
the appropriate sky components from each image row. 
These steps produced frames with extremely flat sky levels.

\section{Catalogs}

Here we describe the procedures and assumptions 
we used to create the galaxy catalog,
the matched filter cluster catalog, the X-ray catalog, and the
cross-identifications of probable detections of the same physical
cluster system.

\subsection{Optical Catalog}

\subsubsection{Galaxy Catalog Construction}

We followed the procedures discussed in P96 to construct
the galaxy catalog. A modified version of the Faint Object Classification
and Analysis System (Jarvis \& Tyson 1981, Valdes 1982) was used to
detect, measure, and classify objects in the calibrated CCD images. 
The detection algorithm estimated a point spread function (PSF)  for
each frame, to compensate for seeing variations from exposure to exposure.
To separate stars from galaxies, FOCAS classification parameters were
chosen to cover PSF variations but to be robust to star-galaxy distinctions.
%The classification rules we used are listed in Table~\ref{FOCAS}. 
%Definitions for Table~\ref{FOCAS} are quoted for convenience from Valdes (1982): 
%$s$ is a dimensionless radial scale factor for the PSF, $f$ is the fraction
%of the source which is best described by a nominal point source scaled
%by a factor $s$. 
Perfection in distinguishing galaxies from stars 
is not critical to our experiment since galaxies
dominate the detected object counts in the magnitude range of interest 
(P96). 

Astrometric calibration was performed by selecting unsaturated bright
stars. The J2000 celestial positions of the stars were measured
from the Digitized Sky Survey (the ``Quick V'' survey, epoch 1983) 
based on Palomar Observatory Schmidt plates (Lasker et al. 1990). 
An astrometric solution was computed based on a 6-term per 
coordinate polynomial for each field. Typical solution uncertainty
is $<0.5"$.   
The absolute photometry for Johnson V and Cousins I fluxes
was obtained using Harris V and I-band 
observations of 
Landolt Standards fields PG1047+003, 1159-035, SA107, and PG1657+078 
(Landolt 1992).  For our figures and tables we use 
Johnson V and Cousins I. Those magnitudes are related to AB 
magnitudes using the following relations.
$V(Johnson)=V(AB)-0.02$ and 
$I(Cousins)=I(AB)-0.45$.

\subsubsection{Cluster Catalog Construction \label{optcat_sect}}

The procedures in P96 were followed in constructing the
cluster catalog. The algorithm based on the matched filter method, tuned
to redshifts in the range $0.2 < z < 1.2$, was used to generate cluster
catalogs in $\Delta z = 0.1$ intervals. $H_0=75 h_{75}$ km s$^{-1}$ Mpc$^{-1}$ and
$q_0=0.5$ were assumed for all quantities. Clusters were identified by searching
for the local maxima within a moving box which was $1.667h_{75}^{-1}$~Mpc across. A
candidate cluster was registered when the central pixel in the box was the
local maximum and it lay above a prescribed threshold. We used the
following cluster detection parameters and cluster assumptions to generate
the cluster catalog. The slope of the radial filter profile was 1.40.  A
cluster core radius of $133.3h_{75}^{-1}$ kpc was assumed, and the smoothed map was sampled
every 0.50 core radii. The cut-off radius for cluster detection was set to
$1333 h_{75}^{-1}$ kpc, with a detection box half-width of $1666.7 h_{75}^{-1}$ kpc.
The profile kernel halfwidth was 20 map pixels, defined by the 0.50 core
radius for each map, the catalog scale
was 1 arcsecond/unit, and the CCD scale was 0.45 arcsecond/pixel. 
For the luminosity function
filter we assumed a Schechter function (Schechter 1976) with
 $M^*=-21.0+5 \log h_{75}$ in the V band and $M^*=-21.90 +
5 \log h_{75}$ in the
I band with a slope of -1.10 for both bands. Each cluster candidate is
assigned a central position, effective radius (corresponding to the area of
detection equal to $ \pi r_{eff}^2$), an estimated redshift based on the
best-fit luminosity function, an average galaxy magnitude ($I_{mag}$ or
$V_{mag}$ depending on the bandpass), and a detection confidence in units
of sigma ($\sigma$). The algorithm also estimates an effective optical
luminosity  $\Lambda_{cl}$, which corresponds to the 
equivalent number
of $L^*$ galaxies in the cluster such that the optical cluster 
luminosity $L_{cl}/L_{\odot}=\Lambda_{cl}L^*/L_{\odot}=
\Lambda_{cl}10^{-0.4(M_\odot - M^*)}$ where $M_\odot$ is the absolute
magnitude of the Sun in the appropriate band
(P96). The quantity $\Lambda_{cl}$ is relatively 
insensitive to $H_0$; variations $\sim 5\%$ were measured 
if $h_{75}=1.33$ rather than 1.0. For reference, a richness class 1 
cluster at $z\leq0.7$ had $\Lambda_{cl}=30-65$ in the I-band (P96). 

These parameters generated  cluster catalogs that are directly
comparable to cluster candidates in the Postman \etal (1998b) 16 square degree
I-band survey (called ``Deeprange"). 
The similarity in optical methods and instrumentation for 
the ROXS and the Deeprange surveys allowed us to compare   
the spurious rates of cluster identification and the reliability of the
estimated redshifts with respect to spectroscopic redshifts 
for the ROXS from the larger Deeprange survey, in addition to  
the P96 survey.
                     	
\begin{figure}
%\plotone{rox_optsel.ps} 
\plotone{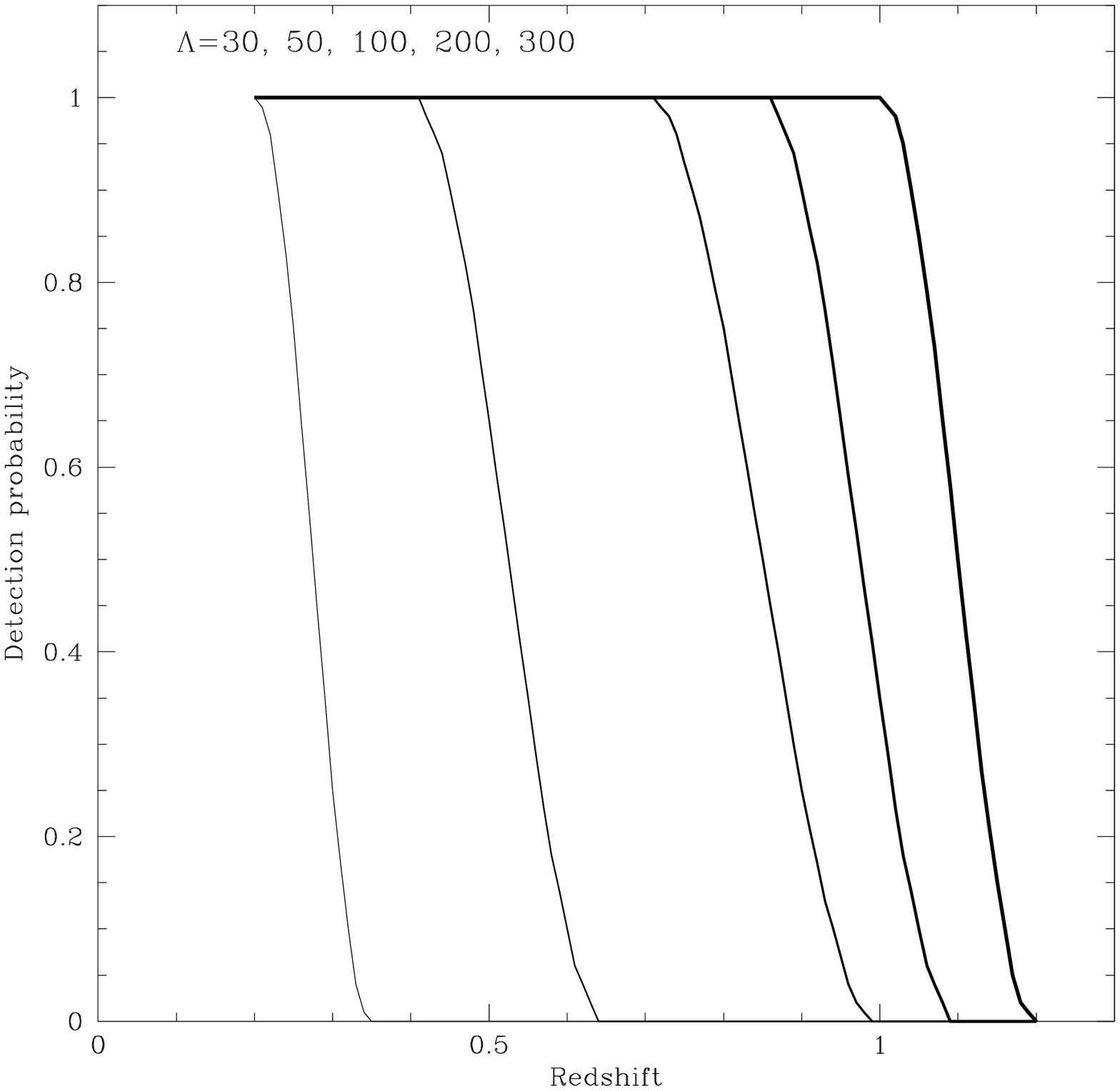}
\caption[]{The optical matched filter
detection probabilities in the I-band as a function of redshift for systems of richness
$\Lambda=$30, 50, 100, 200, 300 (in order of increasing line weight). The
probabilities were determined using the methods of P96.
\label{optsel} } \end{figure}

Table~\ref{Optical} lists the cluster candidates for 23 fields observed.
We have extracted the optical selection function as a function of
redshift in Figure~\ref{optsel}. These probabilities were computed
following P96, assuming a K-correction appropriate
to a galaxy population dominated by ellipticals.

\subsection{V and I band Cross-Identifications}

Here we describe the comparison of the cluster candidate
catalogs resulting from the matched
filter algorithm for the I-band data and the
V-band data that we obtained for five of the ROXS fields. We find
that 60-65\% of the clusters identified in the I band are also
detected in the V-band with similar estimated redshifts if we
exclude all candidates with high redshift estimates ($z>1$).

For five fields (Table~\ref{Fields}), 
we obtained V-band images, with a source detection limit $V_{AB}\sim24$ magnitudes
within an 
exposure time of $600$ seconds. We used the matched filter algorithm, 
adjusted for the V-band to provide appropriate estimated redshift 
estimates and $\Lambda_{cl}$, 
to provide an independent sample of optical candidates. In 
the five fields, we found 46 V-band candidates with a detection
confidence of $>3\sigma$ and
33 I-band candidates with a $>3\sigma$ confidence. 

When we cross-identify the candidates
from the I-band and the V-band, we find that  of the 33 I-band candidates, 
there are positional 
matches for 15 candidates with one or more V-band candidates (Table~\ref{vi}). 
All but one of these 15 candidates also match in estimated 
redshift ($\Delta z \leq 0.1$).  Regions in the V-band images
containing 6 I-band candidates were seriously affected by scattered
light or bright stars -- a more serious observational effect at V than at
I. Therefore, a maximum of 27 of the I-band clusters could have been detected   
in V band. Of those
27, 4 were high redshift candidates ($z \geq 1$), leading to a V-band detection
rate of $z<1$ I-band candidates of 15/23 or 65\% (Table~\ref{crossid_tab}).
The failure to detect high-redshift
I-band candidates in the V-band is not unexpected since the
4000-Angstrom break spectral feature would make the elliptical galaxies 
in such clusters, even if the clusters are real, 
very difficult to see in the V-band. 
Additionally, 
the matched filter algorithm reliability degrades rapidly for our I-band data
at redshifts greater than $\sim1$ 
because that is where the faint end of the luminosity function
is severely truncated by the survey magnitude limit. In this
regime, corrections for the missing optical light are only mildly reliable and the
result is a significantly higher error in the derived $\Lambda_{cl}$ value.

Correspondingly,
of the 46 V-band candidates, 18  had one or more counterparts
in the I-band (20 with a detection significance $>2.9\sigma$), 
and only 2 candidates were undetectable because of stray light
in the I-band image, for a detection fraction of 18/44 or 41\%. This
detection fraction has been diluted by many high redshift
V-band candidates with no I-band counterparts (15 with $z>1$). 
We suspect that all of the 17 
$z>1$ V-band candidates may be spurious. Of the two high redshift
V-band candidates with plausible
I-band counterparts, one has a loose association with
a $z=1$ I-band candidate (120436.7+280520, OC4 in the 1202+281 field)
and the other (154923+212325, OC3 in the 3C324 field)
has an estimated redshift of 0.7 from the I-band data, compared to
the V-band estimate of 1.2. 

We expect
that the reliability of the matched filter method may become poor for the
highest estimated redshifts, especially in the V-band for the following
reasons. At the highest redshifts, the luminosity function
filtering algorithm in the matched filter cluster detection
method is only sensitive to the brightest galaxies in 
a cluster, especially for the V-band where K-correction 
effects are strongest.  Moreover, the contrast of the most
distant clusters against the foreground/background galaxy
population becomes very weak, especially at bluer wavelengths
due to the combination of K-correction effects and the steeper
number counts of faint field galaxies (themselves potentially 
clustered) at bluer wavelengths.

We can therefore provide plausible explanations for many if not 
all of the failures to 
cross-correlate between I-band identifications and V-band
identifications (Table~\ref{vinot}) 
by a cluster candidate (either in I or V) with a high estimated
redshift which may be a warning flag for unreliable or weak
detections, 
or by obscuration or confusion with scattered light or
spikes from bright stars.

Our resulting efficiency in cross-identifying clusters in either 
V-band or I-band is $\sim60-65\%$, for candidates with estimated
redshifts of $z<1$ and if 
we correct for the candidates which are detected through one
filter but undetectable in the other because
of scattered light. 
We present a summary of our numbers in Table~\ref{crossid_tab}.

We cannot explain the failure to detect 8 I-band cluster candidates 
of moderate redshift in V. Four of these candidates have possible  
V-band counterparts with redshifts from the V-band data similar to
those derived from the I-band data, but 
with low detection 
significance or with centroids separated by $>3-4'$. 
These possible counterparts are listed in Table~\ref{vi}.
We also cannot explain the lack of an I-band counterpart for  
11 V-band cluster candidates of moderate redshift  
(as in I band, 3 of these candidates 
have low significance counterparts or widely separated  
I-band counterparts with similar redshift estimates). We note that
while the detection significance of these unmatched cluster candidates
is similar to the rest of the sample (not much dynamic range), 
the estimated richnesses tend to be
rather low $\Lambda_{cl}~20-50$, mostly $\leq 30$. We could be 
seeing incompleteness effects rather than spurious effects at lower
$\Lambda_{cl}$, but   
it is possible that some of these unconfirmed cluster candidates are false. 
 
If we count only the most
secure cross-identifications, the I- and V-band cross identification
statistics imply a spurious fraction
of $<35\%$ for matched filter candidates with 
estimated cluster redshifts of less than one 
-- consistent with the estimates of the spurious fraction of 25-30\% 
from the Deeprange survey (Postman et al 1998b), with simulations
done in P96 and in Postman et al (in preparation), and with
spectroscopic observations (for $z<0.6$) 
in Postman et al. (in preparation). 
The spurious fraction
for candidates with 
estimated redshifts larger than one is completely unknown since
no candidate with $z\geq1$ in one bandpass 
has a secure counterpart in the other
bandpass.

Comparison of matched filter results from the I and V band do not address 
important sources of spurious cluster candidates, such 
as unbound aggregates of galaxies seen in a ``pole-on'' filament.
Both the I- and V-band data would be contaminated by such  
structures. For this reason, comparision of the matched filter
method with X-ray observations and
other cluster detection methods, such as Sunyaev-Zel'dovich or
weak-lensing  observations and spectroscopy of member galaxies 
is critical. With ROXS, we make the direct 
comparison with the X-ray observations; in this paper, we provide 
the catalogs of optical and X-ray candidates for follow-up with 
other cluster hunting techniques.

\section{X-ray Catalog \label{xraycat_sect}}

Here we describe the creation of the catalog of X-ray cluster candidates
and the cross identification of X-ray candidates with optical candidates
detected by the matched filter algorithm in the I-band images.

\subsection{X-ray Source Detection and Upper Limits for Optical Sources
 \label{xdetupplim}}

A wavelet-based technique, described by Rosati \etal (1995), was used to
create an catalog of X-ray clusters of galaxy candidates for each field
observed at Kitt Peak. Several of these
fields overlapped with the orginal ROSAT Deep Cluster Survey  (RDCS; 
Rosati \etal 1995, 1998) sample, and
thus the X-ray cluster candidates in some of the fields already have 
been confirmed and have spectroscopic redshifts. The flux limits for 
ROXS  are
approximately the same as those of the RDCS ($F_x > 10^{-14}~ \flux$), but
of course vary with exposure time in both surveys.
Table~\ref{Xray} lists the 57 X-ray candidate clusters and their associated
X-ray parameters for each ROSAT 
field, and the spectroscopic redshift, when available. The X-ray 
parameters (Table\ref{Xray}) 
measured are a centroid position (RA and Dec J2000), the 
net number of X-ray photons in the source (Counts), the off-axis angle
of the source in arcminutes (Theta), the FWHM of the source in 
arcseconds (FWHM), the confidence level of the extended nature of
the source in $\sigma$ (Sig-ext), and 
the X-ray flux and error in the 0.5-2.0 keV 
bandpass ($F_x$ and $eF_x$). The comment field includes the spectroscopic
redshift, if available. A notation of "d" indicates the source may be
a double source. Occasionally, a source has a large flux error, reflecting
not the confidence in the detection, but the uncertainty in measuring the
flux from an extended, complex object.

A by-product of our analysis was a catalog of all the X-ray point sources
in the field down to the flux limit of
each ROSAT observation. 
These sources matched very nearly one-to-one to sources 
available in the WGACAT (White, Giommi, Angelini 1995), so we do not list them here. 
We will provide this list on request. In Table~\ref{Optical}, 
we identify the optical candidates without an  
associated extended X-ray source but with an X-ray 
point source within 30-60".

\begin{figure}
%\plotone{sect32.ps}
\plotone{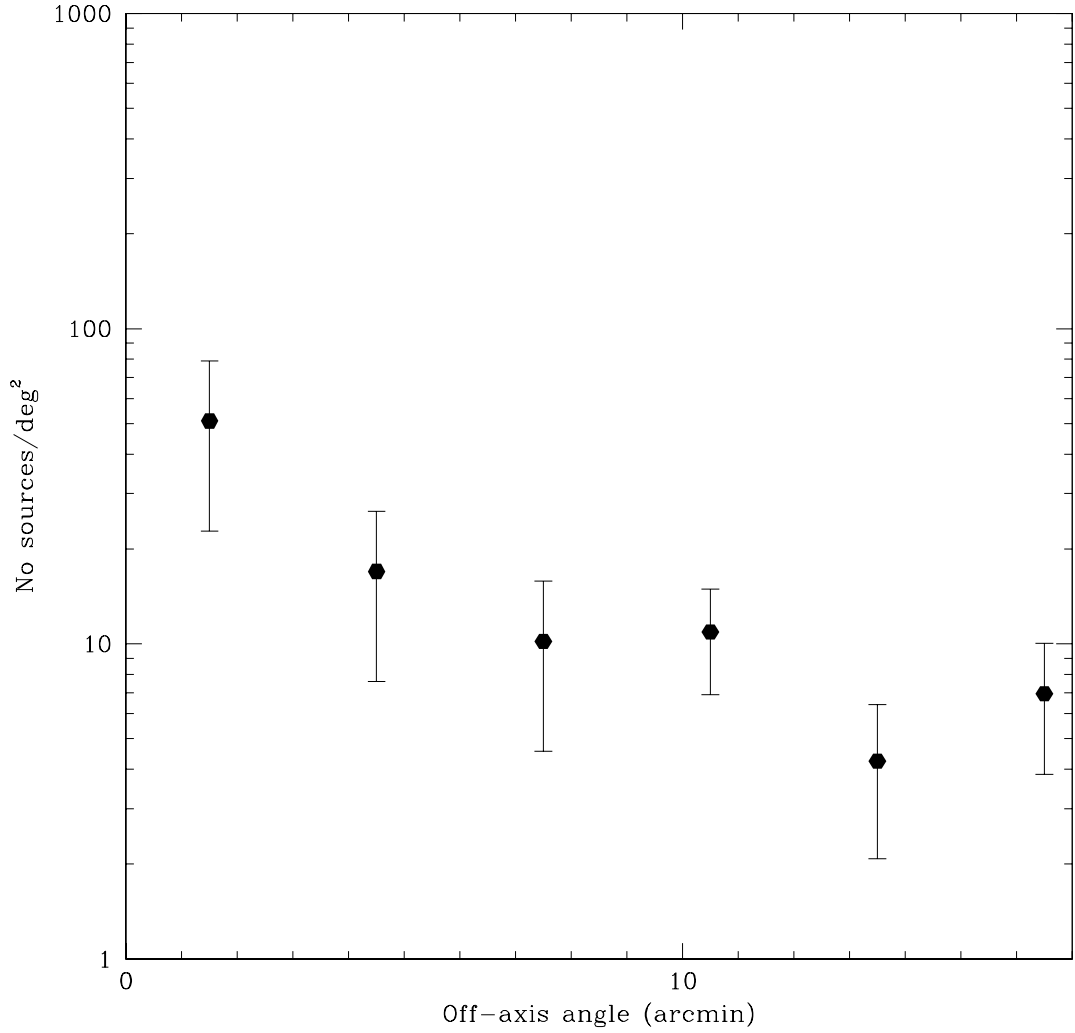}
\caption[]{Mean number of extended and pointlike 
X-ray sources per unit area as a function
of distance from the center of the ROSAT field. \label{nox}}
\end{figure}

For every optical cluster candidate,  
we estimated the X-ray detection threshold defined by an  
upper limit to the observed $0.5-2$ keV flux as the flux 
corresponding to a $4\sigma$
fluctuation above the limiting surface brightness, within a $r=1'$ aperture, 
as a function of position in the ROSAT field of view.  
The variation of X-ray source detection efficiency, arising from
the degradation of the Point Source Function (PSF), as a function
of off-axis angle can be seen in Figure~\ref{nox} where we plot the mean number
of X-ray sources per unit area in radial bins. There is almost a factor of
ten reduction in the source counts at the largest off-axis angles.  This
reduction 
corresponds to as much as a factor of $4-5$ change in the flux limit, which
is taken into account in our upper-limit estimates in Table~\ref{Optical}. 
Using the known, deep,
cumulative source number counts in the $0.5-2$ keV band (e.g. Hasinger \etal 
1998) we can map the trend seen in Figure~\ref{nox} 
into flux-limit vs off-axis
angle. A simple, approximate, linear fit to this yields a relationship of
(in units of $10^{-14}$ erg s$^{-1}$ cm$^{-2}$ ($0.5-2$ keV)):
$f_{lim}\simeq A + 0.078\theta$, where A is the flux limit at $\theta=0'$ 
and $\theta$ is the off-axis angle in units of arcminutes.
Since our X-ray detection algorithm provides us with an estimate of the
background surface brightness we can normalize the above relationship with
the flux of a $4\sigma$, on-axis source in a given aperture, and for a
given field. The detection thresholds $F_{lim}$ in units of $10^{-14}$ erg s$^{-1}$ 
cm$^{-2}$ ($0.5-2.0$ keV) are reported in Table~\ref{Optical}. 

\begin{figure} 
%\plotone{rox_xraysel.ps}
\plotone{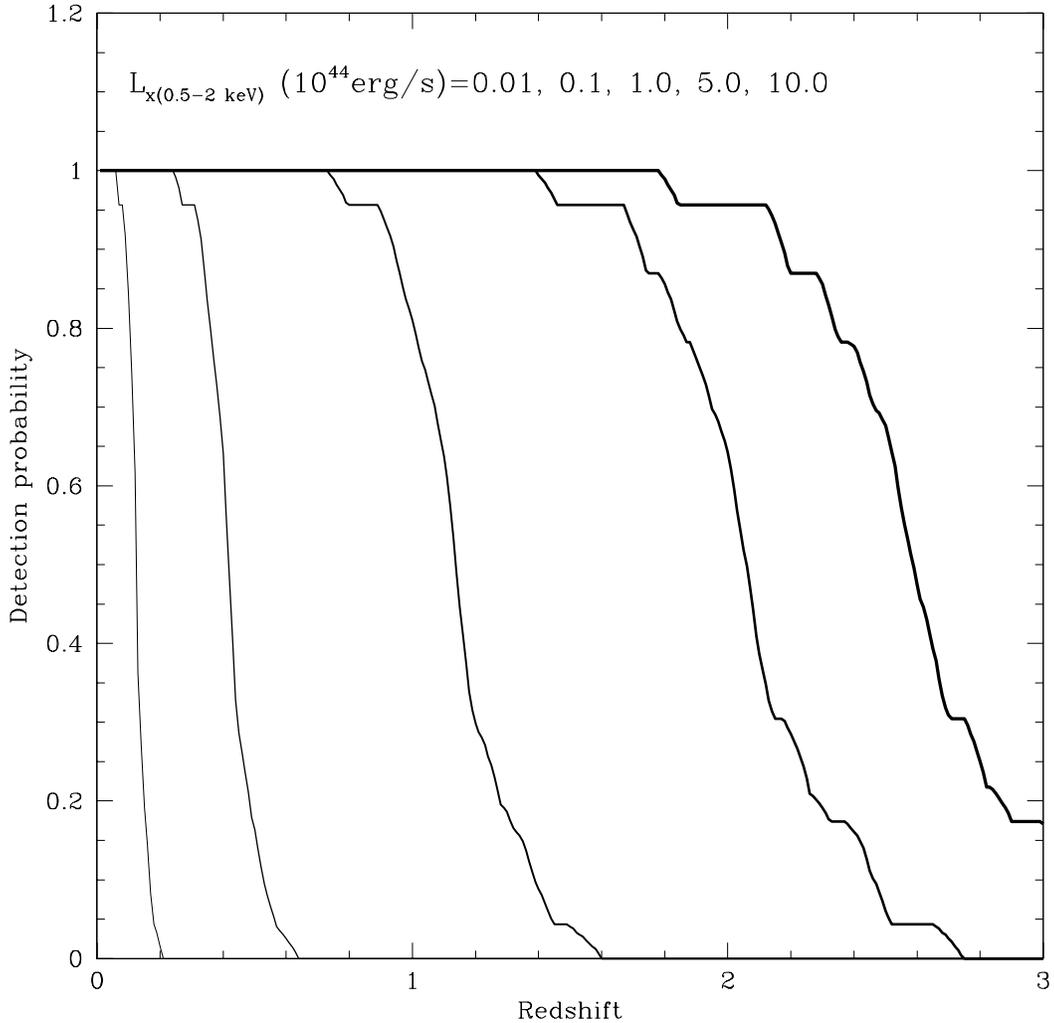} 
\caption[]{The X-ray detection
probability in the 0.5-2 keV band as a function of redshift for intrinsic
X-ray luminosities (0.5-2 keV) of 0.01, 0.1, 1.0, 5.0 and 10 $\times 10^{44}$erg
s$^{-1}$ ($H_0=75$, $q_0=0.5$) (in order of increasing line weight,
from left to right).  
The detection probability is plotted as
the net probability over all 23 ROXS fields, based on exposure time, background
surface brightness, and radial detection efficiency due to PSF variations 
for a cluster with a given X-ray luminosity and redshift and a standard
radial surface profile (Jones et al. 1998) and $\beta=2/3$ 
(Scharf et al. 1997; Rosati et al. 1995). \label{X-raysel} 
} \end{figure}

We convert the detection thresholds for the ROSAT fields into 
X-ray detection probabilities, as plotted in Figure~\ref{X-raysel}, as
a function of X-ray luminosity ($L_x$) and redshift, by
assuming a canonical
$L_x$-core radius relationship for the cluster surface brightness
profiles (e.g. Jones et al. 1998). The cluster surface brightness 
profiles are assumed to be standard
$\beta$-profiles with $\beta=2/3$. The detection probability is computed as
the net probability to detect a cluster with a given redshift and
$L_x$ in the entire ROXS coverage, based on exposure time, background
surface brightness, and radial detection efficiency due to PSF variations
discussed in the previous paragraph (Scharf et al. 1997; Rosati et al. 1995).

\subsection{X-ray-Optical Cross-Identification}
\begin{figure}
%\plotone{/data/atalanta2/mack/ROXS/postscript/Pancake_NE.ps}
\plotone{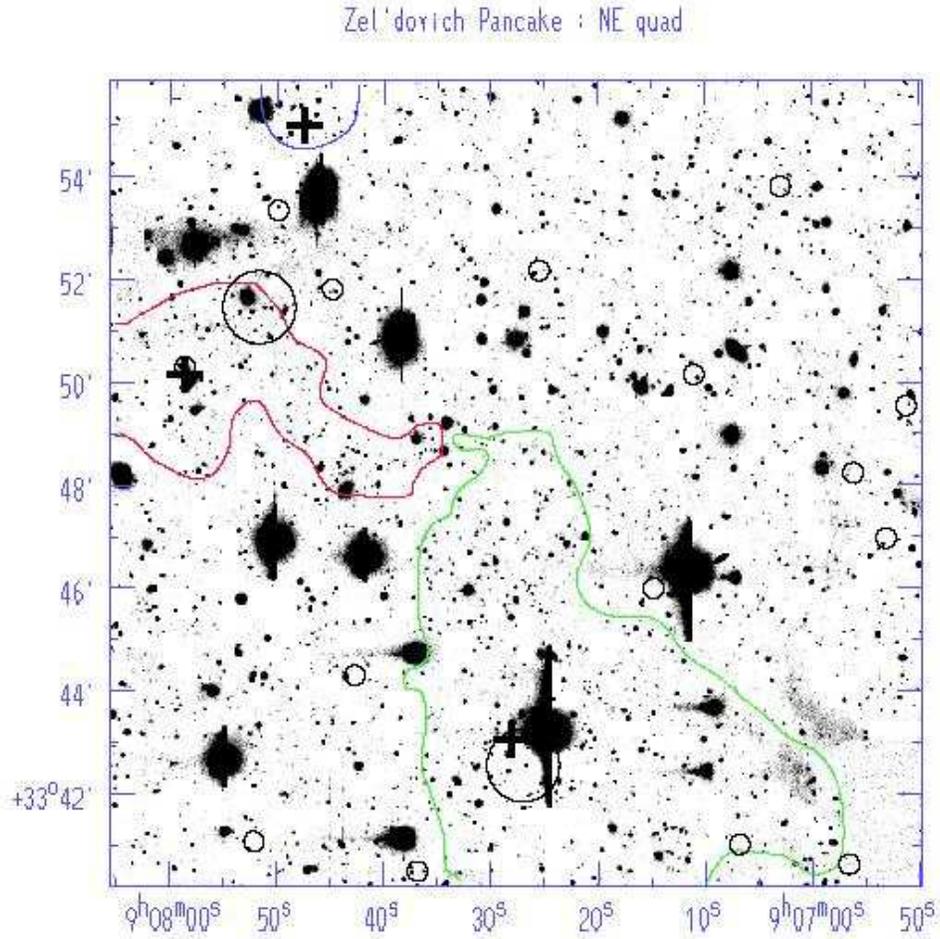}
\caption{The rich northeast quadrant of the Zel'dovich Pancake field. X-ray 
point source detections are plotted as small circles. The large circles are extended
X-ray sources. The optical cluster candidates are identified with crosses
and single contours. In this quadrant there are 3 optical cluster candidates,
two of which match an X-ray cluster candidate. \label{Pancake} }
\end{figure}

We cross-identified each X-ray candidate and I-band 
optical candidate by
visually inspecting the overlap of the X-ray surface
brightness contours and the optical contours of the detection thresholds for 
filtered optical maps at the
nominal detected estimated redshift. Optical and X-ray contour shapes did
not match in detail. Cluster candidates which overlapped
significantly were identified as cross-identification candidates. 
We have noted the separation of the nominal optical
and X-ray centroids in the tables. The centroid of the optical candidate
could be shifted from the true position because of regions
contaminated by bright stars, which must be masked from
the data. The centroids of both X-ray and optical candidates are likely only 
to be good to $\pm0.5'$; cluster diameters at any redshift are usually
$>2'$. Typical centroid separations are $<2'$. An example I-band image quadrant
with the X-ray and optical sources labelled is shown in Figure~\ref{Pancake}.

As reported in Paper I, of the 57 X-ray candidates, 43 would have
been visible on our optical frames (that is, not obscured by bright 
stars or scattered light). 
Of the 43, 29 were visually 
identified with potential cluster candidates with centroid 
separations of $\leq 3'$ or with significant contour overlap, 
of which 26 are 
very secure (separations $\leq 2'$ or visual confirmation of
X-ray and optical correspondence of somewhat complex structure). 
The other three less secure identifications 
are more questionable affiliations of
sprawling optical galaxy systems or filaments with a more compact X-ray cluster
candidate. One such X-ray candidate, a confirmed cluster at $z=0.167$, 
MS1201+283 or Abell 1455, does not have a formal optical counterpart,
but has up to 3 optical possible counterparts based on a visual inspection
of the field. The optical field around this cluster 
 was cut-up by bright stars, 
and the matched filter seems to have 
detected individual cluster candidates in the remnants.

The remaining 14 X-ray cluster candidates can be divided into
3 categories. Six candidates are
bona fide, optically faint candidates, three of which have
low-significance counterparts ($<3\sigma$) that are 
interesting because of their high estimated redshift, but they are not counted
as true X-ray/optical coincidences. 
The other two categories contain sources that have a lower
probability of a true cluster: double sources and sources with very uncertain
X-ray fluxes ($F_x/\sigma_{F_x} < 3$). There are two double sources without
optical counterparts, and six sources with uncertain X-ray fluxes.

Only two of the optically-blank X-ray sources have yet been classified
or confirmed at other wavelengths. One source (RXJ1256.9+4720) is
tentatively associated with 3C280, the target of the original observation.
Here we count this X-ray detection as an X-ray cluster
candidate but not an optical cross-identification. It was the original target; this correspondence
is the only example of the original target coinciding with a cluster 
candidate. The statistics are not
skewed by the inclusion of this one very interesting cluster candidate.
There was a $2\sigma$ detection at this position in the optical data.
The estimated optical redshift ($z=1.0$)
was so remarkably close to that of 3C280's spectroscopic redshift ($z=0.997$)
that we quote the matched-filter parameters of the optical candidate 
despite our extremely low confidence in its reality. 
 We note that the
3C280 cluster only has one redshift in-hand, and a preliminary investigation of the Chandra
image of this source shows X-ray structure associated with the radio source but 
no obvious cluster emission 
(M. Donahue, private communication.) The other candidate,
RXJ1120.0+2115, was
flagged as a double source; Rosati and his collaborators have found a 
cluster with $z=0.75$ (P. Rosati, private communication).

Since the spurious fraction typical of the X-ray
surveys is $\sim 10\%$ (Rosati et al. 1998; Vikhlinin et al. 1998), 
approximately 6 of the 57 sources
are expected to be  
false -- X-ray sources which are not really extended  
or are constellations  
of point-like X-ray sources. Some of these candidates may be bona fide,  
albeit optically-faint clusters. Truly high redshift
clusters are very faint in the I-band because of the 4000\AA~ break
in elliptical galaxy spectra. Near IR imaging is probably the
best tool to reveal the
presence of any high redshift clusters.

Eleven X-ray candidates have confirmed spectroscopic redshifts from
Rosati's followup of clusters in the RDCS (Rosati et al., 
in preparation). We list
the candidates in Table~\ref{z}. All X-ray spectroscopic 
confirmations which were
not obscured by bright stars  
had an optical counterpart with a detection confidence of
$\geq3\sigma$, except for the X-ray candidates associated
with 3C280 at $z=1$
and with RXJ1120.0+2115, which now has a confirmed redshift 
of 0.75 (Rosati, private communication).

The redshift  estimated by the matched filter algorithm has
proven to be surprisingly good for the clusters of galaxies, at least for the
X-ray selected clusters (Table~\ref{z}). For 7 out of 11 
extended X-ray sources with spectroscopic redshifts, 
the discrepancies between the photometric and the spectroscopic
redshift are less than $\Delta z=0.1$, finer than our redshift grid.
From Deeprange spectroscopic followup (Postman, private communication), 
the mean difference between
spectroscopic 
and estimated redshifts for candidates with $0.3 \leq z_{est} \leq 0.4$ 
is 0.04 with an rms scatter of 0.07, based on $\sim25$ 
clusters. The scatter 
for all Deeprange candidates out to $z \sim 1$ is closer to 0.15
(Postman, private communication).

Of the five X-ray cluster candidates contained in the fields with both V-band
and I-band data, all five were detected in both the V-band and
I-band.

\section{Cluster Candidate Properties}
In this section we discuss the cluster properties such as X-ray
luminosity,  estimated for
the cluster candidates in our samples. We find that while the cluster candidates
in the ROXS sample have X-ray and optical properties consistent
with the range of properties in other cluster samples, the
clusters are somewhat less X-ray luminous, optically poorer, and have much less prominent
sequences of red galaxies in color-magnitude diagrams than the typical
rich and massive clusters of galaxies found in all-sky surveys. This
sample thus may include cluster candidates that may be missed in pure X-ray,
optical, or color searches for clusters. We make what may seem a 
pedantic distinction here: These clusters may  be 
missed by  a search not because the  sample is incomplete, but 
because their properties fall outside the boundary of the sample
selection function. 

\subsection{X-ray Luminosities}

Here we show that the cluster luminosities are typical, with estimated
bolometric 
$L_x\sim 10^{43}-10^{44}~ \lum$, and that the upper limits are
all $L_x \ge 10^{42}~ \lum$. However, a couple of rich optical 
candidates are found with very low X-ray upper limits.

\begin{figure}
%\plotone{lx_dist.eps}
\plotone{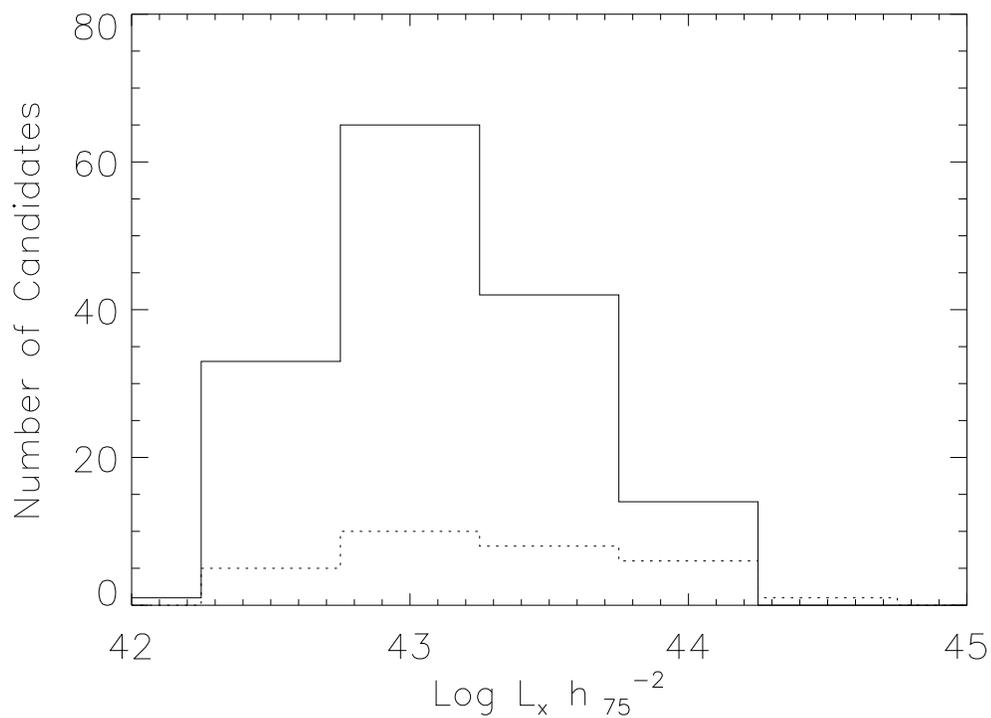}
\caption[]{The black line shows the distribution of the upper limits to the 
bolometric X-ray luminosities of the cluster candidates which were
not detected in the X-ray. The dashed
lines show the distribution of the optical cluster candidates which were
cross-identified with X-ray sources. The lower luminosity candidates
and optical candidates with low luminosity upper limits lie at 
low estimated redshifts $z\sim0.2-0.3$.\label{lx_dist}}
\end{figure}

\begin{figure}
%\plottwo{/data/atalanta2/mack/ROXS/postscript/OC6_1024+4707_bw.ps}
%        {/data/atalanta2/mack/ROXS/postscript/OC1_1118+2107_bw.ps}
\plottwo{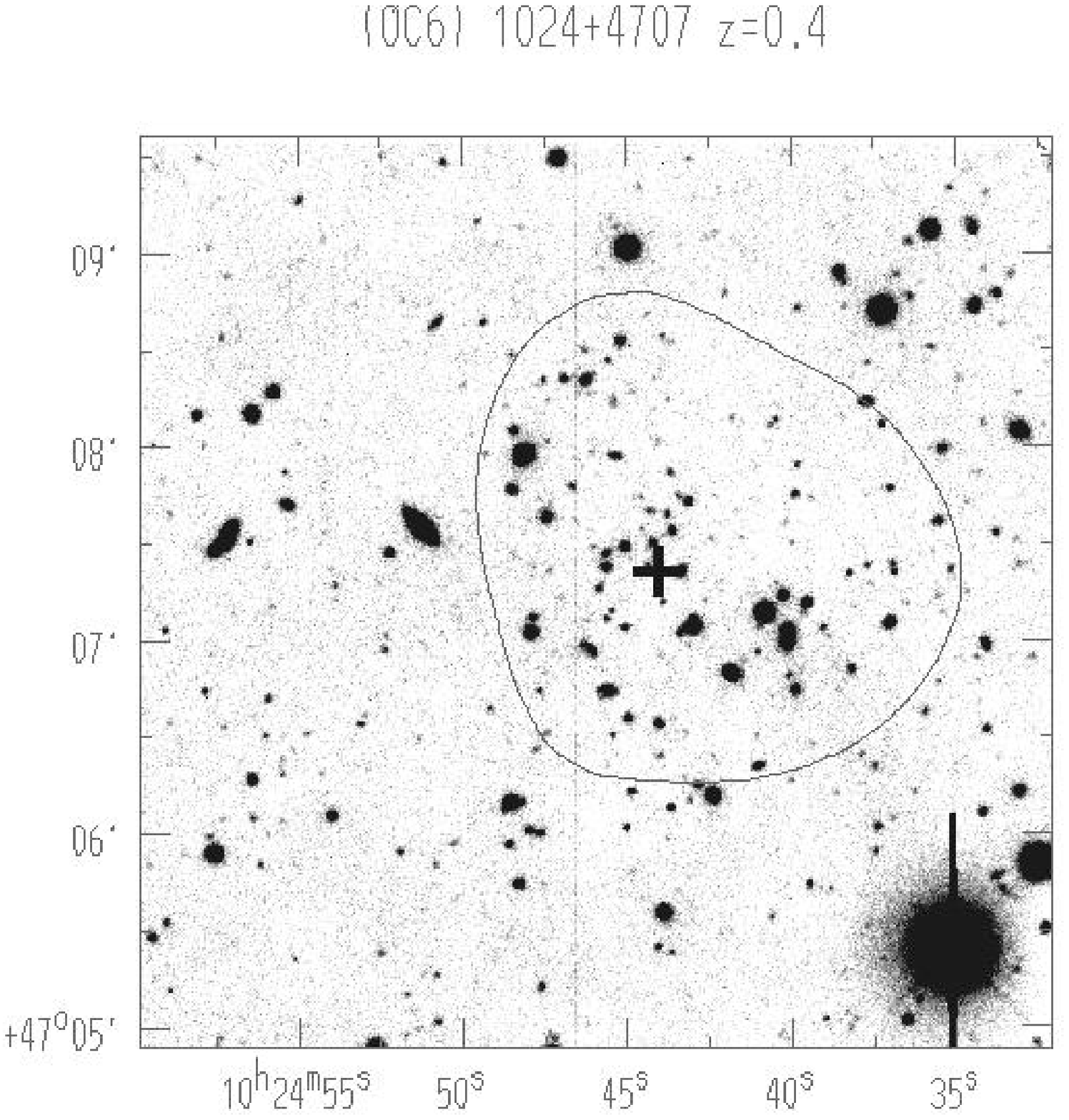}{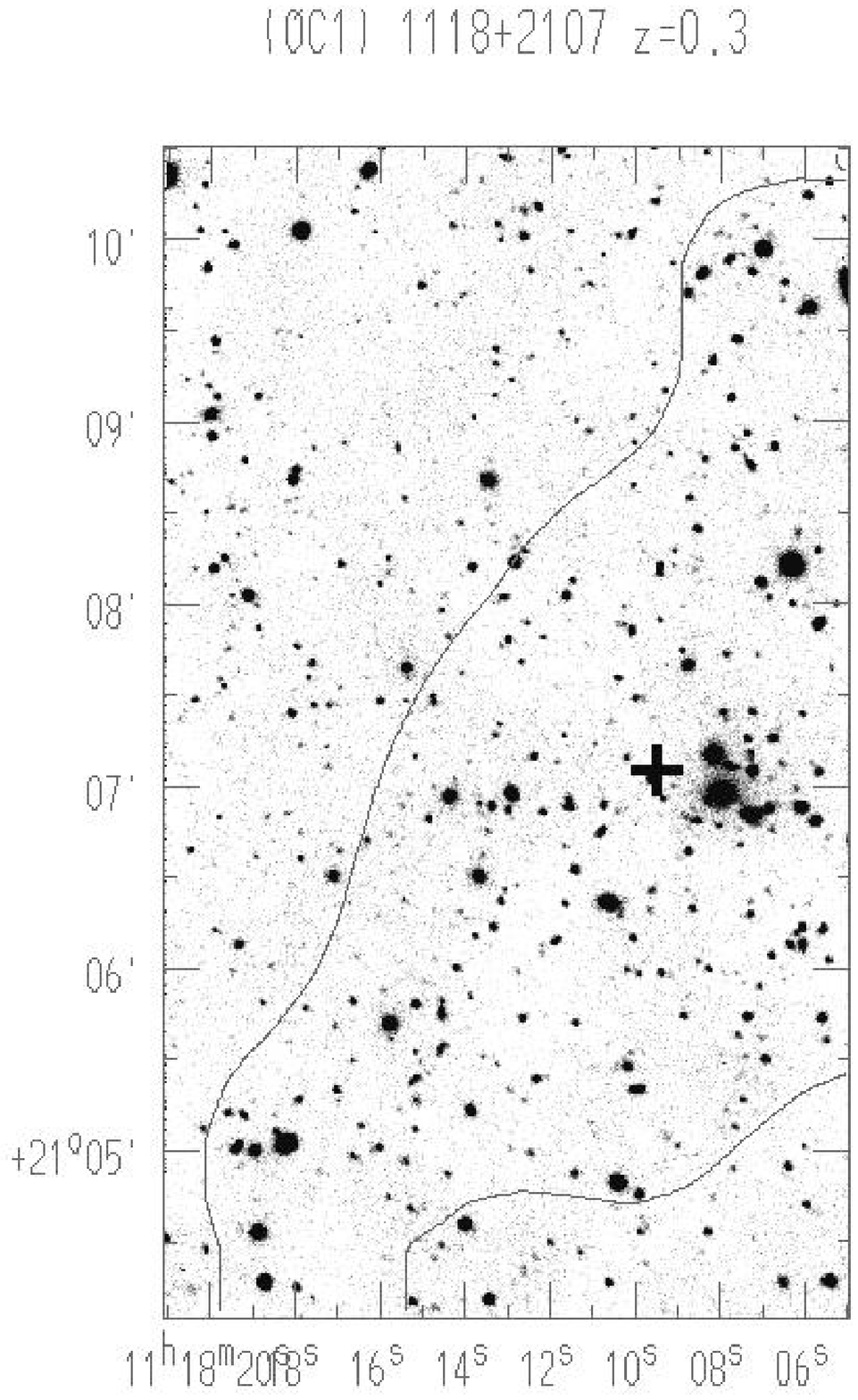}
\caption[]{Two optical candidates with high detection significance from
the matched filter algorithm ($>4\sigma$) but no associated X-ray flux, 
OC6 1024+4707 from the 1024+4707 field ($z=0.4$) and   
OC1 1118+2107 in the 1116+215 field ($z=0.3$).
\label{lowXhighO}} 
\end{figure}

The X-ray luminosity for each cluster candidate with an optical
counterpart was computed assuming the matched-filter estimate
for the cluster redshift, the X-ray flux from the wavelet detection
algorithm, and a self-consistent bolometric correction commensurate  
with the X-ray temperature $T_x$ implied by the $L_x-T_x$ relation
from Markevitch (1998). The $L_x-T_x$ relation 
is nearly constant with redshift (Donahue et al. 1999; Borgani
et al. 2001). 
The optical cluster candidates which were not detected in the
X-ray band all have X-ray upper limits from a $r=1'$ aperture
(see \S~\ref{xdetupplim}), ranging
from $2-10 \times 10^{-14} \flux$ (the bolometric corrections for
the upper limits were computed assuming an 
X-ray temperature of $kT=4$ keV and the estimated redshifts).
A mean $N_H=2.33\times 10^{20}$cm$^{-2} $ column density for
high Galactic latitudes was assumed for all luminosities and upper limits. 
In Figure~\ref{lx_dist}, we have plotted the 
distribution of estimated X-ray luminosities
of cross-identified clusters and the upper limits of the 
optical candidates without X-ray counterparts. 
Note that none of the clusters
are extremely luminous ($>6\times10^{44}h_{75}^{-2}\lum$) and that 
none of the X-ray candidates has an X-ray luminosity 
upper limit significantly below a $2 \times 10^{42} h_{75}^{-2} \lum$. Since
this limit is approximately the luminosity of poor clusters, 
our X-ray data are 
not quite sensitive enough to rule out the 
presence of  
an X-ray group or low-luminosity X-ray cluster.  
However, a significant number of the optical candidates (54) have flux
limits and estimated redshifts which imply X-ray cluster luminosities
which are between $10^{42-43} h_{75}^{-2} \lum$, typical of luminosities 
of poor clusters of galaxies and even groups (Figure~\ref{lx_dist}).
Two such cluster candidates, with redshifts of 0.4 and 0.3 respectively, 
have $\Lambda_{cl} \geq 50$ (relatively rich) and upper limits on 
$L_x<10^{43} h_{75}^{-2} \lum$, clusters 
(OC1) 1118+2107 and (OC6) 1024+4707 (See Figure~\ref{lowXhighO} for
the I-band images of these candidates.) 
The optical detection significance for each of these
two clusters was $\sim5\sigma$. If these clusters turn out to be
X-ray faint ($L_x$ significantly
lower than that predicted by the $L_x-T_x$ or $L_x - \sigma_v$ relations
for clusters), that would argue that X-ray surveys could miss some
relatively massive systems. Such a hypothesis would be straightforward
to test with XMM or Chandra X-ray imaging observations, alongside 
ground-based galaxy spectra to confirm the redshift and establish
a velocity dispersion.

%\begin{figure}
%\plotone{flux_z.eps}
%\caption[]{Plot of the bolometric X-ray flux upper limits
%and detections for the cluster candidates as a function of estimated
%redshift. Lines of constant bolometric X-ray luminosity 
%are drawn for $10^{45}$, $10^{44}$, $10^{43}$, and $10^{42} h^{-2} \lum$
%respectively. Note that none of the limits fall below the
%$10^{42} h^{-2}~\lum$. \label{X-rayflux_z}} 
%\end{figure}

\subsection{X-ray Luminosity - Optical Richness}

Here we show that the X-ray luminosities and optical
richness of the ROXS clusters are in the expected range and
ratios typical of clusters of galaxies; however, as expected, 
the ROXS clusters are on average fainter in X-rays than
those clusters selected from a very large area X-ray survey
at much shallower X-ray flux depths.

In Paper I (Donahue et al. 2001) we evaluated the relationship 
between the X-ray luminosity $L_x$ and the estimate of optical
luminosity $\Lambda_{cl}$ and while we found a marginally 
statistically significant correlation
between these two properties, independent of estimated 
redshift, we also demonstrated that this relationship has 
significant scatter beyond that indicated by the estimated 
observational uncertainties.

To further investigate the relationship between the
X-ray and optical properties of the ROXS we have compared the X-ray
luminosities or upper limits and optical richnesses with those obtained
from the Abell clusters composing the X-ray Brightest
Abell-type Cluster Survey (XBACS) ROSAT survey of Ebeling et al
(1996).  The XBACS survey is composed of the Abell clusters detected in 
the northern portion 
of the Rosat All-Sky Survey, and therefore contains many high luminosity
clusters of galaxies. Since such clusters are extremely rare, they are not
expected in much smaller surveys such as ours. However, the X-ray 
luminosities  spanned in ROXS are also spanned in the XBACS
coverage, albeit for lower-redshift clusters. 

We have combined the 283 XBACS X-ray luminosities with the
Abell richness number counts $N_{R}$ from the Abell/ACO catalogue.
For comparison to the ROXS, we have converted the 
matched filter richness
parameter $\Lambda_{cl}$ (approximately equivalent to the number of $L^*$
galaxies in the system (P96)) to Abell richness
counts $N_R$. However, as demonstrated in
P96  the scatter between $\Lambda_{cl}$ and $N_{R}$,
(where $N_{R}$ is defined as per Abell's galaxy number counts
within a $1.0$h$^{-1}$Mpc radius rather than Abell's 1.5h$^{-1}$Mpc) is
large, albeit with a positive correlation. For our 
purposes we use a relatively crude 
conversion from $\Lambda_{cl}$ to Abell
counts; $N_{R}^{Abell}=(\Lambda_{cl}\pm 23)/0.72$,
where the factor 0.72 converts galaxy counts from within a 
$1.0$h$^{-1}$Mpc radius to a 1.5h$^{-1}$Mpc radius (P96). 

\begin{figure}
%\plotone{rox_xbacs75.ps}
\plotone{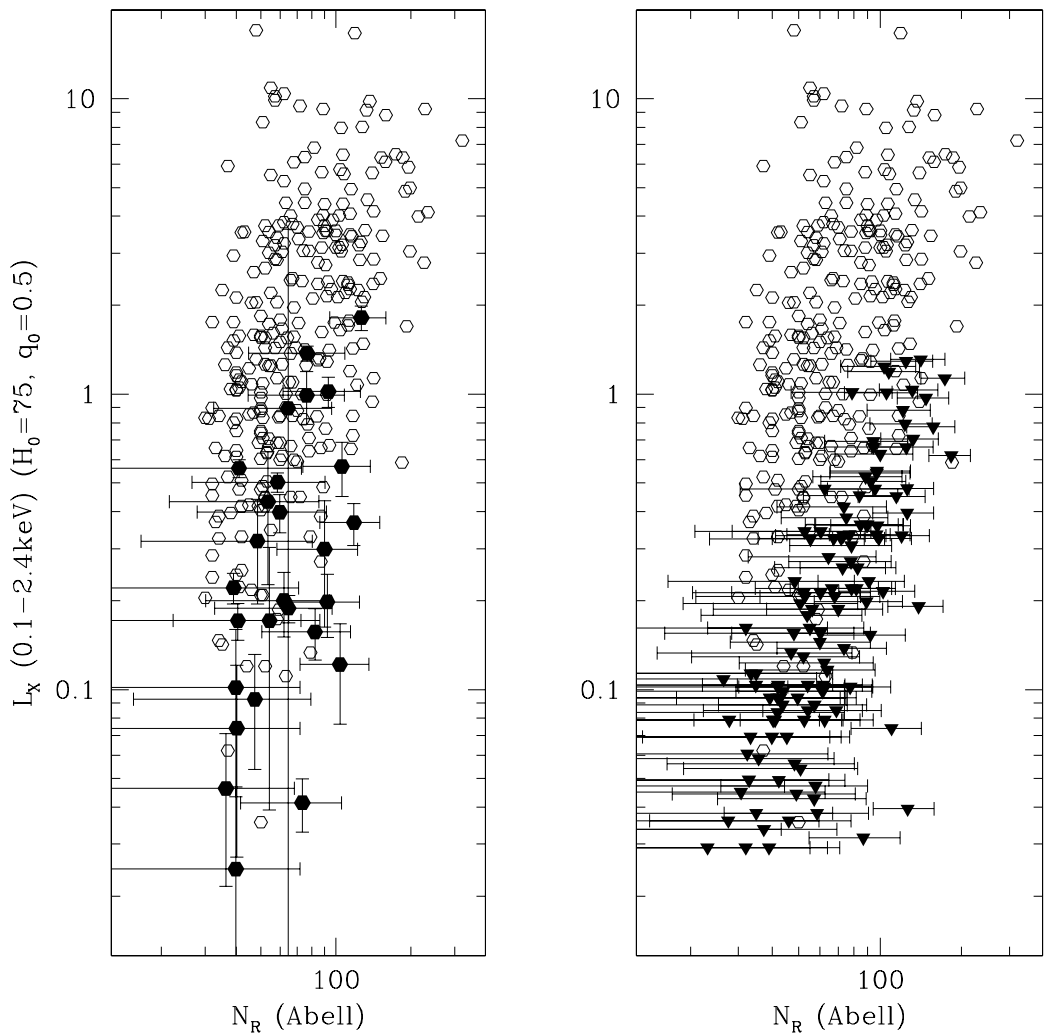}
\caption[]{X-ray luminosities vs. optical richness for XBACS clusters 
(open circles, both panels) and
ROXS optical cluster candidates with X-ray counterparts (filled circles, 
left panel)
and the upper limits on the X-ray luminosity for ROXS optical candidates
without X-ray counterparts (filled triangles, right panel). The size of the error bars
for the ROXS optical richness reflects the uncertainty in the conversion
from $\Lambda_{cl}$, the matched filter optical luminosity parameter,
to $N_R$, optical richness. 
\label{LXrich} }
\end{figure}

In Figure~\ref{LXrich} panel (a) the X-ray luminosity (0.1-2.4 keV, 
converted to $H_{0}=75$ km s$^{-1}$ Mpc$^{-1}$, $q_{0}=0.5$) 
is plotted against $N_{R}^{Abell}$ for all XBACS clusters (open circles) 
and for all ROXS optical candidates with X-ray counterparts 
(filled circles). The
uncertainties in the X-ray luminosities of ROXS clusters  ranges from $\sim
7\%$ to $\sim 80\%$. 
In Figure~\ref{LXrich} panel (b) the XBACS clusters are again plotted, but now with
the upper limit on X-ray luminosities for all non-X-ray detected ROXS cluster
candidates. 
The resulting relationship between Abell richness number counts
and X-ray luminosity for ROXS plotted in Figure~\ref{LXrich} 
has large scatter, a result consistent with
our conclusions from Paper I (Donahue et al. 2001) and also 
with results from Borgani \& Guzzo (2001), which we will discuss later. 

It is also clear for both the X-ray detected and non-detected objects that
the ROXS criteria tend to include more X-ray-poor systems  
than the Abell/ACO catalogue intersection with the XBACS.
The XBACS half-sky survey has sufficient sky coverage and
depth to find more luminous clusters, but it is 
 not deep enough to detect proportionally 
as many X-ray faint clusters as are
found in the ROXS survey.
As we will also show in our red sequence analysis in
\S~\ref{redseq}, the cluster candidates
that we sample here are all likely to be somewhat lower in mass than
their massive, X-ray luminous cousins found in surveys of larger sky
area. The general form of the $L_{X}-N_{R}$
relationship of the ROXS cluster candidates is qualitatively similar to 
that found for the low-$L_x$ clusters in the XBACS,

\subsection{Cluster-Point Source Correlation} 

One goal of our survey was to see whether a criterion of
requiring cluster X-ray candidates to have significant extent
in the ROSAT PSPC image filtered out clusters with X-ray 
emission too compact to be unambiguously resolved. Such a criterion
could screen out high redshift sources or sources with prominent
cooling flows. Since we were imaging complete fields we could
investigate whether X-ray point sources, ignored
by some cluster surveys, could actually be high-redshift or
compact clusters.
 
Of the 142 optical candidate clusters with a detection
confidence of $>3\sigma$, 
27 have extended X-ray counterparts (one has two 
counterparts). Up to twenty-nine additional 
optical candidates have possibly 
interesting X-ray point sources either near the central core 
of galaxies (within 1'-2') or
near a possible cluster member. When an X-ray point source is near an
optical candidate, we have
indicated so in the Comments column of Table~\ref{Optical}.
For example, in the field of MKN~78, OC2 has a point source very close
to it, RXJ0741.7+6525, an active galaxy with a redshift of 1.65 
(RIXOS F234\_001, with Mg II and CIII emission seen by 
Puchnarewicz et al. 1997). The estimated
redshift of the cluster, however, is 0.4. In the field of
10214+4724, OC10 ($z=0.3$) is a compact optical candidate that is
exactly coincident with the point source RXJ1025.4+4716; such 
exact alignments are worth investigating. 

However, most of the point sources with positional coincidence with
optical cluster candidates were not exact; coincidental alignments 
are likely, given the size of a typical cluster candidate.
The number of coincidences between point sources and optical cluster
candidates in our survey can be estimated by assuming that a typical
optical cluster candidate has a radius of $\sim100"$, and that the
X-ray point source density at the 0.2-2.0 keV flux limit of $F_x \sim
10^{-14} \flux$ is 100 sources deg$^{-2}$ (Rosati et al. 1998). 
With 142 optical candidates,
the estimated number of chance coincidences is 34, comparable to the number
that we identify. Therefore we do not detect a significant excess
of X-ray point sources near optical cluster candidates.

In order to see whether the correspondence with point sources
increased when we applied our judgement 
as to the reality of a cluster candidate,
we visually inspected each optical
candidate and assessed a subjective believability index to it, 
of ``probable'', ``blend'', or ``unlikely''. 
This subjective assessment showed 
that many of the optical candidates are likely to be blends with
other optical candidates at different redshifts. Lacking a purely
objective means of sorting out blends, we have listed all
blend candidates with notations in the Comments column (Table~\ref{Optical}). 
Some 57 out of the 142 optical candidates without X-ray counterparts 
passed this subjectivity test; 
17 of these may have associated point sources. The
fraction of optical candidates without extended X-ray counterparts but
with possible X-ray point source counterparts did not increase 
significantly with this subjective assessment. Therefore, we 
suspect that most of these associations are likely to be 
random projections of background AGN with the optical candidates.

\begin{figure}
%\plottwo{red_hist.eps}{lambda_hist.eps}
\plottwo{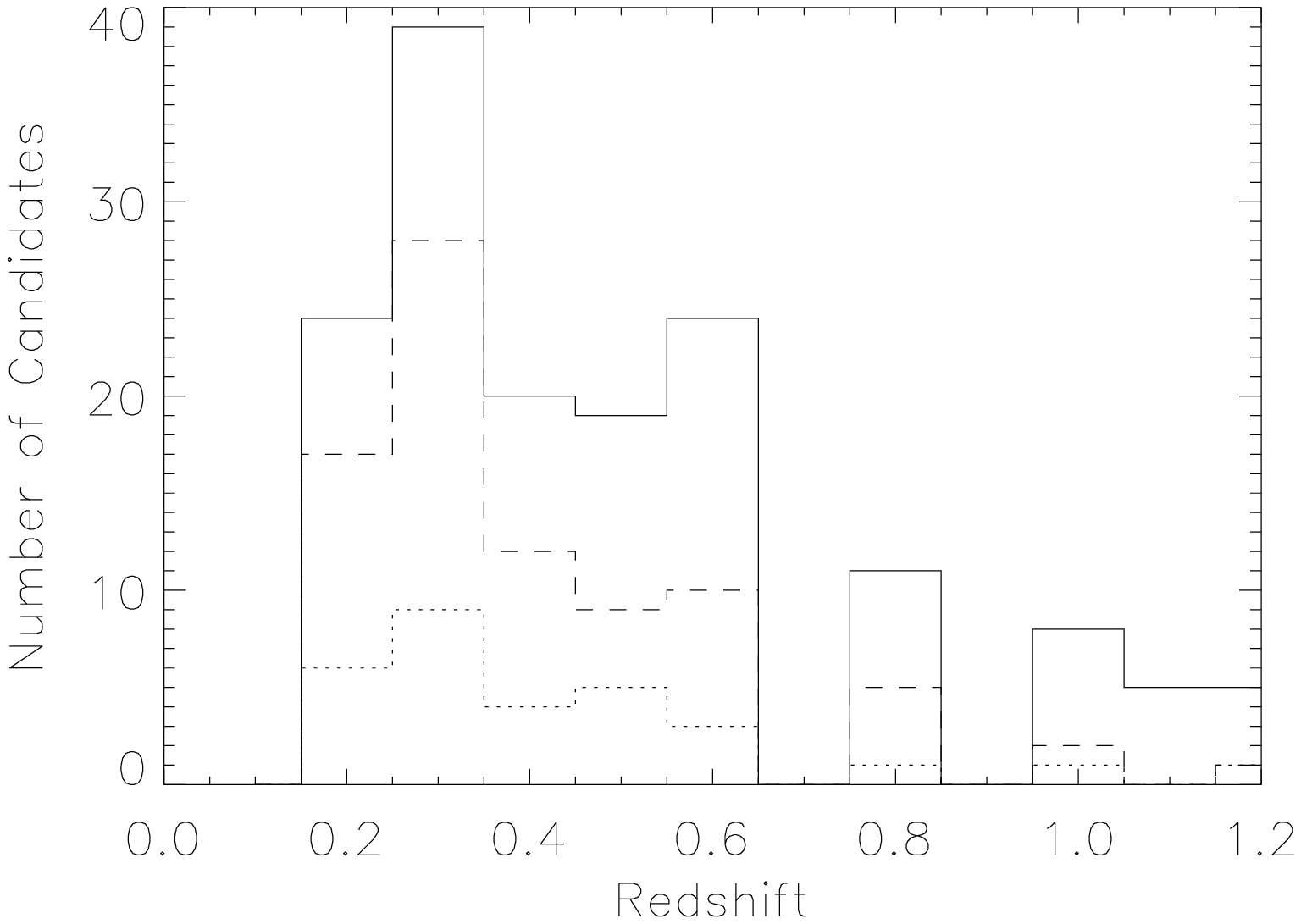}{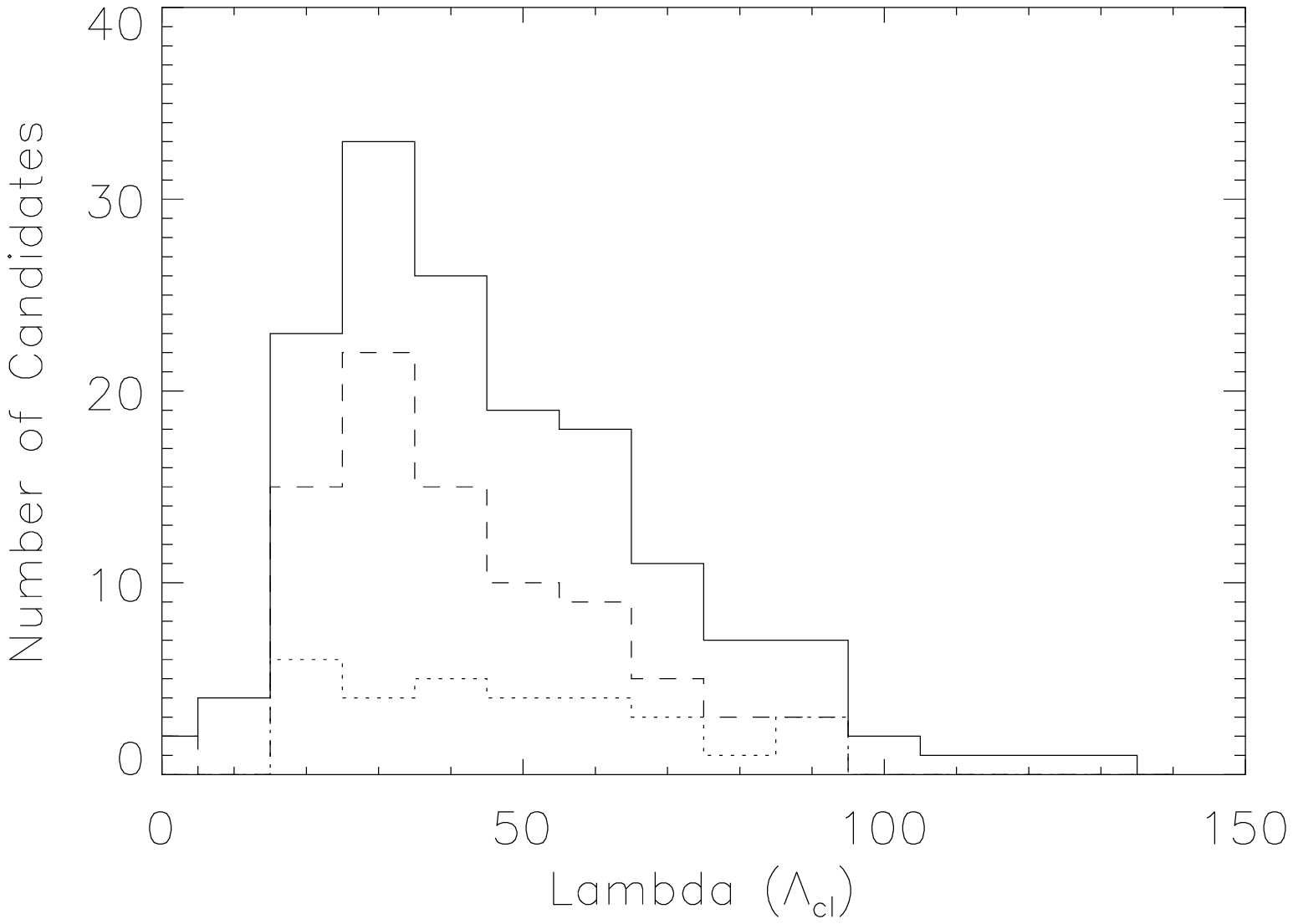}
\caption[]{The redshift (left) and $\Lambda_{cl}$ (right) histograms of all 
155 optically selected cluster candidates, of which 3 have $<2.9\sigma$. 
The dotted line indicates those clusters with
X-ray counterparts. The dashed line indicates the
histogram of candidates with X-ray
counterparts and optical candidates deemed ``probable'' in a subjective
assessment.  \label{redshifts}}
\end{figure}

\begin{figure}
%\plotone{sigma_dist.eps}
\plotone{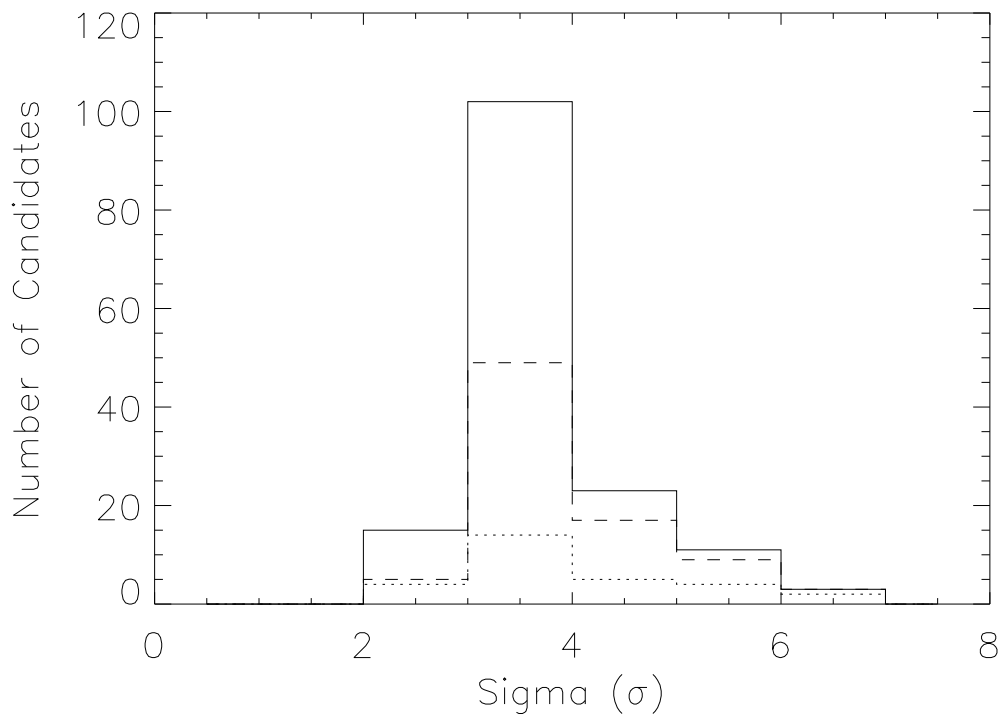}
\caption[]{The distribution of the optical detection confidence in
units of $\sigma$ for all 
of the optical candidates is plotted in a solid line. The same
distribution for the X-ray/optical cross-identifications is plotted
with a dotted line. Those clusters which were additionally identified
as ``likely'' clusters are shown in a dashed line. Note that
the majority of the X-ray/optical clusters lie in the bin with
detection significance 3-4$\sigma$. \label{sigdist}}
\end{figure}

We note the fraction of optical candidates which have X-ray
counterparts or which are deemed believable in a subjective
assessment is a strong function of detection confidence ($\sigma$)
and estimated redshift. The percentage of X-ray clusters plus ``probable''
optical clusters is 70\% of the total at $z=0.2$, dropping to 
45\% at $z=0.8$, and $0-20\%$ at $z=1.0-1.2$ (Figure~\ref{redshifts}a
and ~\ref{sigdist}). 
The same fraction as a function of $\Lambda_{cl}$ (Figure~\ref{redshifts}b) 
does not vary as strongly as it does with detection confidence ($\sigma$) and 
estimated redshift, presumably since $\Lambda_{cl}$ is a 
measure of cluster richness (and richer clusters are in general
more powerful X-ray sources and may subjectively
look more ``probable'') but candidates with higher $\Lambda_{cl}$ also tend
to have higher estimated redshifts (and thus are harder to detect by
X-ray methods and are less likely to be assessed as ``probable'' in 
a subjective review of the optical data.) 
However, only a few optically selected 
clusters with $\Lambda_{cl}<30$ were
considered ``probable'' (Figure~\ref{redshifts}b). 
Our statistical analysis in Paper I was 
only for cluster candidates with $\Lambda_{cl}>30$. 
Only one cluster
was judged to be ``probable'' with $z\geq1$ and detection
confidence of $>3\sigma$. The 
presence or absence of this high redshift candidate does not affect
our analysis in this paper.

\subsection{Redshift, $\Lambda_{cl}$, and Detection Confidence Distributions \label{redlam}}

Here we provide an analysis of the detection limitations of
our dual band surveys. We compare the properties of the X-ray detected cluster
candidates with those which had no X-ray counterparts; we find
that the optical and X-ray surveys were well-matched in terms of
depth and sensitivity. 

As described previously and in Paper I, each optical cluster 
candidate was assigned an estimated redshift and a $\Lambda_{cl}$ 
from the best-fit luminosity function by the matched filter algorithm.  
The detection efficiency of the matched filter, 
when confronted with the limitations placed by the depth of the I-band
images and the spectrum of cluster galaxies, 
drops dramatically at $z>0.5-0.6$ for clusters with $\Lambda_{cl} \lesssim 50$.
Optical clusters with $\Lambda_{cl}$ of 100 or more are detectable
beyond $z\sim1$ (Figure~\ref{optsel}). This detection efficiency can
be demonstrated in the distribution of estimated redshifts for our sample.
The redshift distribution of all of the candidates is 
shown in Figure~\ref{redshifts}.

\begin{figure}
%\plotone{lambda_corr_hist.eps}
\plotone{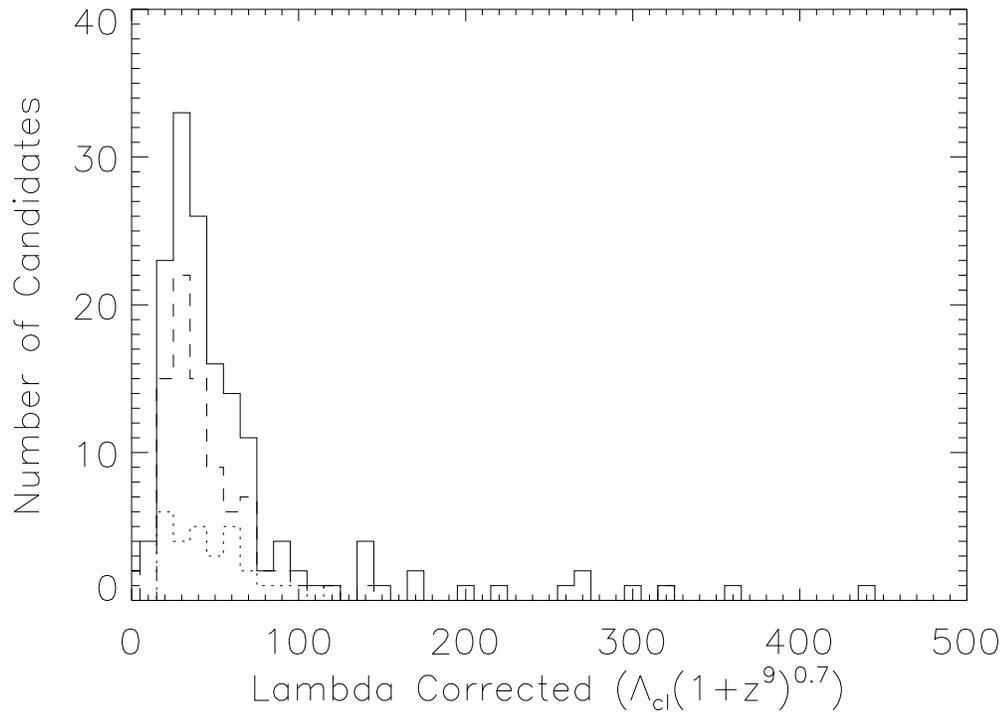}
\caption[]{The distribution of aperture-corrected $\Lambda_{cl} (1+z^9)^{0.7}$
(correction from P96), with the same line codes
as Figure~\ref{redshifts}. The majority of the optically selected cluster
candidates with corrected $\Lambda_{cl}>100$ have estimated $z>1$, and
are not probable clusters, according to our subjective assessment.
\label{lambda_corrs}}
\end{figure}

The estimated richness parameter from the matched filter
algorithm, $\Lambda_{cl}$, may
be corrected for aperture effects by multiplying the original
$\Lambda_{cl}$ by $(1+z^9)^{0.7}$ (P96). We do not
use this correction since it does not significantly 
correct the $\Lambda_{cl}$ for cluster candidates at $z<1$.
For comparison, the distribution of cluster
candidates with $\Lambda_{cl}$ is plotted in Figure~\ref{redshifts}b. The
distribution of redshift-corrected $\Lambda_{cl}$ is plotted in
Figure~\ref{lambda_corrs}.

\begin{figure}
%\plotone{siglambda.eps}
\plotone{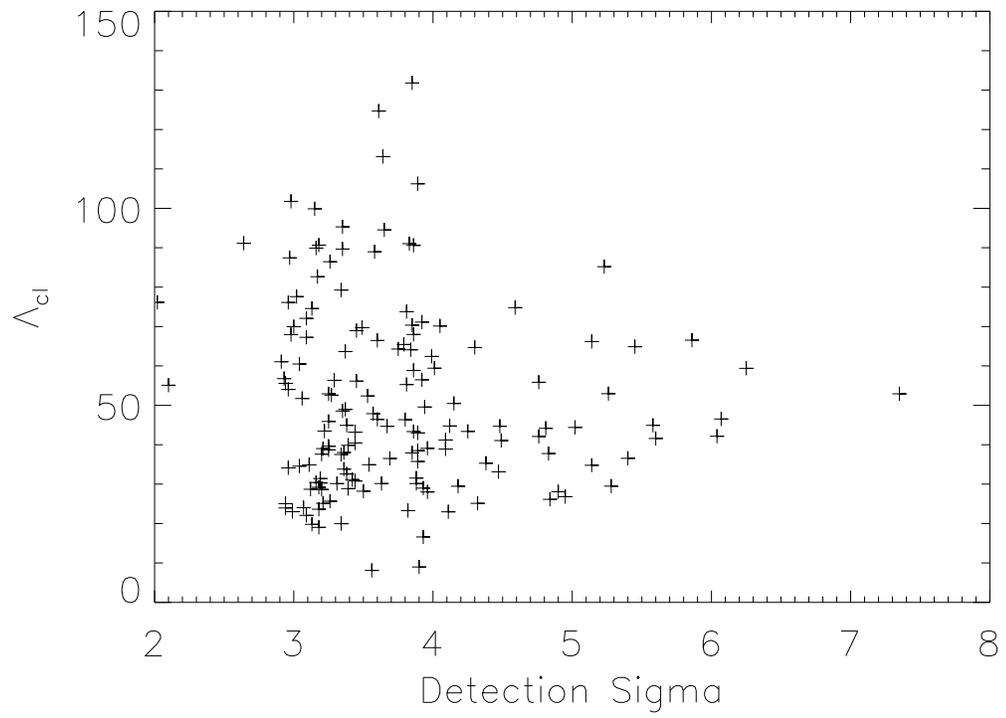}
\caption[]{The optical detection confidence is plotted here against $\Lambda_{cl}$,
an estimate of the cluster's optical luminosity. The two quantities
are not strongly correlated.\label{siglambda}}
\end{figure}

The relative depth of the X-ray and matched filter selection 
techniques and respective flux sensitivities can be assessed by looking
at the distribution of the optical detection significance from the matched filter 
algorithim in Figure~\ref{sigdist}. If the optical detection
threshold were not 3 but 4$\sigma$, nearly half of the optical/X-ray
detections would be excluded. Therefore, in terms of depth, the 
detection constraints of the two surveys are compatible. In 
Figure~\ref{siglambda}, we plot the relationship between the
detection significance and the estimated $\Lambda_{cl}$. This Figure
demonstrates that $\Lambda_{cl}$ and detection significance 
$\sigma$ are not correlated. 

%[FOR CO-AUTHORS ONLY (THIS PARAGRAPH AND ACCOMPANYING FIGURE
%ARE NOT INTENDED FOR THE SUBMITTED VERSION OF THE PAPER.) 
%Figure~\ref{detfrac} plots for each field the fraction
%of X-ray sources with optical counterparts as a function of
%X-ray exposure time. The matched filter algorithm finds more or
%less a constant fraction of the X-ray sources within the optical
%coverage as function of X-ray exposure time (or equivalently,
%increasing sensitivity in the X-ray bandpass.) ]

In order to see how much the depth of the X-ray exposure affected
the fraction of clusters detected in the X-ray (or in the optical), 
we divided the ROSAT
exposures into two subsamples 
with roughly equal number of fields in each, one with 
$t_{exp} \leq 25,000$ seconds and the other with $t_{exp} > 25,000$ seconds.
The fraction of optical sources which are detected in the
X-ray was statistically equal in both samples, 
at $0.21\pm0.05$ and $0.24\pm0.07$
respectively. The fractions of X-ray cluster candidates within the optical
coverage (not obscured by bright stars or scattered light) 
which were also detected in the optical in those two samples
are  $0.75\pm0.23$ and $0.68\pm0.25$ respectively. The uncertainties
reflect only the Poisson statistics ( $\sqrt{N}$ uncertainties in
each sample). We conclude that 
the exposure length of the X-ray observation did not 
make a statistically significant effect in the fractions  
of cross-identified sources, at least for the limited
dynamic range of exposure
times in our survey coverage.

Even for the 
deepest X-ray images, the optical survey methods failed to find  
fewer than 1-2 X-ray cluster candidates in a field, a number
that could be further reduced by spurious 
or false X-ray clusters ($\sim10\%$ of the X-ray cluster candidates in our
sample are probably spurious, being overlapping point sources). We
suggest that the matched filter algorithm applied to 
images with flux sensitivies of $m_I \sim 23$ is fairly well-matched
to discovering the $z \leq 1$ X-ray clusters detected by   
wavelet analysis of Rosat PSPC exposures
between 8,000 and 60,000 seconds.

\subsection{Galaxy Color Distributions - Red Sequence~\label{redseq}}

The identification of a red sequence of old, evolved galaxies in a color-
magnitude diagram has been proposed as another method 
to identify cluster
candidates. The application of this method has resulted in
the detection of high redshift clusters (e.g. Gladders \& Yee 2000;
Stanford et al. 1997).
Here we examine the clusters detected in both
I- and V-band to see how many of these candidates show a distinct
red sequence and whether the clusters with X-ray counterparts
showed strong red sequences. We reserve a direct application of
the red sequence selection of clusters for a future paper.
Utilizing the V and I band imaging, we determined 
the limiting magnitude for each field 
and then matched galaxies by their sky positions in
the FOCAS catalogues. We then extracted magnitudes for
each galaxy, measured in a fixed 4.7" radius aperture, and we calculated 
the V-I colors. 

If there was a corresponding X-ray counterpart, we used the X-ray 
contours to define the boundary of the cluster, and we plotted 
colors for galaxies in this region. If no matching X-ray cluster 
existed, we compared the cluster centroid positions in V and I 
to determine the optical cluster location. We also compared the $3\sigma$ 
optical detection contours, since this threshold was the 
criteria for optical cluster detection. Often, however, this 
threshold encompassed too large an area ($\sim15$ to 50 square
arcminutes). When available, we also compared the areas defined
by the 4 and $5\sigma$ 
contours to improve the location of the core of
the optical cluster. Generally galaxy colors were measured in a sky area of 
$\sim5$ square arcminutes or smaller, depending on the redshift of the 
optical cluster candidate.

\begin{figure}
%\plotone{/data/atalanta2/mack/ROXS/megan_VI2/plot1.ps}
\plotone{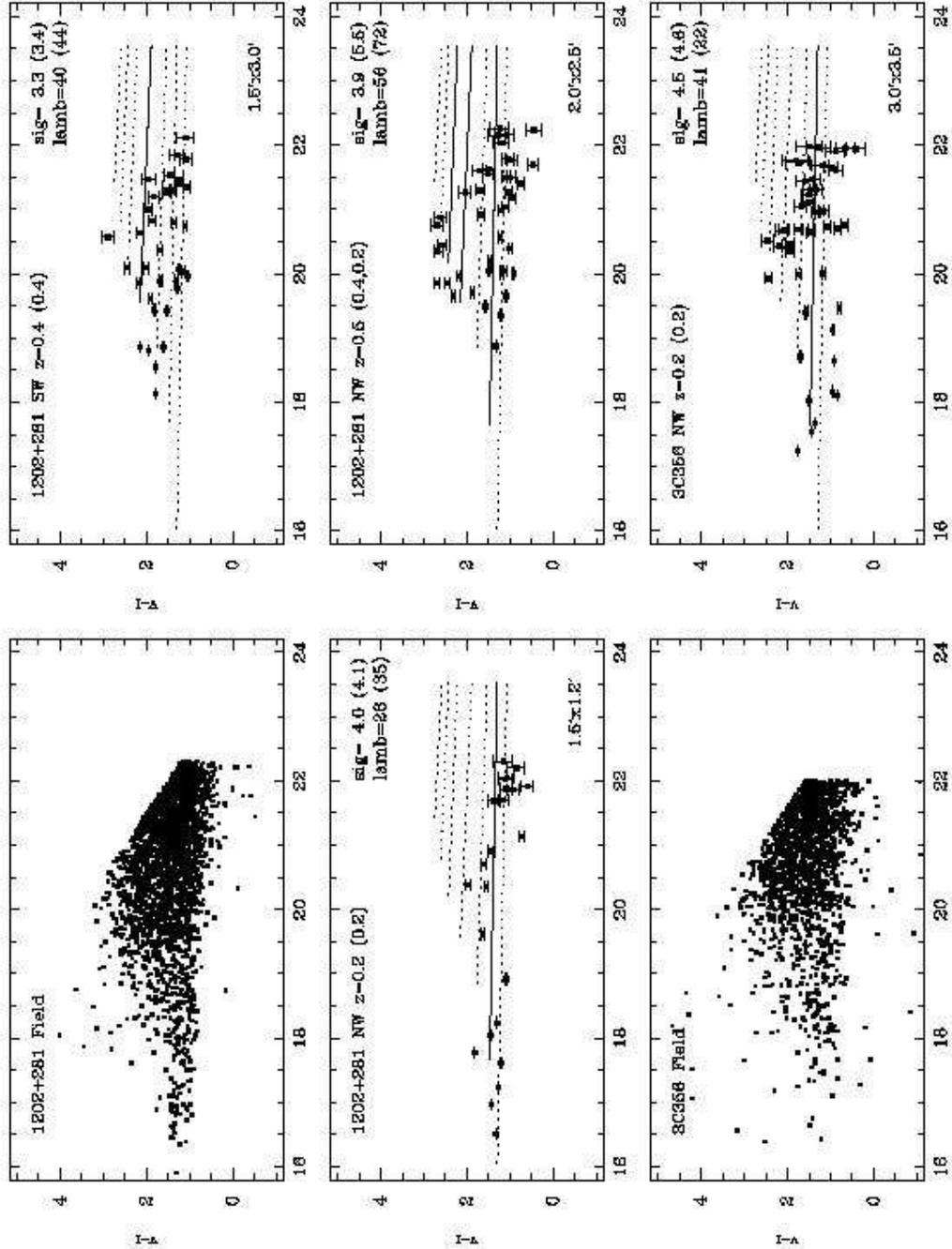}
\caption[]{V-I vs. I distributions for mutual V and I cluster candidates 
in 1202+281 ROSAT field and for 1 of the 2 candidates in the 3C356 ROSAT field. 
The solid line(s) plot  
the sequence at the estimated redshift(s) of the cluster candidates.  
The dashed lines show the red sequences predicted
for $z=0.1-0.7$ in increments of $\Delta z=0.1$. The upper left label indicates
the field, the quadrant, the estimated redshift from the I(V). The upper right
label lists the detection confidence (sig) in I(V), and $\Lambda_{cl}$ (lamb) 
from the I(V) data. Lower right label lists the cluster core size.\label{vi1}} 
\end{figure}

\begin{figure}
%\plotone{/data/atalanta2/mack/ROXS/megan_VI2/plot2.ps}
\plotone{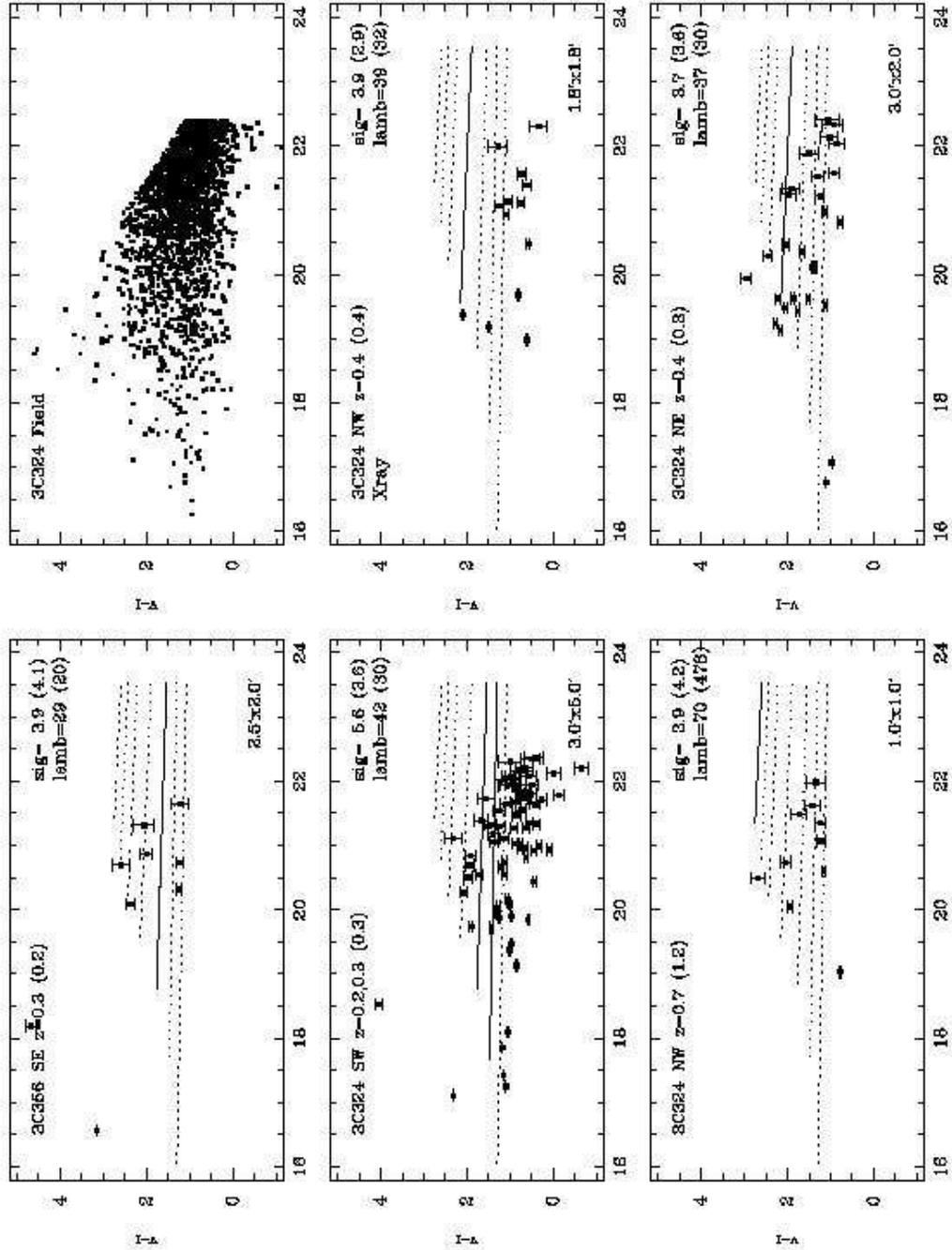}
\caption[]{V-I vs. I distributions for mutual V and I cluster candidates 
in the 3C356 (1 of 2) and 3C324 ROSAT fields. Plots are annotated as
in Figure~\ref{vi1}. \label{vi2}}
\end{figure}

\begin{figure}
%\plotone{/data/atalanta2/mack/ROXS/megan_VI2/plot3.ps}
\plotone{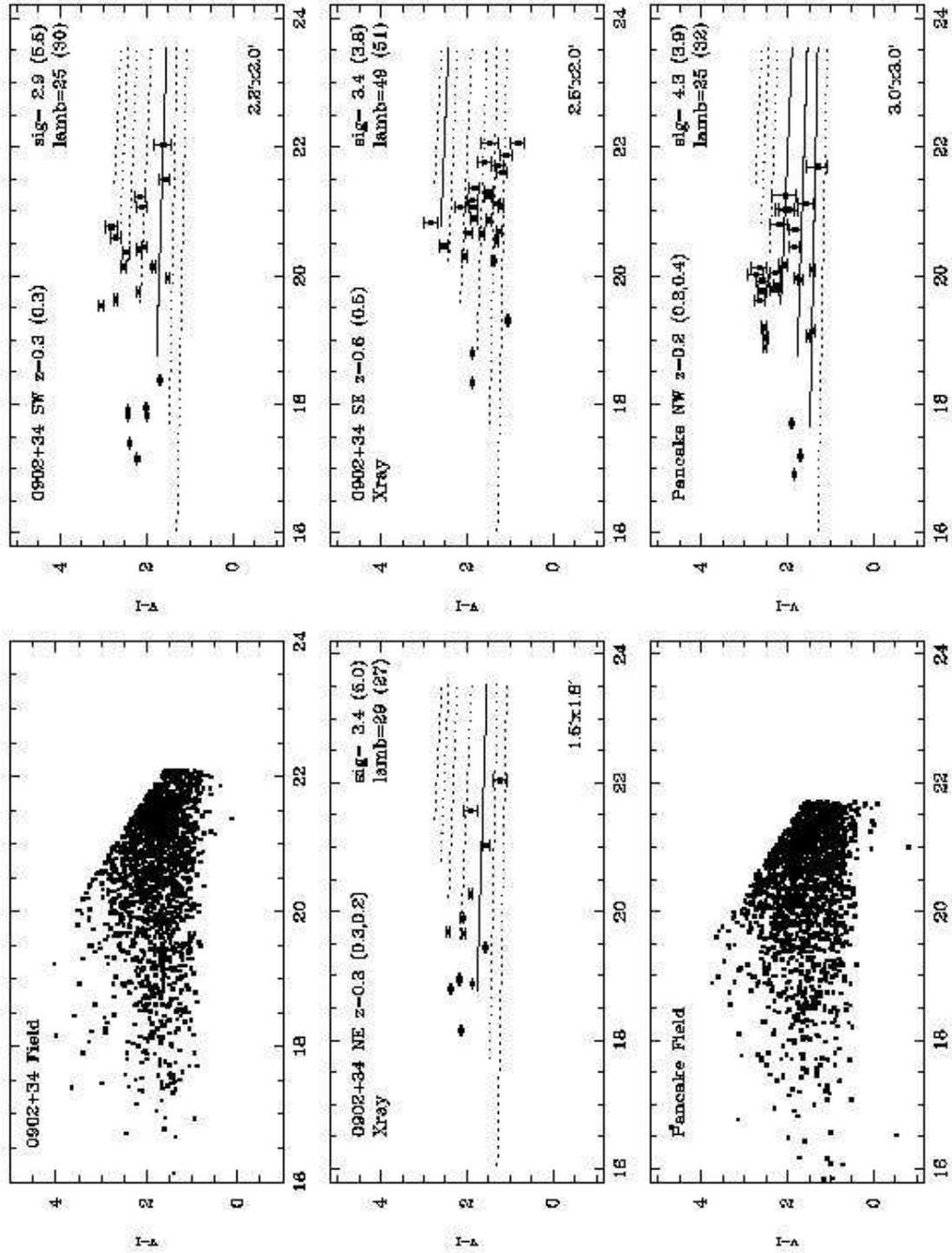}
\caption[]{V-I vs. I distributions for mutual V and I cluster candidates 
in the Zel'dovich Pancake and the 0902+34 ROSAT fields. 
Plots are annotated as
in Figure~\ref{vi1}. \label{vi3}}
\end{figure}

\begin{figure}
%\plotone{/data/atalanta2/mack/ROXS/megan_VI2/plot4.ps}
\plotone{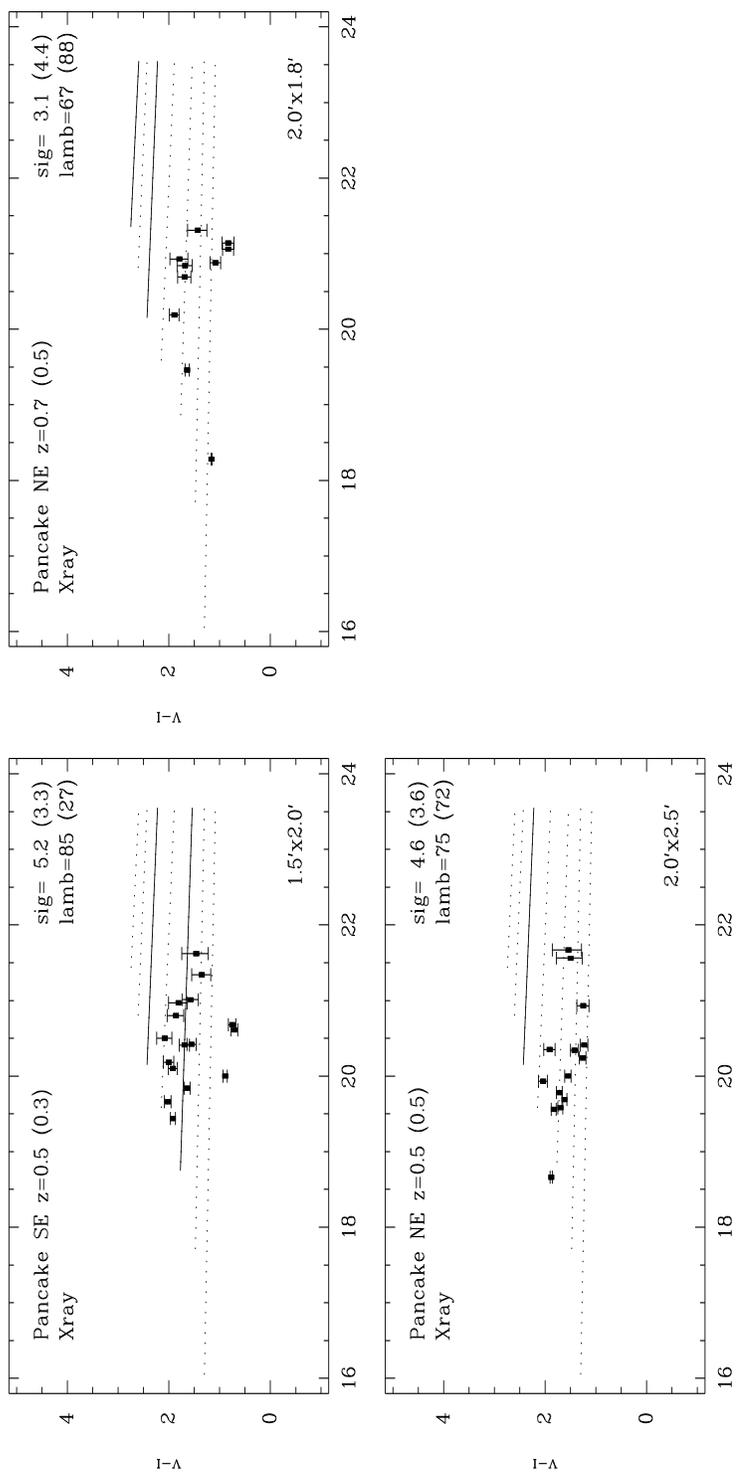}
\caption[]{V-I vs. I distributions for 3 mutual V and I cluster candidates 
in  Zel'dovich Pancake ROSAT field. 
Plots are annotated as
in Figure~\ref{vi1}.
\label{vi4}}
\end{figure}

In Figures~\ref{vi1}-\ref{vi4}, we present the galaxy color-magnitude diagrams
for the background galaxies and the candidate clusters,
along with the size of the cluster aperture, the estimated redshifts
from the I (V) band data, the optical detection significance in
I (V), and the corresponding value of $\Lambda_{cl}$ in I (V).
We overplot the color magnitude relations for 7 different 
redshifts between $z=0.1-0.7$, corrected to the Landolt (1992) 
system (Johnson V and Cousins I) from the AB V and I magnitudes
used in Gladders \& Yee (2000).
The darkest line represents the estimated redshift from the
matched filter output for the V and I data. 
The uncertainties in I and V-I are relatively small
for $I<22$ - over a magnitude above our completeness limit, 
so we are confident that our photometry is
sufficiently deep and accurate to show a red sequence at 
least for clusters with $z<0.7-0.8$. However, 
visual inspection shows that very few of the clusters has a 
particularly strong red sequence - even the confirmed
X-ray clusters
have fewer than 10 galaxies near the expected red sequence 
lines.

To quantify statistically our visual assessment of the
color-magnitude plots, we constructed
a two-dimensional two-sample Kolmogorov Smirnov test 
(Press et al. 1997) in order to estimate the probability that
the photometric V-I and I values of the possible cluster galaxies and the
field galaxies were simply drawn from the same parent distribution.
We list the results in Table~\ref{vi}, for all cluster
candidates where analysis was possible.  For each field 
we list the number of galaxies in the field population, 
typically $\sim2000$ galaxies in each. 
The column $N_{gals}$ lists the numbers
of galaxies in the cluster candidate sample and the column 
$P(KS)$ contains the probability
that the V-I and I values in the cluster candidate sample were randomly
drawn from the same parent population as the field galaxies, as plotted
in the first panel for each of the 5 ROSAT fields. 
Only 7 out of the 16 I-band cluster candidates showed a KS probability 
lower than 2\%. The K-S test is not a test of the presence of
a red sequence; but a lack of a significant K-S result for over
50\% of these candidates, including 50\% of the candidates with
X-ray counterparts, reflects the qualitative difficulty of discerning a 
red sequence in these data. One of the 5 X-ray candidates had only limited
galaxy photometry data; 
2 of the remaining 4 X-ray candidates had high and thus
insignificant K-S probabilities distinguishing the colors and
magnitudes of galaxies in
the core from those of background galaxies. 
The result of the K-S test confirms our sense that not many of the
cluster candidates exhibit a strong red sequence on the color-magnitude
plots.

This result suggests that for many of the ROXS candidates with both
a V and an I detection, a
red sequence is not particularly prominent, even for cluster 
candidates where both I and V-band counterparts exist nearby an
X-ray cluster candidate. If these  
candidates are true clusters -- and it seems unlikely that these
triply-identified candidates are spurious -- 
this result implies that searches for a prominent red
sequence may miss 
clusters of galaxies at any given redshift, at least at the
mass scales sampled by this survey. The
red-sequence method and its creators do not claim to create a sample of  
clusters with galaxy populations representative of all clusters
at a given redshift, since the method selects clusters 
by their old galaxy populations. (Furthermore, the red sequence
method may not require a prominent red sequence.)
Therefore the ROXS result 
may not be surprising since none of the ROXS candidates 
are particularly X-ray luminous. The observed 0.5-2.0 keV X-ray
luminosities of the X-ray candidates with V and I counterparts 
lie between $3-70 \times 10^{42} h_{75}^{-2}$ erg s$^{-1}$, which
places them in the category of richness class 0 clusters and
even groups; we note however that the estimated $\Lambda_{cl}$ for
these cluster candidates are as high as 70-85, not particularly
consistent with groups. 

A prominent red sequence may preferentially exist in the most X-ray luminous
clusters and groups, which are the most likely to be
dominated by elliptical galaxies in their cores. X-ray luminous
groups with central ellipticals and prominent elliptical 
galaxy populations are more X-ray luminous than their spiral-dominated
counterparts (Mulchaey \& Zabludoff 1998).  
Since the ROXS cluster candidates are not very X-ray luminous (whether
they were detected in the X-ray or not), they might 
have a higher spiral component and are more like the Virgo
cluster ($L_x \sim 10^{43} \lum$) than
they are like Coma ($L_x \sim 10^{44} \lum$). 
Stanford, Eisenhardt \& Dickinson (1998; SED98) looked
at various properties of the color-magnitude relation for early-type
galaxies in high-redshift clusters, including the scatter of those
properties, and found that for most of the SED98 clusters 
there was little variation in either the properties or their scatter.
However, all of the SED98 clusters are more luminous than the
ROXS clusters in our color-magnitude study. Therefore if the explanation
for the ROXS results is that there is an
X-ray luminosity threshold below which the color-magnitude properties
break down, the SED98 study would not have seen it.
A red sequence may be the most distinct at the highest
masses and at the lowest redshifts, but at present it is uncertain
at what mass and at what redshift the sequence may begin
to be indistinct and difficult to see in clusters.

\subsection{Simulations of ROX}

In Paper I, we produced a $\Lambda_{cl}$ function based on our data
and, using the $L_x-\Lambda_{cl}$ relation, we compared that function
to the X-ray luminosity function for clusters of galaxies. We found that
a steep $L_x-\Lambda$ relation
(steeper than the predicted $L_x \propto \Lambda_{cl}^2$)
 was needed to explain the observed 
 X-ray and optical luminosity functions for the ROXS clusters.
The proportionality, $L_x \propto \Lambda^2$ derives from the 
observed $L_x \propto T_x^3$ relation for clusters, an assumption
of a constant $M/L$ ratio for clusters, and  
 $T_x \propto
M^{2/3}$ from hydrodynamics simulations.  We also demonstrated in Paper I
that unless our observation errors are underestimated, the scatter in the
observed $L_x-\Lambda$ relation is large and intrinsic.

We have performed a numerical simulation of the ROXS, accounting for all major
selection and measurement effects. We confirm the basic empirical results, namely
that the instrinsic cluster population must have both a moderately steep richness
to X-ray-luminosity relationship and significant physical scatter between richness
and X-ray luminosity in order to reproduce the survey results. In this section we
briefly describe the simulation methodology and analysis.

Using Monte-Carlo techniques we first generate a large X-ray cluster population
($5\times 10^4$ members) occupying a volume from $z=0$ to $z=2$. We draw members
according to the local X-ray luminosity function (0.1-2.4 keV) of Ebeling et al
(1997), which we assume to be unevolving with $z$. Current constraints show
at most weak (less than 10\%) evolution in the cluster population to redshifts of
at least $z=0.5$ (cf. Gioia et al 2001; Lewis et al. 2002). Since ultimately very few clusters will be selected at $z>1$
by the ROXS criteria we consider a non-evolving X-ray luminosity
function to be a valid assumption for our first-order goals.

Assuming a optical to X-ray luminosity relationship of the form $\Lambda = A
L_{x}^{\alpha}$ and an instrinsic scatter expressed by a normally distributed
random variable $\Delta A$ where the amplitude is either fixed or a fractional
value of $A$, we assign a $\Lambda$ to each cluster. Via a Monte-Carlo approach we
then sample this cluster population and apply {\em both} the X-ray selection
criteria {\em} and optical selection criteria of the ROXS until a total of $N$
clusters have been ``observed''. The X-ray detection criteria are slightly
complicated by the known dependency on the combination of cluster angular extent
and net flux. Using an empirical relationship between cluster luminosity and X-ray
core radius (assuming a standard King profile for cluster density) of:
$r_c=0.125(L_x$h$^{-2}/1.25 \times 10^{44}~\lum)^{0.2}$ Mpc h$^{-1}$ 
(Jones et al 1998), we derive the
angular size of the X-ray cluster. The X-ray selection functions for the simulation are 
essentially the same as the X-ray selection functions in Figure~\ref{X-raysel}
in \S~\ref{xraycat_sect}, except for the simplifying notion 
that all clusters were observed on-axis by the ROSAT PSPC. 
Cluster luminosities are converted 
to the X-ray detection band (0.5-2 keV)
assuming a constant 4~keV thermal spectrum, which is a reasonable 
approximation for our observed sample.

Clusters are entirely discarded which fall outside of the $0.2<z<1.2$ range of the
ROXS. Clusters which are only detected in either the X-ray or optical, but not
both, are flagged. Those detected only in the optical are then assigned an X-ray
flux upper-limit (and X-ray luminosity upper limit) based on the mean X-ray flux
limit of the ROXS ($\sim 10^{-14}$ erg s$^{-1}$ cm$^{-2}$ 0.5-2 keV). Finally, the
$\Lambda_{cl}$ for each cluster is assigned an observational error as a function of
redshift, based on the known error distributions of the matched-filter algorithm,
which correspond to $\sim 10$\% errors 
for $z<0.7$ and $\sim 50$\% errors for $z>1$
(P96).

We are therefore left with four datasets: joint X-ray and optical detected
clusters, optically detected clusters with X-ray upper limits, X-ray detected but
not optically detected clusters, and entirely undetected clusters.

\begin{figure}
%\plotone{rox_sim_figureA.ps}
\plotone{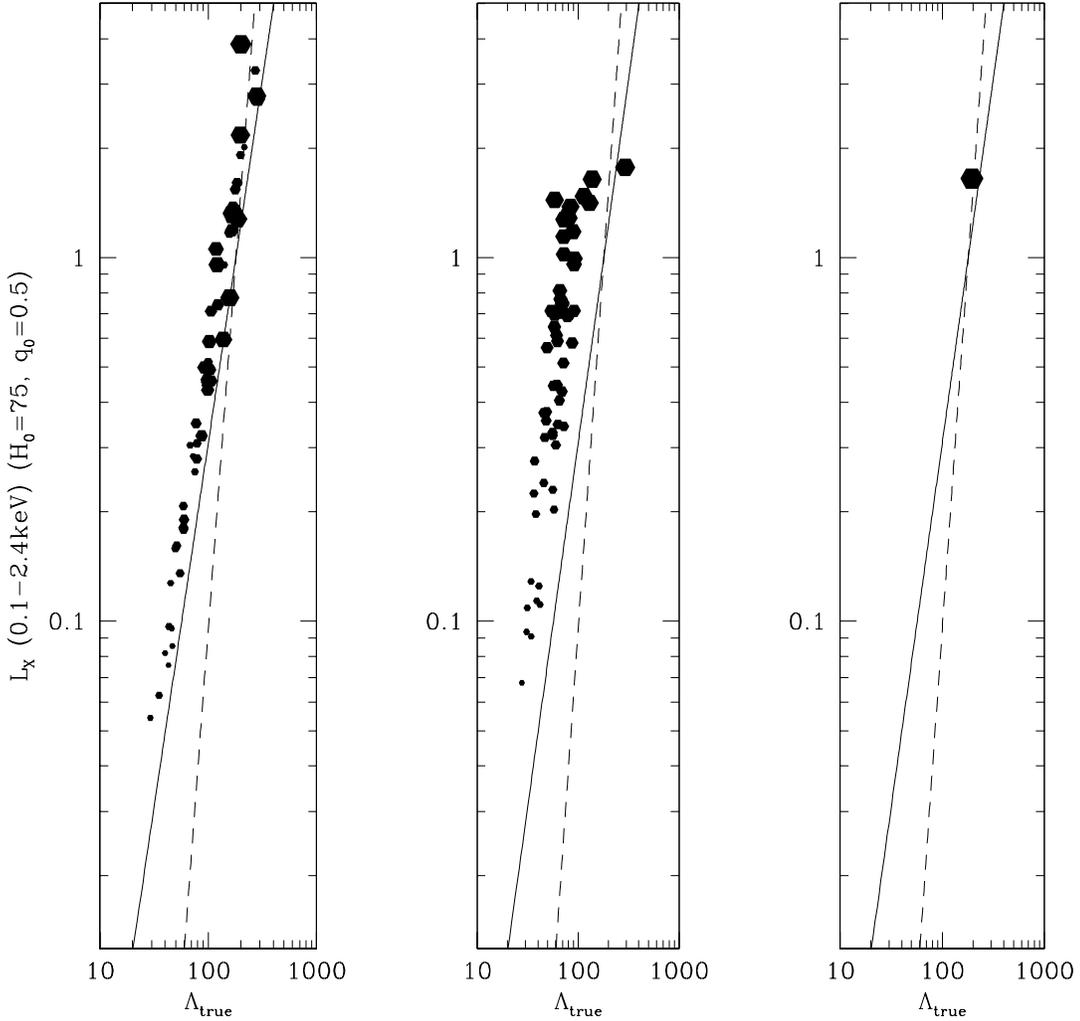}
\caption[]{
$L_x$ vs. $\Lambda$ is plotted for the three cluster detection datasets in a ROXS
simulation with $L_x \propto \Lambda^2$, normalized to approximate the observed
relationship. Leftmost panel plots clusters {\em jointly} detected in the X-ray and
optical, point size is proportional to cluster redshift ($0.2<z<1.2$). Solid curve
is the $L_x \propto \Lambda^2$ relationship used to generate the clusters, dashed
curve is an $L_x \propto \Lambda^{4}$ relationship. Center plot is of X-ray upper
limits for clusters only detected optically. Note that since these are $4\sigma$
upper limits to $L_x$ and the clusters are typically at higher $z$ for a given
$\Lambda$ than the detected systems, these points occupy a region above the
instrinsic curve. Rightmost panel plots those clusters which are X-ray detected but
{\em not} optically detected (1 object in this simulation).
\label{simA}
$L_x$ is in units $10^{44} \lum$. }
\end{figure}

In Figure~\ref{simA} we plot the results for a simulated dataset where $L_x\propto
\Lambda^{2}$ (the fiducial relationship expected under basic assumptions) and that the
$\Lambda-L_x$ scatter is zero. Observational errors are still included in
$\Lambda$, so some spread is still seen. The total number of systems is arbitrary,
but chosen here for clarity of presentation. It is immediately apparent that this
particular simulation does not qualitatively agree with the observations, both in
the shallowness of the relation and in the scatter between $L_x$ and $\Lambda$.
The relative numbers of joint, only optical, or only X-ray detections are also in
strong disagreement with observations. We observe approximately 4 cluster
candidates in the ROXS which are only detected optically for every 1 cluster which
is jointly detected, and no more than 3\% of all candidates are solely X-ray
candidates. In the simulation of Figure~\ref{simA} 
the mean numbers of detections (over 10
runs) yields a straight 1:1 ratio between jointly detected and optically detected
clusters, at odds with the real ROXS, although the fraction of the purely X-ray detected numbers
are consistent and are at about the 3\% level. Adding a significant
intrinsic scatter ($1\sigma=$28\% of the normalization 
$A$) to the $L_x-\Lambda$ relationship does
not alter these results.

\begin{figure}
%\plotone{rox_sim_figureB.ps}
\plotone{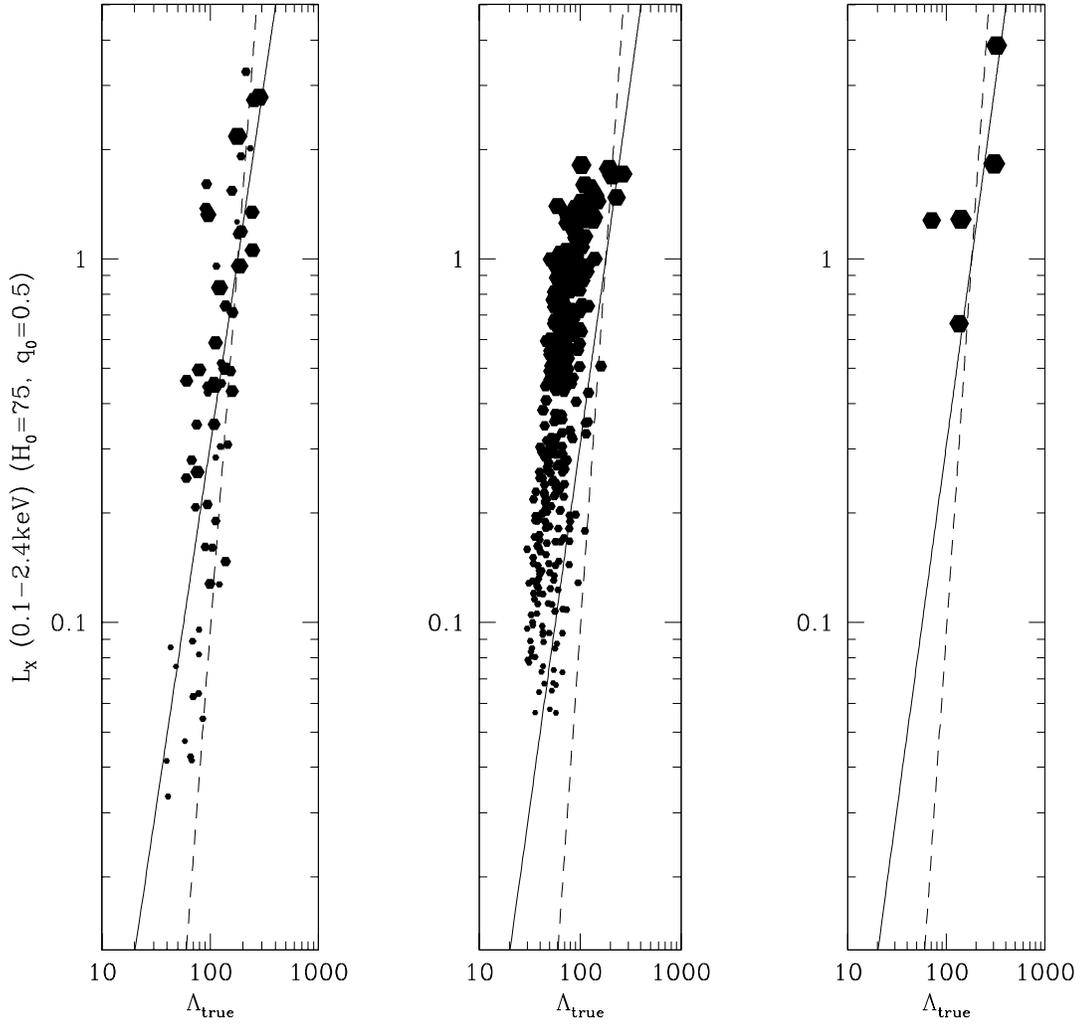}
\caption[]{
As for Figure~\ref{simA}, but simulation now uses $L_x\propto \Lambda^3$ and
an instrinsic scatter with $1\sigma=$39\% of the amplitude. \label{simB} }
\end{figure}

In contrast, in Figure~\ref{simB} we plot the results for $L_x\propto \Lambda^3$ and
an intrinsic scatter of $1\sigma=$39\%. 
It is apparent that there is much better qualitative 
agreement between this simulation and the ROXS results. Furthermore, the relative
numbers of X-ray vs. optical cluster detections are now in much better agreement.
The simulation yields a 4 to 1 ratio for optical detections vs joint detections,
and a pure X-ray detection rate of 1-2\%. 

Interestingly, a steeper relationship ($L_x\propto \Lambda^4$) appears to be less
successful at reproducing the ROXS observations, even when we 
adjust the allowed scatter.
Typically, the number of optical-only detections are overproduced by at least a
factor of 2 in such tests.

We note here that the simulations are seriously limited in their ability to
provide a faithful reproduction of the actual ROXS data. This limitation 
is for a variety of
reasons. Without much greater complexity we cannot mimic the details of the X-ray
selection, namely the variation in detection sensitivity as a function of off-axis
position in the ROSAT fields, and the detailed variations in field exposure times
and backgrounds. We are also reliant on the $\Lambda$ selection functions, which 
are known to suffer increasing uncertainties with $z$. The basic findings are however clear. The $\L_x-\Lambda$
relationship must be steeper than $\Lambda^2$, and the intrinsic scatter between
$L_x$ and $\Lambda$ must be large - these are {\em not} due to systematics in the
observations.

\begin{figure}[hp]
%\plotone{roxsim_figure_c.ps}
\plotone{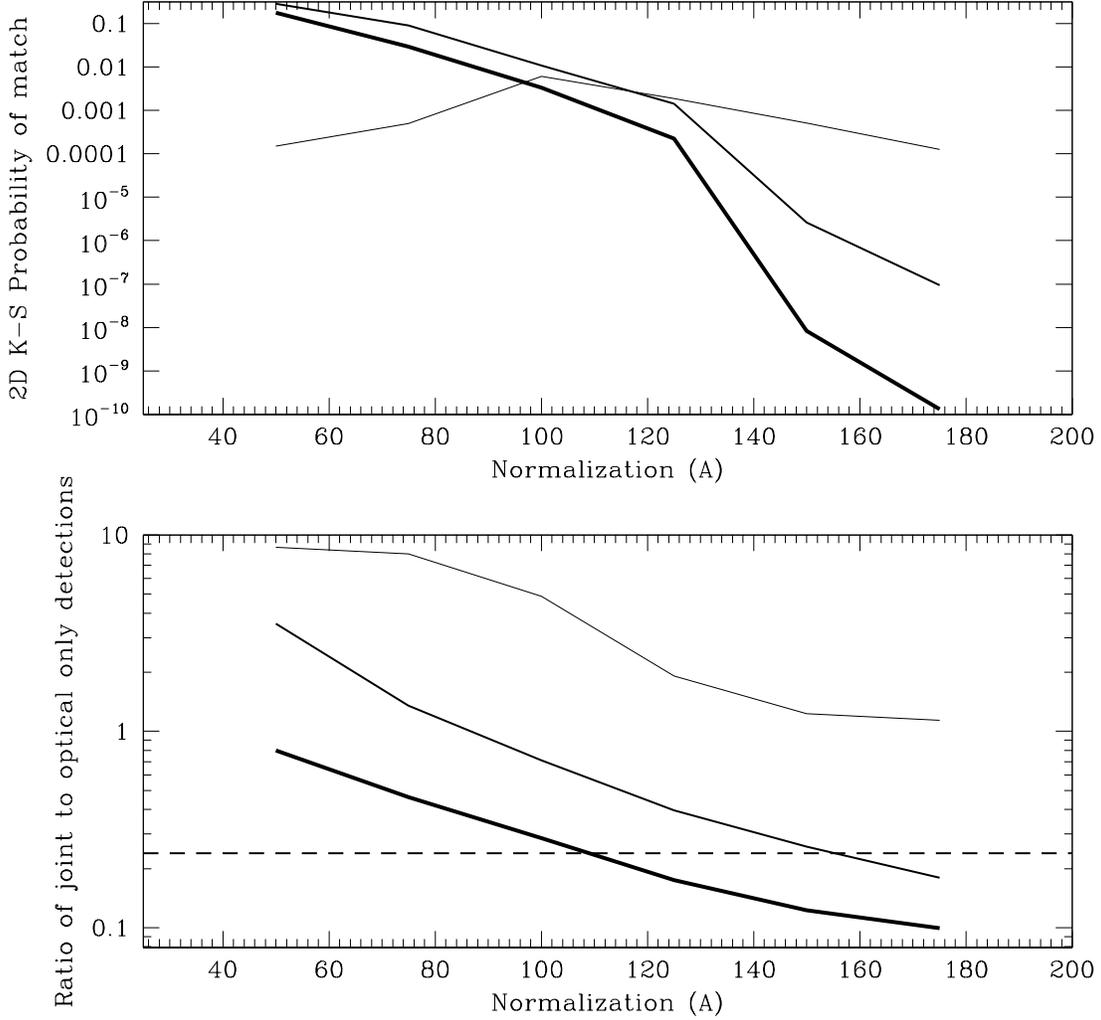}
\caption[]{
Summary of 2D K-S test and simulation results. In the upper panel the 2D
K-S probabilities (of a significant match between the ROXS joint
X-ray/optical detection $\Lambda$, $L_x$ data and the simulation joint
detections) are plotted as a function of the normalization $A$ in the
simulation $\Lambda=A L_x^{\alpha}$ relationship. Simulation data is taken
from single Monte Carlo runs which generate at least 50 joint detection
data points, scatter between runs is small compared to the true variations
in these curves. Lightest curve is for $\alpha=1/2$, medium is
$\alpha=1/3$, and the heaviest is $\alpha=1/4$. Lower panel shows the number
ratio of jointly detected clusters to those only detected optically,
curves are as for the upper panel. The dashed horizontal line denotes the
observed ratio for the ROXS. \label{simC} }
\end{figure}

A somewhat more quantitative evaluation of the simulation results is
summarized in Figure~\ref{simC}. We have applied a 2D K-S test 
(Press et al 1997)
to the ROXS joint detections (X-ray and optically detected clusters) and
the equivalent simulation outputs for a range of parameters in the
$\Lambda=AL_x^{\alpha}$ relation. Specifically, we have chosen a fixed
instrinsic fractional dispersion in this relation such that the $1\sigma$
width is 100\% when the normalization $A$ is 50 and approximately 30\%
when $A=175$. We have then varied $A$ and set $\alpha$ to $1/2$, $1/3$ and
$1/4$. In Figure~\ref{simC} (upper panel) we plot the resultant 2D K-S
probabilities. These should be considered only in relation to each other
rather than as absolute indicators of a `goodness-of-fit'. The higher the
probability the more significant the agreement between distributions. It
is apparent that when $\alpha=1/2$ the agreement between the ROXS and the
simulations is never very good, but does seem to peak around $A=100$. In
constrast, for $\alpha\leq 1/3$, the agreement appears best for low $A$,
comparable to $\alpha=1/2$ for $a\simeq 100$ and significantly worse for
$A>120$. Variations as large as 50\% in the intrinsic scatter used in the
simulations do not significantly alter these results.

However, as described above, visual inspection of the data in the
$L_x-\Lambda$ panel reveals that the 2D K-S test may be somewhat misleading,
and in fact neither the low-$\alpha$ results at low $A$ or the
$\alpha=1/2$ results at $a\sim 100$ would be classed as very good fits
`by-eye'. For example, when $\alpha=1/2$ the simulated and observed
distributions overlap only at the lower $L_x$ end, which boosts the K-S
probability relative to $\alpha=1/3$, although the latter distribution is
better aligned, its overlap is smaller.

The situation is considerably clearer when we also take into account the
number ratios of joint detections versus optical only detections (as
discussed above). In Figure~\ref{simC} (lower panel) we plot the ratio 
of the number of joint detections to the number of optical-only detections 
for the 3 cases described above. We also plot the actual ROXS ratio (dashed line). 
It is immediately apparent that only the $\alpha=1/3-1/4$ curves intersect the
observations for reasonable $A$. Weighing the 2D K-S results by this
additional constraint we conclude that the simulations are in best
agreement with the data when $\alpha \simeq 1/3-1/4$ (since the KS
probabilities are overall higher) and when $A\simeq 100-150$. $L_x\propto
\Lambda^2$ is therefore excluded as a plausible intrinsic relationship.

\section{Discussion}

The optical matched-filter selection technique works well to find
candidate clusters of galaxies, but the spurious
fraction is high at $\sim30\%$, and it is demonstrably 
not 100\% complete - it misses clusters at a rate of 
10-20\% at least. The selection window for a matched filter cluster sample 
is more difficult to quantify because of its sensitivity to the 
spectral energy distribution of
the galaxies for which it is searching.

On the other hand, the X-ray selection technique, while generating fewer
spurious cluster candidates ($\sim10\%$) and reliably finding the 
(apparently) more massive clusters,  
misses some high $\Lambda_{cl}$ cluster candidates.
We do not know yet whether these are true massive clusters or whether they 
are fortuitous projections of less massive systems. Redshifts and X-ray 
observations are required to determine the nature of these candidates.
Zabludoff \& Mulchaey (1998) and Mulchaey \& Zabludoff (1998) showed in their sample of 12 optically
selected groups that groups without
detected X-ray emission tend to be lower velocity dispersion systems 
with few or no elliptical galaxies. The undetected groups 
in the Zabludoff \& Mulchaey sample may not even be bound. Some X-ray 
observations of optically-selected, high redshift clusters have 
revealed such clusters to be ``underluminous'' in the X-ray
(Castander et al 1994.) The high-significance, optically rich 
ROXS candidates without X-ray counterparts
are prime targets for observational tests 
whether X-ray selection indeed selects
on the basis of cluster mass. The correlation of the presence of
a dense, X-ray emitting intracluster media and the presence of a
significant elliptical galaxy population is another testable hypothesis
with further observations of the ROXS sample. With the limited color
information we have in hand, we were not able to show that
X-ray cluster candidates at the moderate X-ray luminosity 
levels available in our sample are more likely to have red elliptical
sequences. But we were only able to make this test for a small 
number of cluster candidates, given the available photometry.

A related result from ROXS is that 
the cluster richness $N_R$ is not well-correlated with cluster X-ray 
luminosity, a result which is supplementary 
to our result in Paper I regarding cluster X-ray
luminosity and the $\Lambda_{cl}$ richness parameter.
The clusters in our sample are on average poorer and less 
massive than those found in a half-sky survey (e.g. XBACS), 
owing to the smaller sky area and greater depth of our survey. 
For $\Lambda > 30$, Donahue et al. (2001) suggested that
the joint redshift and $\Lambda_{cl}$
distribution arises from the population of X-ray clusters with a 
steep dependence between $L_x$ and $\Lambda_{cl}$. Equivalently, we show 
here that we can also reproduce this joint distribution with a large 
scatter between $L_x$ and $\Lambda_{cl}$ in Monte-Carlo simulations 
of the ROXS.

A correlation between cluster X-ray luminosity and cluster richness
has been suspected at least since Jones \& Forman (1978) plotted 
cluster richness class against X-ray luminosity for nearby ($z<0.07$)
clusters. But even in their sample, a $3 \times 10^{44} ~\lum$ 
cluster is equally as likely
to have a cluster richness of 0 as it is 2 or 3. Only the most 
luminous clusters ($L_x > 10^{45} ~\lum$) had cluster richnesses reliably
of 2 or 3. The XBACS clusters (Figure~\ref{LXrich}) show this correlation
and very large scatter with cluster richness. However, richness has
long been a suspect observational parameter for X-ray astronomers
(e.g. Mushotzky et al 1978), since the measurements rely 
on number counts that only become more
difficult to ascertain at higher redshift with the associated  higher
contamination levels. The matched filter method provides a possibly
more robust and objective means of estimating a cluster candidate's luminosity 
in the form of $\Lambda_{cl}$, yet, this measure
too is not strongly correlated with cluster X-ray luminosity. 

The
lack of strong correlation between X-ray and optical luminosity, and
the large scatter of those two quantities, are confirmed by the ROX
Survey. A review by Borgani \& Guzzo (2001) showed that the velocity
dispersions and cluster optical luminosities of an optically-selected
sample of clusters are not as well-correlated
as the velocity dispersions and the optical luminosities of 
a subsample of clusters selected for their X-ray fluxes from the 
original sample. The velocity dispersions of these 
X-ray selected clusters are even better correlated with their X-ray
luminosities, suggesting that X-ray luminosity is better correlated
with cluster mass than is optical luminosity. The ROX Survey has no
such independent test of correlation with mass; however, followup
observations of this sample in the X-ray and the optical will further
test the Borgani \& Guzzo suggestion at higher redshifts. 

The ROXS,
since it is a uniformly selected sample at both X-ray and optical
wavelengths, has examples of both X-ray candidates without optical
counterparts and optical candidates without X-ray counterparts. Both
subsamples probe the extremes of the $L_x/L_{opt}$ distribution. 
Explaining the extremes will go a long way towards explaining why and how
the emissivity of the intracluster gas is related to the light emitted
by the stars in the cluster. It is particularly important 
to see if there is a 
``third parameter" such as galaxy formation 
efficiency, evolutionary stage of the cluster, or galaxy morphology
populations, that creates the large spread in the $L_x$ versus
richness relationship for clusters, or if the large spread is due
to projection effects significantly impacting the optical selection
technique. One possible clue is the existence of a difference in
the conclusions based on cluster contents from an optically-selected, rather 
heterogenous sample of clusters (the MORPHS sample, in Smail et al 1997)
and those found in an X-ray selected sample of clusters by
Ellingson et al (2001). The X-ray selected clusters show cores that
exhibit very little evolution between $z=0.5$ and the present, whereas
the MORPHS clusters show evidence that the S0 population may be 
turning into ellipticals during that same time. It is possible that
the presence of a well-developed intracluster medium is ubiquitous in
a massive cluster, but in less massive clusters, the galaxy populations
are still evolving in the cluster cores. Uniform morphological 
studies of a cluster sample diverse in X-ray and optical 
properties are needed  to test such a statement, and such 
a test could be accomplished with
a sample such as the ROXS.

Rosati et al. (1995, 1998) selected clusters for the RDCS 
not only for their X-ray emission, but used an extent criteria as well. 
There was some worry that this selection criteria may have missed  
the most compact X-ray clusters which remain unresolved by the ROSAT PSPC, 
particularly those at high redshift. 
However, our survey does not reveal an obvious population of 
high redshift cluster candidates with
X-ray point sources. The correspondence between an optical cluster and an
X-ray point source is not more than we would expect by chance - therefore
we do not believe we have found a population of cluster sources which 
would go undetected by the ROSAT cluster surveys which select for extent, 
a result consistent with the findings of the Wide Angle ROSAT 
Pointed Survey (WARPS; Jones et al 1998), 
a ROSAT serendipitous search which did not filter for extent.
We have some examples of optical clusters with X-ray
point source counterparts which could be AGN in those clusters. 
Approximately 25 of our candidates have
X-ray point sources within 1'-2' of the cluster
centroid; optical identification of these sources is required to 
ascertain whether the candidate and the point source are physically related.

Probably the most surprising result is that 
the ROXS cluster candidates do not, as a rule, show distinct or 
strong red sequences in the color-magnitude diagrams of their galaxy 
populations. Only about 50\% of the optical cluster candidates detected in
both I- and V-band show galaxy colors distinct from those of the
field population. And only $\sim50\%$ of the X-ray clusters so 
identified show such sequences.  It is possible that the red
sequence is ubiquitous in the most massive, virialized clusters. But 
perhaps at
some unknown threshold in mass or level of virialization or luminosity, 
the red sequence ceases to be distinct and prominent. We plan to use the
red sequence detection method (Gladders \& Yee 2000), which does not 
rely on an obvious red sequence to select cluster candidates, to see what cluster
candidates it finds in the ROXS galaxy photometry data.

The ROXS sample identifies  several problems that warrant further
investigation:
\begin{itemize}
\item A red sequence is not prominent in all X-ray
selected clusters. At
what X-ray luminosity (or equivalently at what mass) does the
red sequence cease to be prominent? Is the presence of a red
sequence dependent on mass?
\item The estimated optical luminosity and richness of a cluster are not
at all well-correlated with the X-ray luminosity. 
There are at least three unsolved issues regarding the correlation of 
the optical luminosity, X-ray luminosity, and cluster virial mass:
	\begin{enumerate}
	\item Is the lack of correlation representative of the
	difficulties of estimating the optical luminosity or is
	it representative of an intrinsic variation or scatter in
	the M/L ratios in clusters?
	\item If the estimated optical or X-ray luminosity, and therefore
	the detectability of a cluster, has enormous scatter with
	respect to the mass of the cluster, selection by mass using
	optical or X-ray light may be very difficult. Knowing 
	the scatter with respect to cluster
	mass with both optical and X-ray luminosity is 
	important. Individual studies
	show good correspondence of both quantities 
	with velocity dispersion (e.g. Girardi et al. 2000 and 
	Mahdavi \& Geller 2001 for $L_B$ and $L_x$ respectively), 
	but a comparative analysis of the masses and luminosities of 
	a well-chosen, homogeneous 
	sample has not been done yet.
	\item If there is such a large intrinsic variation in the optical
	M/L for clusters, what evolutionary process controls this variation?
	Can galaxy formation efficiency or evolutionary history be so radically
	different from one cluster environment to another to produce this effect?
	Does M/L correlate with any other property of a cluster of galaxies?
	\end{enumerate}
\item While galaxies and hot gas may trace the gravitational potentials
of the most massive clusters, mergers and other physics may disrupt that
correspondence. True optical condensations of galaxies 
with low levels of X-ray emission may
be galaxies which have been temporarily separated from the hot intracluster
gas while undergoing a merger event; galaxies experience the merger as a
non-collisional fluid while the gas experiences hydrodynamic effects of 
shocks and cooling. An observational example of such a merger is Abell~754
(Zabludoff \& Zaritsky 1995). Studies of the ROXS ``X-ray poor''
cluster candidates may reveal their true nature, whether they are mergers, 
minor systems embedded in filaments, or mere projection effects.
\end{itemize}
	
Because the ROXS has greater depth and smaller 
sky area than surveys like the EMSS or XBACS, 
the ROXS clusters must be relatively low-mass clusters compared
to typical clusters in those surveys. 
The low-mass end of the cluster mass function may  be where the 
physics of the intracluster medium collides with the physics of star
formation and galactic winds. The energy injected by stellar processes
is closer to the specific energy per particle in the gravitationally
bound gas of a low mass cluster of galaxies. Mergers may be more
common or at least more significant in the low-mass clusters -- 
when the relative 
velocities of the galaxies are not much greater than the internal
velocities of the galaxies, collisions and interactions 
are more
likely to
induce star formation. The least massive clusters may be the most recent members to
the cluster hierarchy, whose galaxies are most recently accreted from the
field. The optical and X-ray properties of such clusters may thus
still be correlated, but only very weakly and with high scatter.
Such recent formation may explain the weakness or lack of a red sequence,
which may only dominate in clusters where the oldest ellipticals formed 
preferentially in high density regions at an early time. 
The other inference possible from ROXS is that at some point below a given mass 
on the mass scale of clusters,
some treasured assumptions about clusters, their M/L constancy,
the invariance of the baryon fraction in clusters, 
and their ubiquitous red galaxy content may break down. 

\section{Summary}

No single cluster selection method is perfect. Multiple, complementary methods
of selecting clusters may be essential for cluster studies, 
particularly at high redshift where all methods start to 
run into completeness limits, incomplete understanding of physical
evolution, and projection effects. Understanding the biases inherent
in any cluster selection method is essential to extracting conclusions 
about cluster evolution or the evolution of galaxies in clusters from any
cluster sample. The ROSAT Optical X-ray Survey (ROXS) for clusters of 
galaxies provides a well-matched, head-to-head comparison of two of the most popular
methods for finding clusters of galaxies: a matched filter in the optical
and extended source detection algorithms such as the wavelet transform
in the X-ray. Both algorithms do seem to agree on the location of at
least some of the cluster candidates in our sky coverage; however, the
overlap was far from perfect.

We find that the correlation is weak between the X-ray and optical luminosity
or richness of a cluster, and it has considerable intrinsic scatter.
Some of this scatter may arise from underestimating our observational
uncertainties, but the scatter is too large for it to be accounted for
by pure observational uncertainties in $L_x$ or $\Lambda_{cl}$ 
(Paper I). 
We confirm these results in a numerical model of the
ROXS. 
While we did not find any 
bias that would affect the conclusions on the evolution
of clusters of galaxies, we did find evidence for 
possible bias that could 
affect conclusions regarding galaxy evolution in clusters.  
X-ray and optical cluster detection techniques sample different
cluster populations, a statement which should worry those who
use cluster samples to make claims about galaxy
evolution.

We find that the red-sequence does not appear to be very strong in
our cluster candidates, regardless of whether they have X-ray 
counterparts. We plan to do a proper red sequence detection with the
ROXS galaxy photometry and position data in an upcoming paper.

All cluster methods have their merits; the cluster community should consider,
however, the effect of cluster selection and the measurement of
cluster properties on their conclusions. The ROXS has provided a number
of follow-up avenues to assess in detail 
the real differences between cluster
candidates which were detected in the X-ray and those which were not. 
Observations to obtain the weak lensing signatures, velocity dispersions,
and/or ICM temperatures  would provide the most fundamental property of
a cluster from a cosmologist's perspective - its virial mass. Confirmation 
of at least a subsample
of the optical clusters with the faintest X-ray limits would more
strongly test our working hypothesis that we have no need to 
invoke the presence a bi-modal population of
X-ray faint and X-ray luminous clusters of galaxies at the mass scales
of $10^{14}-10^{15}~\rm{M}_\odot$. Clusters whose X-ray/optical
luminosities lie outside the normal distribution could reveal
clusters in early evolutionary states or clusters embedded in
unusual projection with respect to filaments and other large scale
structure, or they may represent a fundamental challenge to our
assumptions about cluster physics and formation.   

\acknowledgements The authors would like to acknowledge  
the support of the NASA LTSA grant NAG5-3257 and a STScI DDRF grant 82208. 
This research has made use of data obtained from the High Energy Astrophysics
Science Archive Research Center (HEASARC), provided by NASA's Goddard
Flight Center, in particular the ROSAT observation catalogs and data, 
and the NASA/IPAC Extragalactic Database (NED) which is operated by the
Jet Propulsion Laboratory, California Institute of Technology, under contract
with the National Aeronautics and Space Administration.
JTS acknowledges support of a NASA/Astrophysical Data Products
grant NAG5-6936.

\clearpage


\clearpage

\clearpage

\begin{deluxetable}{llrrrrrr}
\tablecaption{ROSAT fields\label{Fields}}
\tabletypesize{\small}
\tablewidth{7.5truein} 

\rotate

\startdata
\tableline\tableline\\
\multicolumn{1}{l}{ROSAT Field} &
\multicolumn{1}{l}{ROSAT Target} &
\multicolumn{1}{r}{ROSAT Exptime} &
\multicolumn{1}{c}{RA} &
\multicolumn{1}{c}{Dec} &
\multicolumn{1}{c}{Area}  &
\multicolumn{1}{c}{KPNO Date} &
\multicolumn{1}{c}{}\\
\multicolumn{1}{l}{} &
\multicolumn{1}{l}{} &
\multicolumn{1}{c}{(sec)} &
\multicolumn{1}{c}{J2000} &
\multicolumn{1}{c}{J2000} &
\multicolumn{1}{r}{deg$^2$}  &
\multicolumn{1}{c}{I-Band} &
\multicolumn{1}{c}{V-Band}\\ \tableline
RP700112	&MKN 78			 	&19844	&07:42:40	&65:10:52     &0.17518      &03/18/96			\\
RP700326N00	&0902+34		 	&14355 	&09:05:31	&34:07:48     &0.24414      &03/17/96      &05/06/97	\\
RP900327	&Zeldovich Pancake		&55611	&09:06:52	&33:40:16     &0.24161      &03/16/96      &03/18/96	\\
WP700540	&10214+4724			&19359	&10:24:33	&47:09:04     &0.24401      &03/17/96			\\
WP201243N00	&PG1034+001		 	&18111	&10:37:05       &-00:08:21    &0.19623      &05/09/97			\\
WP700228	&1116+215			&24585	&11:19:09	&21:19:16     &0.24343      &03/17/96			\\
WP201367M01	&PG1159-035		 	&32219	&12:04:25       &-03:40:16    &0.23667      &03/18/96			\\
WP700232	&1202+281			&27859	&12:04:43	&27:54:04     &0.23531      &03/18/96      &05/06/97  	\\ 
RP700864A01	&3C 270.1			&19290	&12:20:33	&33:43:16     &0.19836      &05/06/97   		\\
RP600242A01	&Giovanelli-Haynes Cl		&24830	&12:27:43	&01:36:04     &0.18036      &05/08/97			\\   
RP700073	&3C 280			 	&48051	&12:56:57	&47:20:28     &0.16501      &03/17/96   		\\
RP700216A00	&B2 1308+32			& 8316	&13:10:29	&32:20:59     &0.19679      &05/09/97  			\\ 
RP700117	&3CR294				&22692	&14:06:43	&34:11:27     &0.19156      &05/07/97   		\\
WP700248	&1411+442			&25319	&14:13:48	&44:00:04     &0.23088      &03/16/96   		\\
RP700122	&Q1413+1143		 	&28066	&14:15:45	&11:29:46     &0.24770      &03/18/96   		\\
RP800401A01	&4C23.37			&11811	&14:15:58	&23:07:12     &0.23037      &03/16/96   		\\
RP700257N00	&MKN 841			&16842	&15:04:02	&10:26:24     &0.18221      &05/09/97   		\\
RP701373N00	&3C 324				&15434	&15:49:50	&21:25:52     &0.22442      &03/17/96      &05/06/97	\\   
RP800239N00	&CL1603				&28763	&16:04:29	&43:13:12     &0.18933      &05/08/97   		\\
RP300021N00	&MS1603.6+2600			&24506	&16:05:46	&25:51:36     &0.19376      &05/08/97 			\\  
WP170154	&WFC/PSPC Calibr		&37490	&16:29:32	&78:04:52     &0.18426      &05/09/97   		\\
%701121N00	&HS1700+6416			&16078	&17:01:00	&64:11:59     &0.18889      &05/08/97   		\\
701457n00	&HS1700+6416		        &27408	&17:01:00	&64:11:59     &0.18889      &05/08/97   		\\

RP800395N00	&3C 356				&18563	&17:24:19	&50:57:36     &0.22243      &03/18/96      &05/06/97 	\\ 
\\
\tableline

\enddata

\end{deluxetable}

%FOCAS PARAMETERS COMMENTED OUT
%\begin{table}
%\caption[]{Object Classification Rules for FOCAS\label{FOCAS}}
%\begin{tabular}{lll} \tableline \tableline
%	Noise	&	$0.1 \leq s \leq 0.7$ &	$0 <  f \leq 1$ \\
%	Star	&	$0.7 <  s \leq 1.3$ &	$0 <  f \leq 1$ \\
%	Fuzzy Star &	$1.3 <  s \leq 10$	&	$0 <  f <  0.25$ \\
%	Galaxy	&	$1.3 <  s \leq 10$	&	$0.25 \leq f \leq 1$ \\
%	Diffuse	&	$10  <  s \leq 100$ &	$0< f \leq 1$ \\ \tableline
%\end{tabular}
%\end{table}

%Optical catalog begins here

\begin{deluxetable}{l|rrrrrrrrrr|l}
\tablecaption{Optical Cluster Candidates \label{Optical}}
\tabletypesize{\tiny}
\tablecolumns{12}
\tablewidth{7.5truein}

\rotate
 
\startdata

\multicolumn{1}{l}{Optical ID} & 
\multicolumn{1}{c}{RA } & 
\multicolumn{1}{c}{Dec } & 
\multicolumn{1}{c}{Sigma} & 
\multicolumn{1}{c}{Lambda} & 
\multicolumn{1}{c}{z} & 
\multicolumn{1}{c}{I-mag} & 
\multicolumn{1}{c}{Radius} & 
\multicolumn{1}{c}{Theta} & 
\multicolumn{1}{c}{F$_{lim}$ 1'} & 
\multicolumn{1}{l}{Xray Match} & 
\multicolumn{1}{l}{Comment} \\ 
\multicolumn{1}{l}{} & 
\multicolumn{1}{c}{} & 
\multicolumn{1}{c}{} & 
\multicolumn{1}{c}{} & 
\multicolumn{1}{c}{} & 
\multicolumn{1}{c}{} & 
\multicolumn{1}{c}{} & 
\multicolumn{1}{c}{(")} & 
\multicolumn{1}{c}{(')} & 
\multicolumn{1}{c}{(10$^{-14}$\tablenotemark{1})} & 
\multicolumn{1}{l}{} & 
\multicolumn{1}{l}{} \\

\cutinhead{RP700112 - MKN 78} 
  OC1 0740+6504& 07 40 57.25  &  65 04 00.9  &  3.42 & 31.17 &  0.30 & 19.46  & 138 & 12.85 & 1.9 &  &  \\*  
  OC2 0741+6524& 07 41 44.06  &  65 24 42.9  &  3.25 & 38.69 &  0.40 & 20.29  & 162 & 15.11 & 2.1 &  &PtSource within 30"\\*
 OC2a 0742+6519& 07 42 14.19  &  65 19 43.7  &  3.31 & 30.13 &  0.30 & 19.60  & 170 & 9.35  & 1.7 &  &  \\*
  OC3 0743+6458& 07 43 45.82  &  64 58 14.9  &  5.45 & 64.88 &  0.40 & 20.13  & 162 & 14.31 & 2.1 &RXJ0743.7+6457&Cluster within 1'\\  

\cutinhead{RP700326N00 - 0902+34} 
  OC1 0904+3411& 09 04 23.88  &  34 11 04.9  &  4.25 & 43.38 &  0.40 & 20.13  &  70 & 14.27 & 2.5 &  &Edge,Compact  \\* 
  OC2 0905+3359& 09 05 10.18  &  33 59 37.6  &  2.94 & 25.04 &  0.30 & 19.24  &  59 & 9.24  & 2.1 &  &Compact,Low Sig  \\* 
  OC3 0905+3421& 09 05 11.11  &  34 21 53.0  &  5.14 & 34.79 &  0.20 & 18.39  & 116 & 14.67 & 2.5 &  &Edge,Pt Source within 1'\\*
  OC4 0905+3403& 09 05 52.95  &  34 03 17.5  &  3.58 & 88.98 &  1.00 & 22.51  & 101 & 6.40  & 1.8 &  &Blend OC5?\\*  
  OC5 0905+3407& 09 05 57.01  &  34 07 00.6  &  3.35 & 48.56 &  0.60 & 21.28  & 177 & 5.44  & 1.8 &  &Blend OC7 and OC4? \\*  
  OC6 0906+3417& 09 06 17.59  &  34 17 16.0  &  3.39 & 28.87 &  0.30 & 19.24  &  79 & 13.50 & 1.8 &RXJ0906.3+3417&Cluster within 15"\\*
  OC7 0906+3405& 09 06 28.25  &  34 05 11.9  &  5.40 & 36.56 &  0.20 & 18.39  & 275 & 12.13 & 2.4 &  &PtSource within 2.0' \\*  
  OC8 0906+3359& 09 06 31.74  &  33 59 51.1  &  4.05 & 70.13 &  0.70 & 21.65  & 165 & 14.89 & 2.3 &  &PtSource within 1.5', blend OC7? \\* 
  OC9 0906+3413& 09 06 32.70  &  34 13 17.4  &  3.35 & 89.67 &  1.10 & 22.71  &  50 & 13.88 & 2.5 &  &  \\  
 
\cutinhead{RP900327 - Zel'dovich Pancake}
  OC1 0906+3354& 09 06 12.55  &  33 54 21.9  &  4.32 & 25.13 &  0.20 & 18.10  & 227 & 16.45 & 1.7 &  &  \\*  
  OC2 0907+3330& 09 07 18.64  &  33 30 44.5  &  5.23 & 85.19 &  0.50 & 20.71  & 174 & 10.88 & 1.3 &RXJ0907.3+3330&Cluster within 30"\\*
  OC3 0907+3343& 09 07 27.69  &  33 43 02.5  &  4.59 & 74.79 &  0.50 & 20.85  & 268 & 7.80  & 1.0 &RXJ0907.4+3342&Cluster within 30"\\*
  OC4 0907+3355& 09 07 47.16  &  33 55 00.4  &  3.50 & 28.21 &  0.30 & 19.47  &  68 & 18.62 & 1.9 &  &Edge,Compact,Near cut area\\*
  OC5 0907+3350& 09 07 58.29  &  33 50 09.6  &  3.09 & 67.27 &  0.70 & 21.57  & 135 & 16.86 & 1.8 &RXJ0907.8+3351&Cluster within 2',PtSource within 15"\\
 
\cutinhead{WP700540 - 10214+4724}
  OC1 1023+4722& 10 23 18.89  &  47 22 59.8  &  3.20 & 37.61 &  0.40 & 20.18  &  74 & 18.86 & 2.5 &  &Edge  \\* 
  OC2 1024+4657& 10 24 06.16  &  46 57 05.7  &  3.09 & 22.08 &  0.20 & 18.26  &  55 & 12.80 & 2.0 &  &Compact \\*  
  OC3 1024+4721& 10 24 08.26  &  47 21 02.4  &  3.53 & 52.37 &  0.50 & 20.72  & 115 & 12.78 & 2.0 &  &  \\*  
  OC4 1024+4714& 10 24 19.52  &  47 14 25.6  &  3.45 & 56.14 &  0.60 & 21.13  & 103 & 5.93  & 1.5 &  &  \\*  
  OC5 1024+4721& 10 24 42.64  &  47 21 41.7  &  3.02 & 77.59 &  1.20 & 22.85  &  56 & 12.79 & 2.0 &  &  \\*  
  OC6 1024+4707& 10 24 43.91  &  47 07 21.1  &  4.76 & 55.85 &  0.40 & 20.14  & 108 & 2.41  & 1.2 &  &  \\*  
  OC7 1024+4656& 10 24 52.51  &  46 56 56.2  &  4.01 & 59.42 &  0.50 & 20.69  & 176 & 12.49 & 2.0 &  &Blend with OC 11  \\*  
  OC8 1025+4701& 10 25 01.77  &  47 01 53.7  &  2.96 & 76.12 &  1.20 & 22.91  &  58 & 8.57  & 1.7 &  &Low Sig\\*  
  OC9 1025+4722& 10 25 07.14  &  47 22 58.8  &  3.44 & 40.46 &  0.40 & 20.12  &  71 & 15.09 & 2.2 &  &Near Edge; Blend with 9A?\\*   
 OC9a 1025+4723& 10 25 26.37  &  47 23 16.4  &  3.25 & 52.92 &  0.60 & 21.33  &  93 & 16.84 & 2.3 &  &Near Edge; Blend with 9?; PtSrc\\*
 OC10 1025+4716& 10 25 22.73  &  47 16 13.9  &  3.63 & 30.16 &  0.30 & 19.04  &  39 & 11.03 & 1.9 &  &Compact,PtSource within 30" \\*  
 OC11 1025+4701& 10 25 35.85  &  47 01 23.7  &  7.35 & 52.92 &  0.20 & 18.25  & 356 & 13.05 & 2.0 &RXJ1025.4+4703&Cluster within 3'; Blend OC7\\

\cutinhead{WP201243N00 - PG1034+001} 
  OC1 1036+0005& 10 36 36.01  &  00 05 24.9  &  4.30 & 64.68 &  0.80 & 21.78  & 137 & 7.79  & 1.2 &  &Near Edge,PtSource within 1' \\*
               &              &              &       &       &       &        &     &       &     &  &Blend 1a, 1b\\*   
 OC1a 1036+0002& 10 36 50.70  &  00 02 36.2  &  3.86 & 43.35 &  0.50 & 20.69  & 131 & 6.78  & 1.2 &  &Near Edge, blend 1, 1b\\*
 OC1b 1037+0005& 10 37 03.35  &  00 05 37.1  &  3.27 & 52.58 &  0.90 & 22.17  &  89 & 2.81  & 0.9 &  &Near Edge; blend 1a, 1\\*
  OC2 1036-0018& 10 36 43.81  & -00 18 52.9  &  4.18 & 29.47 &  0.30 & 19.26  &  98 & 11.72 & 1.6 &  &Near Edge  \\*  
  OC3 1037-0007& 10 37 22.54  & -00 07 59.0  &  3.34 & 37.53 &  0.50 & 20.55  & 119 & 4.45  & 1.0 &  &Blend with OC3a \\* 
 OC3a 1037-0010& 10 37 37.70  & -00 10 25.0  &  3.93 & 16.60 &  0.20 & 18.27  & 200 & 8.47  & 1.3 &  &PtSource within 1'; Blend with OC3 \\*
  OC4 1038-0008& 10 38 03.24  & -00 08 22.4  &  3.11 & 34.88 &  0.50 & 20.66  &  67 & 14.61 & 1.8 &  &  \\  
 
\cutinhead{WP700228 - 1116+215 }
  OC1 1118+2107& 11 18 09.36  &  21 07 04.7  &  5.26 & 52.96 &  0.30 & 19.42  & 122 & 18.56 & 2.8 &  &Near Edge \\* 
  OC2 1118+2126& 11 18 29.45  &  21 26 48.0  &  2.93 & 56.79 &  1.00 & 22.54  &  39 & 12.04 & 2.3 &  &Low Sig \\*
  OC3 1119+2116& 11 19 18.80  &  21 16 54.9  &  3.80 & 46.36 &  0.50 & 20.79  & 184 & 3.13  & 1.6 &RXJ1119.2+2117&Cluster within 30", Blend OC3a\\*
 OC3a 1119+2118& 11 19 31.88  &  21 18 54.8  &  4.09 & 41.21 &  0.30 & 19.46  & 219 & 5.20  & 1.8 &  & Blend OC3, OC5; PtSrc\\*
  OC4 1119+2107& 11 19 26.77  &  21 07 08.9  &  6.07 & 46.53 &  0.20 & 18.18  & 189 & 12.70 & 2.4 &RXJ1119.4+2106&Cluster within 30"\\*
  OC5 1119+2124& 11 19 35.04  &  21 24 19.2  &  3.38 & 44.92 &  0.70 & 21.63  &  76 & 7.83  & 2.0 &  &PtSource within 1'; Blend OC6, 3a\\*
  OC6 1119+2127& 11 19 35.95  &  21 27 18.2  &  3.89 & 43.00 &  0.40 & 20.14  & 192 & 10.16 & 2.2 &RXJ1119.7+2126&Cluster within 1.5',PtSource within 1'\\*
               &              &              &       &       &       &        &     &       &     &  &Blend OC5\\*
  OC7 1119+2131& 11 19 55.78  &  21 31 36.6  &  3.36 & 33.80 &  0.30 & 19.48  & 158 & 16.41 & 2.7 &  &  \\*  
  OC8 1119+2124& 11 19 59.29  &  21 24 34.1  &  3.13 & 74.57 &  1.10 & 22.72  &  76 & 12.75 & 2.4 &  &  \\  

\cutinhead{WP201367 - PG1159-035} 
  OC1 1203-0344& 12 03 27.66  & -03 44 25.5  &  3.44 & 43.18 &  0.40 & 20.24  &  88 & 15.07 & 1.7 &  &Near Edge,PtSource within 30"\\*
  OC2 1203-0350& 12 03 28.17  & -03 50 34.3  &  3.04 & 60.49 &  0.70 & 21.57  &  79 & 17.66 & 1.9 &  &Near Edge,PtSource within 1'\\*
  OC3 1203-0333& 12 03 29.23  & -03 33 10.3  &  3.00 & 69.96 &  0.80 & 21.89  &  36 & 15.82 & 1.8 &  &Compact \\*  
  OC4 1204-0342& 12 04 00.70  & -03 42 45.8  &  3.84 & 64.10 &  0.60 & 21.15  & 137 & 6.72  & 1.1 &  &Blend with OC4a, OC4b\\*
 OC4a 1204-0342& 12 04 09.87  & -03 42 12.5  &  3.26 & 86.45 &  0.90 & 22.20  & 119 & 4.40  & 0.9 &  &PtSource within 30"; Blend OC4, OC4b\\*
 OC4b 1204-0348& 12 04 14.18  & -03 48 16.5  &  3.85 &131.82 &  1.10 & 22.70  & 174 & 8.47  & 1.2 &  &Blend with OC4, OC4a \\*
  OC5 1204-0351& 12 04 21.80  & -03 51 15.3  &  4.90 & 28.06 &  0.20 & 18.38  & 229 & 11.00 & 1.4 &RXJ1204.3-0350&Cluster within 30" \\*  
  OC6 1204-0330& 12 04 41.21  & -03 30 22.3  &  3.38 & 32.56 &  0.30 & 19.33  & 100 & 10.70 & 1.4 &  &  \\*  
  OC7 1204-0338& 12 04 45.34  & -03 38 17.0  &  3.16 & 30.45 &  0.30 & 19.54  &  99 & 5.28  & 1.0 &  &  \\*  
  OC8 1204-0330& 12 04 55.19  & -03 30 52.9  &  2.10 & 55.10 &  0.90 & 22.21  &  11 &   --  &  -- &?RXJ1205.0-0332&Cluster within 2' (note low sigma) \\  

\cutinhead{WP700232 - 1202+281} 
  OC1 1203+2758& 12 03 47.81  &  27 58 36.5  &  3.92 & 56.48 &  0.50 & 20.72  &  91 & 13.07 & 1.7 &?RXJ1204.1+2807  & A1455? \\*  
  OC2 1203+2743& 12 03 53.23  &  27 43 54.8  &  3.25 & 39.62 &  0.40 & 20.13  & 130 & 14.97 & 1.9 &  &  \\*  
  OC3 1204+2744& 12 04 26.02  &  27 44 12.8  &  3.16 & 89.91 &  1.00 & 22.52  &  53 & 10.50 & 1.5 &  &  \\*  
  OC4 1204+2805& 12 04 36.71  &  28 05 19.9  &  3.35 & 95.30 &  1.00 & 22.51  & 117 & 11.42 & 1.6 &?RXJ1204.1+2807  & A1455? \\*  
  OC5 1204+2801& 12 04 38.01  &  28 01 52.0  &  3.96 & 28.01 &  0.20 & 18.25  & 145 & 7.95  & 1.3 &?RXJ1204.1+2807  & A1455? \\*  
  OC6 1204+2757& 12 04 53.71  &  27 57 59.1  &  3.21 & 39.02 &  0.40 & 20.32  &  92 & 4.61  & 1.0 &  &  \\*  
  OC7 1204+2753& 12 04 56.99  &  27 53 04.8  &  3.18 & 90.66 &  1.00 & 22.48  &  49 & 3.18  & 0.9 &  &  \\*  
  OC8 1205+2805& 12 05 18.49  &  28 05 17.1  &  3.19 & 31.43 &  0.30 & 19.51  &  73 & 13.71 & 1.8 &  &Sprawling\\*  
  OC9 1205+2740& 12 05 43.83  &  27 40 30.2  &  4.15 & 50.47 &  0.40 & 20.05  & 130 & 19.04 & 2.2 &  &Near Edge; PtSource 1' \\
 
\cutinhead{RP700864A01 - 3C270.1}
  OC1 1219+3331& 12 19 35.76  &  33 31 44.0  &  3.44 & 30.79 &  0.30 & 19.21  &  65 & 16.41 & 1.9 &  &Edge; PtSource \\*
  OC2 1219+3345& 12 19 41.34  &  33 45 20.6  &  4.95 & 26.82 &  0.20 & 18.33  & 257 & 11.07 & 1.5 &  &Blend OC2a\\*
 OC2a 1219+3343& 12 19 57.94  &  33 43 54.2  &  3.67 & 44.68 &  0.40 & 20.20  & 189 & 7.45  & 1.2 &  &Blend OC2 \\*
  OC3 1220+3334& 12 20 08.42  &  33 34 02.7  &  2.64 & 91.13 &  1.20 & 22.86  &  34 &  --   &  -- &?RXJ1220.0+3334&Cluster within 1.5' (note low sigma)\\*
  OC4 1221+3344& 12 21 11.30  &  33 44 31.3  &  3.20 & 28.66 &  0.30 & 19.47  &  85 & 7.95  & 1.3 &RXJ1220.8+3343&Both Xray clusters within 4',PtSource within 15"\\*
	       &              &              &       &       &       &        &     &       &     &RXJ1220.9+3343& \\*
  OC5 1221+3352& 12 21 35.57  &  33 52 28.4  &  3.61 &124.68 &  1.20 & 22.87  &  84 & 15.86 & 1.9 &  &Edge\\ 

\cutinhead{RP600242A01 - Giovanelli-Haynes Cloud} 
  OC1 1227+0143& 12 27 48.26  &  01 43 38.8  &  2.96 & 34.13 &  0.30 & 19.47  & 114 & 7.76  & 1.7 &RXJ1227.8+0143&Cluster within 45";Blend OC1a\\*
 OC1a 1228+0142& 12 28 07.53  &  01 42 36.5  &  3.79 & 65.44 &  0.50 & 20.76  & 246 & 9.01  & 1.8 & &Blend OC1, OC4; PtSources  \\*
  OC2 1227+0136& 12 27 46.77  &  01 36 01.0  &  2.91 & 61.06 &  0.80 & 21.86  &  42 & 0.94  & 1.2 &  &Compact,Low Sig \\*
  OC3 1228+0136& 12 28 08.53  &  01 48 17.5  &  3.37 & 63.64 &  0.70 & 21.56  & 110 & 13.85 & 2.2 &  &Edge \\*
  OC4 1228+0136& 12 28 11.07  &  01 36 08.7  &  3.13 & 19.81 &  0.20 & 18.66  & 193 & 7.01  & 1.6 &  &Blend OC1, OC4  \\ 
 
\cutinhead{RP700073 - 3C280}
  OC1 1257+4719& 12 57 03.89  &  47 19 09.2  &  2.02 & 76.15 &  1.00 & 22.52  &   9 &   --  &--   &?RXJ1256.9+4720&Cluster within 1.5'(note low sigma)\\*
  OC2 1257+4709& 12 57 13.31  &  47 09 01.3  &  3.17 & 82.64 &  0.80 & 21.90  &  85 & 11.69 & 1.5 &  &PtSource within 1.5'\\*
  OC3 1257+4708& 12 57 53.29  &  47 08 39.3  &  6.04 & 42.15 &  0.20 & 18.48  & 325 & 15.09 & 1.7 &  &  \\  
 
\cutinhead{RP700216A00 - B2 1308+32}
  OC1 1309+3218& 13 09 54.66  &  32 18 25.4  &  3.86 & 90.60 &  0.90 & 22.10  & 140 & 7.51  & 1.3 &  &Blend with OC2 \\*  
  OC2 1310+3221& 13 10 04.30  &  32 21 17.8  &  4.48 & 44.68 &  0.30 & 19.47  & 124 & 5.03  & 1.1 &RXJ1309.9+3222&Cluster within 2' (MS1308.8+3244,z=0.245); Blend OC1\\*  
  OC3 1310+3214& 13 10 44.83  &  32 14 04.3  &  4.11 & 22.97 &  0.20 & 18.39  & 117 & 7.77  & 1.3 &  &  \\*  
  OC4 1311+3228& 13 11 12.72  &  32 28 47.3  &  5.28 & 29.49 &  0.20 & 18.62  & 274 & 12.23 & 1.6 &RXJ1313.2+3229&Cluster within 30"; Blend OC4a\\*
 OC4a 1311+3228& 13 11 28.36  &  32 28 52.9  &  5.14 & 66.19 &  0.40 & 20.14  & 180 & 14.97 & 1.8 &  &PtSource within 1'; Blend with OC4\\  
 
\cutinhead{RP700117 - 3C294}
  OC1 1406+3401& 14 06 05.35  &  34 01 57.5  &  3.22 & 43.47 &  0.60 & 21.05  &  63 & 12.23 & 1.9 &  &PtSrc \\*  
  OC2 1406+3411& 14 06 07.62  &  34 11 01.2  &  3.56 &  8.10 &  0.20 & 18.86  & 110 & 7.33  & 1.5 &  &PtSource within 1.5'\\*
  OC3 1406+3403& 14 06 30.63  &  34 03 05.3  &  3.34 & 19.97 &  0.30 & 19.47  & 114 & 8.68  & 1.6 &  &  \\*  
  OC4 1406+3407& 14 06 36.45  &  34 07 33.3  &  3.04 & 34.56 &  0.50 & 20.56  &  64 & 4.06  & 1.2 &  &Compact\\*  
  OC5 1406+3402& 14 06 59.14  &  34 02 04.2  &  3.09 & 72.07 &  0.90 & 22.13  &  67 & 9.89  & 1.7 &  &  \\*  
  OC6 1407+3420& 14 07 41.45  &  34 20 45.9  &  3.90 &  8.97 &  0.20 & 18.77  &  76 & 15.28 & 2.1 &  &Compact; PtSrc \\*  
  OC7 1407+3415& 14 07 41.81  &  34 15 24.4  &  5.86 & 66.54 &  0.50 & 20.70  & 144 & 12.8  & 1.9 &RXJ1407.6+3415&Cluster within 30",PtSource within 15'\\*
  OC8 1407+3400& 14 07 44.94  &  34 00 17.3  &  3.18 & 19.03 &  0.30 & 19.47  &  35 & 16.96 & 2.2 &  &Compact,Near Edge\\
 
\cutinhead{WP700248 - 1411+442}
  OC1 1412+4347& 14 12 50.50  &  43 47 57.6  &  3.89 &106.22 &  1.00 & 22.48  & 176 & 15.90 & 2.2 &  &Blend OC3\\*  
  OC2 1412+4353& 14 13 02.49  &  43 53 16.8  &  3.29 & 56.36 &  0.60 & 21.10  & 169 & 10.60 & 1.7 &  &Blend with OC2a\\*  
 OC2a 1413+4347& 14 13 26.52  &  43 47 20.9  &  3.75 & 64.29 &  0.60 & 20.98  & 150 & 13.23 & 2.0 &  &Blend with OC2\\*  
  OC3 1413+4359& 14 13 12.83  &  43 59 10.9  &  3.49 & 69.71 &  0.70 & 21.55  & 148 & 6.38  & 1.4 &  &Blend OC1\\*  
  OC4 1413+4351& 14 13 30.58  &  43 51 28.6  &  4.83 & 37.77 &  0.30 & 19.19  & 318 & 9.08  & 1.6 &  &  \\*  
  OC5 1414+4347& 14 14 12.61  &  43 47 23.8  &  3.45 & 68.96 &  0.70 & 21.44  &  84 & 13.36 & 2.0 &  &  \\*  
  OC6 1415+4348& 14 15 05.58  &  43 48 16.0  &  3.21 & 25.13 &  0.30 & 19.27  & 104 & 18.26 & 2.3 &  &Near Edge,PtSource within 1' \\*  
  OC7 1415+4352& 14 15 09.62  &  43 52 13.3  &  2.98 &101.76 &  1.20 & 22.87  &  39 & 16.64 & 2.2 &  &Edge,Low Sig\\*
  OC8 1415+4409& 14 15 10.84  &  44 09 34.5  &  3.07 & 24.02 &  0.30 & 18.96  &  34 & 17.68 & 2.3 &  &Compact,Off Edge\\

\cutinhead{RP700122 - Q1413+1143} 
  OC1 1414+1123& 14 14 50.97  &  11 23 35.9  &  6.25 & 59.40 &  0.30 & 19.09  & 264 & 14.71 & 1.9 &RXJ1415.2+1119&Cluster within 7',PtSource within 15"\\*
  OC2 1414+1118& 14 14 55.24  &  11 18 22.2  &  2.97 & 87.43 &  1.00 & 22.52  & 122 & 16.76 & 2.0 &  &Low Sig \\*
  OC3 1414+1143& 14 14 58.61  &  11 43 54.4  &  4.09 & 38.91 &  0.30 & 19.14  & 155 & 18.28 & 2.1 &  &Near Edge\\*  
  OC4 1415+1140& 14 15 21.90  &  11 40 06.4  &  4.47 & 33.13 &  0.20 & 18.22  & 196 & 11.92 & 1.6 &  & Blend with OC4a  \\*  
 OC4a 1415+1141& 14 15 30.56  &  11 41 28.2  &  3.57 & 47.88 &  0.50 & 20.88  & 176 & 12.33 & 1.7 &  & Blend with OC4 \\*  
  OC5 1416+1141& 14 16 41.07  &  11 41 06.5  &  3.19 & 30.29 &  0.30 & 19.32  & 102 & 17.74 & 2.1 &  &  \\  
 
\cutinhead{RP800401A01 - 4C23.37}
  OC1 1415+2259& 14 15 14.04  &  22 59 36.9  &  5.02 & 44.39 &  0.30 & 19.17  & 240 & 12.46 & 2.1 &  &PtSource within 30"\\*
  OC2 1415+2317& 14 15 40.08  &  23 17 08.5  &  3.96 & 39.08 &  0.40 & 20.16  & 105 & 10.67 & 1.9 &RXJ1415.8+2316&Cluster within 2.5',PtSource within 1'\\*
  OC3 1415+2307& 14 15 55.85  &  23 07 26.1  &  4.76 & 42.09 &  0.30 & 19.25  & 215 & 0.35  & 1.1 &RXJ1415.9+2307&Cluster within 30"\\*
 OC3a 1415+2311& 14 15 26.72  &  23 11 40.9  &  3.86 & 58.86 &  0.50 & 20.62  & 120 & 8.28  & 1.7 &  &PtSource within 30"\\*  
  OC4 1417+2255& 14 17 01.06  &  22 55 41.8  &  3.26 & 25.66 &  0.20 & 18.23  &  68 & 18.70 & 2.6 &  &Near Edge\\*
 
\cutinhead{RP700257N00 - MKN841}
  OC1 1503+1014& 15 03 16.51  &  10 14 16.9  &  3.60 & 46.43 &  0.40 & 20.24  & 159 & 16.48 & 3.3 &  &Edge  \\*
  OC2 1503+1021& 15 03 33.48  &  10 21 32.5  &  3.82 & 23.28 &  0.20 & 18.67  & 290 & 8.52  & 2.7 &  &Blend with OC2a, OC2b \\*
 OC2a 1503+1018& 15 03 33.92  &  10 18 39.7  &  3.60 & 66.46 &  0.60 & 21.13  & 149 & 10.36 & 2.8 &  &PtSource within 30"; Blend with OC2, OC2b\\*
 OC2b 1503+1021& 15 03 45.45  &  10 21 35.4  &  4.12 & 44.77 &  0.30 & 19.51  & 217 & 6.29  & 2.5 &  &Blend with OC2, OC2a\\*
  OC3 1504+1022& 15 04 00.49  &  10 22 13.3  &  2.98 & 67.95 &  0.80 & 21.93  &  73 & 4.19  & 2.3 &  &Low Sig\\*
 
\cutinhead{RP701373N00 - 3C324}
  OC1 1548+2113& 15 48 50.06  &  21 13 54.3  &  3.39 & 39.86 &  0.50 & 20.76  & 106 & 18.42 & 2.5 &  &Blend with OC2/2a/2b?\\*  
  OC2 1548+2127& 15 48 55.88  &  21 27 21.6  &  3.89 & 38.51 &  0.40 & 20.14  & 171 & 12.78 & 2.1 &RXJ1548.8+2126&Cluster within 1.5',PtSource within 1'\\
  		&		&		&    &	     &       &        &     &      &	&Blend w/ OC1?,OC2a, OC2b\\*
 OC2a 1549+2119& 15 49 02.22  &  21 19 53.2  &  5.60 & 41.59 &  0.20 & 18.02  & 304 & 12.68 & 2.1 &  &PtSource within 1'; Blend with OC1?,OC2,OC2b\\*
 OC2b 1549+2122& 15 49 02.37  &  21 22 26.0  &  3.88 & 31.53 &  0.30 & 19.15  & 198 & 11.68 & 2.0 &  &Blend with OC1?,OC2,OC2a \\*
  OC3 1549+2123& 15 49 32.13  &  21 23 25.0  &  3.85 & 70.35 &  0.70 & 21.55  &  81 & 4.88  & 1.5 &  &  \\*  
  OC4 1549+2129& 15 49 55.45  &  21 29 15.3  &  2.94 & 23.95 &  0.30 & 19.28  &  64 & 3.65  & 1.4 &  &Low Sig,Compact\\*
  OC5 1550+2112& 15 50 02.46  &  21 12 37.5  &  3.18 & 23.64 &  0.20 & 17.65  & 136 & 13.47 & 2.2 &  &Near Edge; PtSrc\\*   
  OC6 1550+2121& 15 50 10.33  &  21 21 13.9  &  2.96 & 54.04 &  0.70 & 21.49  &  59 & 6.51  & 1.6 &  &Low Sig\\*
  OC7 1550+2125& 15 50 13.35  &  21 25 21.6  &  3.64 &113.13 &  1.10 & 22.71  &  87 & 5.36  & 1.5 &  &  \\*  
  OC8 1550+2134& 15 50 44.69  &  21 34 35.8  &  3.69 & 36.51 &  0.40 & 20.08  & 148 & 15.38 & 2.3 &  &  \\  
 
\cutinhead{RP800239N00 - CL1603} 
  OC1 1603+4323& 16 03 29.85  &  43 23 52.2  &  3.92 & 71.12 &  0.60 & 21.14  & 140 & 15.02 & 1.8 &  &  \\*  
  OC2 1604+4313& 16 04 11.46  &  43 13 53.8  &  3.25 & 45.90 &  0.50 & 20.85  & 121 & 3.11  & 0.9 &  &PtSource within 1'\\*  
  OC3 1604+4310& 16 04 39.05  &  43 10 19.7  &  4.38 & 35.31 &  0.30 & 19.34  & 172 & 3.50  & 0.9 &  &  \\* 
  OC4 1604+4324& 16 04 51.73  &  43 24 16.4  &  5.58 & 44.93 &  0.30 & 19.38  & 175 & 11.88 & 1.6 &  &  \\  
  
\cutinhead{RP300021N00 - MS1603.6+2600}
  OC1 1604+2558& 16 04 55.58  &  25 58 54.5  &  2.99 & 23.00 &  0.40 & 20.26  &  63 & 13.3  & 1.9 &  &Compact,Low Sig\\*
  OC2 1604+2554& 16 04 56.65  &  25 54 22.6  &  3.54 & 34.93 &  0.60 & 21.21  &  77 & 11.22 & 1.7 &?RXJ1605.0+2552&Cluster within 2.5',PtSource within 45"\\*
  OC3 1605+2541& 16 05 18.39  &  25 41 33.8  &  3.86 & 67.98 &  0.90 & 22.35  & 117 & 11.70 & 1.8 &  &PtSource within 15"; QSO w/z=1.07\\*
  OC4 1605+2553& 16 05 38.20  &  25 53 00.8  &  4.84 & 26.15 &  0.20 & 18.69  & 292 & 2.08  & 1.0 &RXJ1605.5+2553& Cluster within 1.5'\\*
 OC4a 1605+2554& 16 05 58.12  &  25 54 12.4  &  3.88 & 30.15 &  0.40 & 20.02  & 131 & 3.94  & 1.2 &  &PtSource within 15"\\*
  OC5 1606+2602& 16 06 27.81  &  26 02 35.1  &  3.65 & 94.53 &  1.10 & 22.75  & 107 & 14.6  & 2.0 &  &  \\*  
  OC6 1606+2556& 16 06 40.05  &  25 56 22.8  &  3.85 & 37.92 &  0.60 & 21.22  & 140 & 13.27 & 1.9 &  &  \\  

\cutinhead{WP170154 - WFC/PSPC CAL} 
  OC1 1626+7817& 16 26 20.25  &  78 17 01.5  &  3.81 & 55.28 &  0.50 & 20.64  & 118 & 15.66 & 2.1 &RXJ1626.3+7816&Cluster within 15", Blend with OC1a\\*
 OC1a 1625+7814& 16 25 54.51  &  78 14 39.2  &  4.81 & 44.17 &  0.30 & 19.14  & 126 & 14.86 & 2.0 &  & Blend with OC1\\*
  OC2 1626+7759& 16 26 52.23  &  77 59 05.3  &  3.18 & 29.19 &  0.30 & 19.27  &  83 & 10.13 & 1.7 &?RXJ1627.1+7756&Cluster within 3',PtSource within 45"\\*
  OC3 1627+7753& 16 27 46.31  &  77 53 38.6  &  3.06 & 51.74 &  0.60 & 21.24  &  75 & 12.48 & 1.8 &  &PtSource within 30"\\* 
  OC4 1629+7811& 16 29 39.77  &  78 11 12.1  &  3.12 & 28.70 &  0.30 & 19.43  & 127 & 6.41  & 1.4 &  &PtSource within 15"; Blend OC4a, OC4b\\*
 OC4a 1629+7803& 16 29 41.64  &  78 03 30.0  &  3.81 & 73.76 &  0.70 & 21.48  & 182 & 1.37  & 0.9 &  &Blend with OC4, OC4b\\*
 OC4b 1629+7806& 16 29 53.81  &  78 06 36.5  &  3.94 & 49.58 &  0.40 & 20.19  & 206 & 2.10  & 1.0 &  &Blend with OC4, OC4a\\*
  OC5 1632+7806& 16 32 23.64  &  78 06 17.1  &  3.37 & 48.96 &  0.50 & 20.68  & 130 & 8.92  & 1.6 &  &PtSource within 1'\\*
  OC6 1633+7808& 16 33 39.75  &  78 08 35.8  &  3.89 & 35.76 &  0.30 & 19.23  & 100 & 13.23 & 1.9 &  &Near Edge,PtSource within 30" \\
\cutinhead{RP800395N00 - 3C356}
  OC1 1723+5059& 17 23 14.13  &  50 59 25.1  &  3.36 & 38.02 &  0.50 & 20.75  &  61 & 10.37 & 1.6 &  &Compact \\*  
  OC2 1723+5108& 17 23 24.79  &  51 08 34.9  &  4.49 & 41.06 &  0.20 & 18.03  & 191 & 13.89 & 1.9 &  &PtSource within 1' \\*
  OC3 1725+5059& 17 25 05.24  &  50 59 00.1  &  2.94 & 55.58 &  0.70 & 21.59  &  70 & 7.41  & 1.4 &  &PtSource within 1.5',Low Sig\\*
  OC4 1725+5055& 17 25 25.99  &  50 55 21.9  &  3.93 & 28.99 &  0.30 & 19.34  &  91 & 10.79 & 1.6 &  &  \\  
\cutinhead{WP701457N00 - HS1700+6416}
  OC1 1700+6413& 17 00 41.94  &  64 13 16.9  &  3.83 & 91.04 &  0.30 & 19.29  & 209 & 2.35  & 0.8 &RXJ1700.6+6413 & Cluster (z=0.22)\\*  
  OC2 1701+6412& 17 01 30.22  &  64 12 05.4  &  3.15 & 99.89 &  0.70 & 21.51  & 109 & 3.29  & 0.8 &RXJ1701.3+6414 & Cluster (z=0.45) \\*  
  OC3 1701+6421& 17 01 48.77  &  64 21 03.0  &  3.99 & 62.42 &  0.20 & 18.20  & 260 & 10.49 & 1.4 &RXJ1701.6+6421 & Cluster \\*  
  OC4 1702+6420& 17 02 13.58  &  64 20 07.3  &  3.34 & 79.25 &  0.30 & 19.26  & 177 & 11.39 & 1.5 &RXJ1702.2+6420 & Cluster (z=0.23) \\  
 
\tableline
\enddata
\end{deluxetable}
\tablenotetext{1}{X-ray flux units for the upper limits are $10^{-14}~\flux$.
Upper limits were computed for all optical candidates 
as a function of off-axis position in the
ROSAT field and of ROSAT exposure time, regardless the presence of an
X-ray counterpart.}

%Optical catalog ends here

\begin{deluxetable}{l|rrllrrrrrrr}
\tablecaption{Cross-identifications in V and I \label{vi}}
\tabletypesize{\tiny}
\tablewidth{0pc}
%\rotate
\tablecolumns{12}
\tablehead{

\colhead{I-band ID}   & \colhead{$\Lambda_{cl}(I)$} & \colhead{z(I)} & \colhead{RA (J2000)} & 
	\colhead{Dec (J2000)} & \colhead{Sigma} & \colhead{$\Lambda_{cl}(V)$}
	& \colhead{z(V)} & \colhead{V-mag} & \colhead{Radius} &
	\colhead{$N_{gals}$\tablenotemark{1}} & \colhead{P(KS)\tablenotemark{2}} } 

\startdata

\cutinhead{0902+34 (1849 galaxies)\tablenotemark{3}}
OC2 090510.2+335938\tablenotemark{a} 
		& 25.04 & 0.30 &    09 05 11.36 & 33 59 30.6 & 5.45  & 29.87 & 0.30  & 20.66      &141 & 22 & 3e-4 \\*
OC5 090557.0+340701 & 48.56 & 0.60 &    09 05 44.01 & 34 08 59.4 & 3.83  & 50.72 & 0.50  & 22.45  &163 & 27 & 0.25 \\*
{\bf OC6 090617.6+341716} & 28.87 & 0.30 &    09 06 14.26 & 34 18 15.3 & 5.01  & 27.45 & 0.30  & 20.58  &124 & 12 & 0.015 \\*
                &        &      &    09 06 13.16 & 34 16 25.5 & 3.62  & 18.44 & 0.20  & 19.44     &138 &    &  \\*
OC7 090628.2+340512\tablenotemark{b}
		& 36.56 & 0.20 &    09 06 14.70 & 33 55 57.1 & 4.95  & 27.11 & 0.30  & 20.34  &114 &   & \\*
		&	&      &    09 06 14.76 & 34 08 45.4 & 3.25  & 17.82 & 0.30  & 20.73  &108 &   & \\
\cutinhead{Pancake (1820 galaxies)}
OC1 090612.5+335422  & 25.13 & 0.20 &    09 06 22.21 & 33 52 14.2 & 3.93  & 32.10 & 0.30  & 20.45  &134 & 29 & 9e-5 \\*
{\bf OC2 090718.6+333044}  & 85.19 & 0.50 &    09 07 35.41 & 33 32 59.4 & 3.26  & 26.62 & 0.30  & 20.88  &183 & 16 & 0.20
\tablenotemark{e}\\*
{\bf OC3 090727.7+334302}  & 74.79 & 0.50 &    09 07 23.35 & 33 38 36.8 & 3.60  & 72.12 & 0.50  & 22.45  &175 & 14 & 0.02\\*
{\bf OC5 090758.3+335010}  & 67.27 & 0.70 &    09 07 44.66 & 33 48 25.6 & 4.39  & 87.85 & 0.50  & 22.57  &161 & 10 & 0.39\\

\cutinhead{1202+281 (2397 galaxies)}
OC1 120347.8+275836  & 56.48 & 0.50 &    12 03 53.36 & 27 53 14.5 & 5.51  & 71.72 & 0.40  & 21.64  &161 & 43 & 0.15\\*
		 &	 &	&    12 03 44.70 & 27 59 32.9 & 4.86  & 41.40 & 0.20  & 19.10  &217     &    &   \\* 
OC2 120353.2+274355  & 39.62 & 0.40 &    12 04  00.27 & 27 43 47.3 & 3.36  & 43.74 & 0.40  & 21.93  &113 & 37 & 0.068 \\* 
OC4 120436.7+280520\tablenotemark{c}
                     & 95.30 & 1.00 &    12 04 19.52 & 28 07 07.2 & 4.41  &155.55 & 1.10  & 24.15  &95  &    &  \\*
OC5 120438.0+280152  & 28.01 & 0.20 &    12 04 36.76 & 28 01 38.8 & 4.11  & 34.99 & 0.20  & 19.28  &179 & 23 & 0.017 \\* 
OC7 120457.0+275305  & 90.66 & 1.00 &    12 04 55.47 & 27 53 35.6 & *2.67 & 100.32& 0.90  & 23.99  &67  &    &    \\*  
OC9 120543.8+274030\tablenotemark{c}
		     & 50.47 & 0.40 &    12 05 26.85 & 27 44  4.9 & 3.49  & 29.69 & 0.20  & 19.51  &104 &    &   \\*

\cutinhead{3C324 (2145 galaxies)}
{\bf OC2  154855.9+212722} & 38.51  & 0.40 &    15 48 55.83 & 21 28 29.5 &*2.85  & 31.81 & 0.40   &21.53 & 83 & 13 & 0.17 \\
OC2a 154902.2+211953 & 41.59  & 0.20 &    15 49  1.13 & 21 21 16.6 & 3.61  & 29.57 & 0.30   &20.64 & 146 & 81 & 0.011 \\*
OC2b 154902.4+212226 & 31.53  & 0.30 &                &            &       &       &        &      &     &    & \\*
OC3  154932.1+212325\tablenotemark{d}
		     & 70.35  & 0.70 &    15 49 31.97 & 21 23 28.4 & 4.23  & 477.83 & 1.20  &24.11 & 136 & 29 & 0.179 \\*
OC8  155044.7+213436 & 36.51  & 0.40 &    15 50 34.25 & 21 31 53.1 & 3.61  & 29.53 & 0.30   &20.56  &192 & 28 & 0.012 \\

\cutinhead{3C356 (1591 galaxies)}
OC2  172324.8+510835  & 41.06 & 0.20 &    17 23 19.59 & 51 09 13.5 & 4.56  & 22.23 & 0.20  & 19.43  &91 & 48 & 0.32 \\*
OC4  172526.0+505522  & 28.99 & 0.30 &    17 25 35.94 & 50 57 34.6 & 4.08  & 19.66 & 0.20  & 19.46  &100 & 9 & 0.14    \\  
\enddata

\tablenotetext{1}{$N_{gals}$ is the number of galaxies near the 
centroid of the optical candidate used for 
the 2-d K-S test.}
\tablenotetext{2}{The KS probability between 0 and 1 that the galaxies were randomly
selected from the same parent population as the rest of the galaxies in the 
same field. Typically $\sim2000$ galaxies were measured in the same field.}
\tablenotetext{3}{Name of original target. (Number of galaxies in the optical field used as
the parent population sample.)}
\tablenotetext{a}{Significance in I-band is $2.94\sigma$, formally below a $3\sigma$ threshold.}
\tablenotetext{b}{The center of the I band candidate was significantly obscured by scattered 
light in the V-band image, but the matched filter algorithm found a couple of
cluster candidates in the unaffected regions of the V image.}
\tablenotetext{c}{These two cluster candidates overlap at the $3\sigma$ contour
level, but the centroids differ significantly. These systems, if a true match,
are filamentary and unlikely to be virialized.}
\tablenotetext{d}{Note significant difference in estimated redshifts between the I and V data.}
\tablenotetext{e}{The joint photometry for this candidate was significantly distant
from the X-ray core because of scattered light limitations.}
\tablecomments{This table lists the matched-filter parameters derived from the V-band for
cluster candidates cross-identified with cluster candidates in the I-band. The units for
the matched-filter quantities are the same as in Table~\ref{Optical}, although $\Lambda_{cl}$
and V-mag refer to V-band quantities. Note though that $\Lambda_{cl}$ from the I and
V bands should be similar, nevertheless. The {\bf boldfaced} entries in column 1 are cluster candidates
with X-ray counterparts (Table~\ref{Xray}). }
\end{deluxetable}

\begin{deluxetable}{lrrrrrrr|l}
\tablecaption{Unmatched Cluster Candidates in V and I \label{vinot}}
\tabletypesize{\tiny}
\tablewidth{0pc}
%\rotate
\tablecolumns{9}
\tablehead{
\colhead{Band name} & \colhead{RA} & \colhead{Dec} & \colhead{Sigma} & \colhead{Lambda}
	& \colhead{z} & \colhead{V-mag} & \colhead{Radius} &  \colhead{Comment}} 
\startdata

\cutinhead{0902+34}
I & 09 04 23.88 & 34 11 04.9 & 4.25  & 43.38 & 0.40  & 20.13  &70  & off V frame \\
I & 09 05 11.11 & 34 21 53.0 & 5.14  & 34.79 & 0.20  & 18.39  &116 & off V frame \\
I & 09 05 52.95 & 34 03 17.5 & 3.58  & 88.98 & 1.00  & 22.51  &101 & ``high z'' \\
I & 09 06 31.74 & 33 59 51.1 & 4.05  & 70.13 & 0.70  & 21.65  &165 & cut from V frame \\
I & 09 06 32.70 & 34 13 17.4 & 3.35  & 89.67 & 1.10  & 22.71  &50  & cut from V frame \\

V & 09 04 33.06 & 34 15 30.1 & 4.07  & 22.32  & 0.30  & 20.25  & 79  &    \\
V & 09 04 54.98 & 34 02 33.6 & 3.64  & 248.79 & 1.20  & 24.15  & 87  & ``high z'' \\
V & 09 05 19.40 & 34 20 04.8 & 4.99  & 313.70 & 1.10  & 24.12  & 162 & ``high z'' \\
V & 09 05 22.27 & 34 11 47.3 & 3.43  & 215.38 & 1.10  & 24.12  & 86  & ``high z'' \\

\cutinhead{Pancake}

I &09 07 47.16 & 33 55 00.4  &3.50   &28.21 & 0.30   &19.47 & 68  & off V frame \\

V &09 05 40.86 & 33 54 05.0 & 3.47 & 188.25 & 1.20 &  24.16 & 57  & Very low sig (2.1,$z=1.1$) I candidate\\
V &09 05 48.45 & 33 32 11.6 & 3.86 &  17.20 & 0.20 &  19.03 & 97 &      \\
V &09 06 11.17 & 33 43 41.5 & 3.19 & 173.15 & 1.20  & 24.14 & 60 &  high z\\
V &09 06 41.30 & 33 27 41.1 & 3.50 & 155.25 & 1.00 &  24.07&  78 &  high z\\
V &09 06 52.22 & 33 34 29.3&  3.43 & 185.96 & 1.20  & 24.14 & 76 &  high z\\
V &09 06 57.94 & 33 54 04.2&  4.58 & 248.20 & 1.20 &  24.15  &75 &  high z\\
V &09 06 59.65 & 33 47 48.3&  3.46  &187.81 & 1.20 &  24.14 & 88  & high z\\
V &09 07 16.43 & 33 50 23.2&  4.30  & 35.09 & 0.30 &  20.51 & 149 &    \\

\cutinhead{1202+281}
I & 12 04 26.02 & 27 44 12.8 & 3.16  & 89.91 & 1.00  & 22.52 & 53  & ``high z'' \\
I & 12 04 53.71 & 27 57 59.1 & 3.21  & 39.02 & 0.40  & 20.32 & 92  &  \\
I & 12 05 18.49 & 28 05 17.1 & 3.19  & 31.43 & 0.30  & 19.51 & 73  &  \\

V & 12 03 48.00 & 28 06 55.5 & 2.97  &111.49 & 0.90  & 23.94 & 48 &  ``high z'' \\
V & 12 04 33.52 & 27 51 24.4 & 4.03  &142.08 & 1.10  & 24.15 & 69  & ``high z'' \\
V & 12 05 16.06 & 28 06 42.6 & 4.97  &175.41 & 1.10  & 24.15 & 113 & ``high z'' \\
V & 12 05 18.25 & 27 46 51.8 & 4.67  &164.63 & 1.10  & 24.16 & 79  & ``high z'' \\

\cutinhead{3C324} 
I &15 48 50.06&  21 13 54.3 & 3.39 & 39.86 &  0.50  & 20.76 & 106 &     \\
I &15 50 02.46&  21 12 37.5 & 3.18 &  23.64 & 0.20  & 17.65 & 136 & off V frame \\
I &15 50 13.35&  21 25 21.6 & 3.64 & 113.13 & 1.10  & 22.71 & 87&   ``high z'' \\

V &15 49 01.37&  21 35 35.8 & 3.44 &  12.68 & 0.20 &  19.20&  59  &  \\
V &15 49 13.73 & 21 32 04.1 & 3.51 & 396.85 & 1.20 &  24.08 & 60  & ``high z'' \\
V &15 49 33.91 & 21 33 20.6 & 3.23 & 117.30 & 0.80 &  23.79&  64& Very low sig (2.3,$z=0.7$) I candidate \\
V &15 50 03.67 & 21 19 01.6 & 3.29 & 371.34 & 1.20 &  24.08 & 57  & cut out, ``high z'' \\
V &15 50 46.75 & 21 15 57.0 & 4.99 & 47.04  & 0.20 &  19.49 & 84  & Very low sig (2.6,$z=0.2$) I candidate \\

\cutinhead{3C356}

I& 17 23 14.13 & 50 59 25.1  &3.36 &  38.02 & 0.50  & 20.75&  61  &  \\
    
V& 17 23 00.86 & 50 57 51.2&  4.57 &  272.56 & 1.20 &  24.16 & 121 & ``high z'' \\
V& 17 23 45.28 & 51 02 55.6 & 3.75 &  45.64  & 0.40  & 21.53 & 98& \\
V &17 24 36.12 & 50 59 19.8 & 3.73 &  45.37  & 0.40 &  21.58 & 74  & cut I region \\
V& 17 24 44.99 & 51 05 14.1 & 6.00 &  357.66 & 1.20  & 24.16 & 135 & ``high z'' \\

\enddata
\end{deluxetable}

\begin{table}
\caption[]{Summary of Cross-Identification Statistics for I- and V-band Matched Filter Candidates
\label{crossid_tab}}
\begin{tabular}{l|rr} \tableline
      							&  I-band       &   V-band \\ \tableline
Total number of candidates detected in 5 fields	($>3\sigma$)
							&	33	&	46 \\
Candidates unavailable in other band (scattered light)	&	 6	&	 2 \\
``High redshift'' candidates undetected in other band	&	 4	& 	15 \\
Net viable candidates in 5 fields\tablenotemark{1}	&	23	&	29 \\
Candidates with one or more counterparts in other band\tablenotemark{2}	&	16	&	18 \\
Cross ID efficiency					&	70\%    &	62\% \\\tableline
\end{tabular}
\tablenotetext{1}{V and I counterparts were counted if and only if both candidates had
detection confidence of $>3\sigma$ and the centroids were closer than $\sim3-4'$ in both
RA and Dec. Overlapping matched filter 
candidates with centroids this close have contours which
overlap significantly.}
\tablenotetext{2}{The net number of viable candidates is the difference between the
total number of candidates and the sum of the candidates unavailable in the other band and
and the ``high redshift'' candidates.}
\end{table}

% X-ray catalog begins here
\begin{deluxetable}{rrrrrrrrrcrl}
\tablecaption{X-ray Clusters of Galaxies \label{Xray}}
\tabletypesize{\tiny}
\tablewidth{9.0truein}

\rotate

\tablecolumns{12}
\tablehead{
\colhead{Xray ID} & \colhead{RA} & \colhead{Dec} & \colhead{Counts} & \colhead{Theta}
	& \colhead{FWHM} & \colhead{Sig-ext} & \colhead{Fx\tablenotemark{1}} &  
	\colhead{eFx\tablenotemark{1}} &
	\colhead{Comment} & \colhead{Optical Match} & \colhead{Comment} \\
	\colhead{} & \colhead{J2000} & \colhead{J2000} & \colhead{} &
	\colhead{(')} & \colhead{(")} & \colhead{($\sigma$)} & 
	\colhead{} & \colhead{} & 
	\colhead{} & \colhead{} & \colhead{} }

\startdata

\cutinhead{RP700112}		
RXJ0743.7+6457&	07 43 45.01& +64 57 19.4&    52.6&    15.2&    90.2&     5.4&    3.51&  1.60&	&OC3 0743+6458&z=0.4,within 1'\\
\cutinhead{RP700326N00}		
RXJ0906.3+3417&	09 06 18.73& +34 17 24.8&    17.4&    13.8&    67.2&     2.0&    1.50&  0.95&  	&OC6 0906+3417&z=0.3,within 15"\\
\cutinhead{RP900327}		
RXJ0907.3+3330&	09 07 18.26& +33 30 28.1&    61.3&    11.2&    54.1&     2.2&    2.84&  0.46&   &OC2 0907+3330&z=0.5,within 30"\\*
RXJ0907.4+3342&	09 07 27.10& +33 42 32.0&    32.9&     7.6&    73.9&     5.9&    0.94& 0.35&(0.470?)&OC3 0907+3343&z=0.5,within 30"\\*
RXJ0907.8+3351&	09 07 51.76& +33 51 26.6&    74.6&    16.6&    74.2&     2.5&    4.27&  0.50&   &OC5 0907+3350&z=0.7,within 2'\\
\cutinhead{WP700540}	
RXJ1025.4+4703&	10 25 25.07& +47 03 44.4&    30.6&    10.3&    69.4&     4.6&    1.84&  0.38&    &OC11 1025+4701&z=0.2,within 3'\\*
RXJ1025.8+4709&	10 25 50.87& +47 09 01.1&    53.6&    13.2&    90.7&     6.3&    3.21&  1.31&  d &  &Half of area excluded\\

\cutinhead{WP201243N00}	
\nodata	&\nodata&\nodata&\nodata&\nodata&\nodata&\nodata&\nodata&\nodata   &\nodata &\nodata   &\nodata 		\\

\cutinhead{WP700228}	
RXJ1118.9+2117&	11 18 59.90& +21 17 56.3&    93.7&     2.6&    53.8&     4.0&    4.55&  4.98&      &  &\\*
RXJ1119.2+2117&	11 19 16.73& +21 17 32.3&   142.5&     2.4&    73.5&     6.6&    6.92&?21.00&      &OC3 1119+2116&z=0.5,within 30"\\*
RXJ1119.4+2106&        11 19 25.41& +21 06 44.3&   162.7&    13.1&    60.0&     2.3&    8.38&  0.89&(0.176) &OC4 1119+2107& z=0.2,within 30"\\*
RXJ1119.7+2126&	11 19 43.21& +21 26 36.9&    93.3&    10.8&    55.0&     2.5&    4.68&  0.69&(0.061)&OC6 1119+2127&z=0.4,within 1.5',(z-spec=f.g.galaxy)\\*
RXJ1120.0+2115&	11 20 02.45& +21 15 10.2&    32.5&    13.0&    63.3&     2.8&    1.68&  0.64&      &  &\\

\cutinhead{WP201367M01}	
RXJ1204.3-0350&	12 04 22.58& -03 50 53.9&   207.1&    10.7&    74.2&     5.1&    9.83&  1.16&(0.261)&OC5 1204-0351&z=0.2,within 30"\\*
RXJ1205.0-0332&	12 05 02.23& -03 32 25.1&    54.0&    12.1&    60.3&     2.8&    2.69&  0.55& 	  &?OC8 1204-0330&z=0.9,within 2',(opt sigma=2.1)\\

\cutinhead{WP700232}	
RXJ1204.1+2807&	12 04 03.15& +28 07 03.1&   740.2&     15.7&    79.9&     3.8&   32.83&  0.75&  	  &OC1,4,5\tablenotemark{3}  &Area excluded;MS1201+2823/A1455 ($z=0.167$) \\*
RXJ1205.2+2752&	12 05 16.06& +27 52 48.9&    64.5&     7.4&    65.3&     4.8&    2.79&  0.31&  	  &  &\\

\cutinhead{RP700864A01}		
RXJ1220.0+3334&	12 20 01.31& +33 34 46.4&    49.2&    10.8&    53.0&     2.2&    3.05&  0.29&   d &?OC3 1220+3334&z=1.2,within 1.5',(opt sigma=2.6)\\*
RXJ1220.8+3343&	12 20 52.82& +33 43 50.6&     8.2&     4.1&    88.4&     8.5&    0.50&  0.45&   d &OC4 1221+3344&z=0.3,within 4'\\*
RXJ1220.9+3343&	12 20 54.48& +33 43 52.2&    33.9&     4.4&    84.2&     7.8&    2.06&  1.18&   d &OC4 1221+3344&same Xray cluster as above?\\

\cutinhead{RP600242A01}		
RXJ1227.8+0143&	12 27 51.93& +01 43 37.9&    39.3&     7.9&    87.8&     7.7&    1.88&  0.79&	  &OC1 1227+0143&z=0.3,within 45"\\*
RXJ1228.5+0134&	12 28 30.02& +01 34 42.7&    39.0&    11.8&    66.1&     3.6&    1.91&  0.38&	  &  & Most of area excluded\\

\cutinhead{RP700073}		
RXJ1255.6+4712&	12 55 36.25& +47 12 02.5&   186.5&    16.1&    89.5&     4.8&    4.60&  0.62&        &  & Off frame\\*
RXJ1256.6+4715&	12 56 38.56& +47 15 29.8&   213.5&     5.9&    76.7&     6.6&    5.46&  0.83& (0.410)&  & Optical area excluded\\*
RXJ1256.8+4727&	12 56 53.16& +47 27 23.5&    40.8&     7.0&    64.2&     4.8&    1.01&  0.67&        &  & Optical area excluded\\*
RXJ1256.9+4720&	12 56 57.02& +47 20 46.7&    53.3&     0.3&    55.3&     4.2&    1.29&  0.27& (0.997)\tablenotemark{2}&?OC1 1257+4719& z=1.0,within 1.5',(opt sigma=2.0)\\*
RXJ1257.0+4738&	12 57 05.17& +47 38 17.1&   223.8&    17.9&   113.8&     7.1&    7.02&  0.86& 	     &  &Off frame\\*
RXJ1257.3+4729&	12 57 22.20& +47 29 56.1&    77.8&    10.4&    69.0&     4.5&    1.97&  0.73&        &  &\\*
RXJ1257.6+4737&	12 57 36.97& +47 37 06.3&   153.4&    18.0&   130.9&     9.4&    4.67&  0.92&        &  & Off frame\\*
RXJ1257.7+4723&	12 57 43.50& +47 23 21.3&    64.6&     8.4&    53.4&     3.0&    1.56&  0.22&        &  &Half of area excluded\\

\cutinhead{RP700216A00}		
RXJ1309.9+3222&	13 09 55.79& +32 22 23.0&    29.0&     7.1&    76.5&     6.4&    4.06&  1.00&  	    &OC2 1310+3221 &z=0.3,within 2' (MS1308.8+3244,z=0.245)\\*
RXJ1310.5+3217&	13 10 35.95& +32 17 36.0&    17.5&     3.7&    54.9&     4.0&    2.44&  0.63&	    &	&Star spike; area excluded\\*
RXJ1313.2+3229&	13 11 12.44& +32 29 07.6&   178.4&    12.3&    66.7&     3.5&   24.95&  1.75&  	    &OC4 1311+3228 &z=0.2,within 30"\\

\cutinhead{RP700117}		
RXJ1407.6+3415&	14 07 39.76& +34 15 11.0&    29.1&    12.3&    58.6&     2.1&    1.53&  0.37&	    &OC7 1407+3415 &z=0.5,within 30"\\

\cutinhead{WP700248}	
RXJ1412.6+4359&	14 12 36.30& +43 59 02.9&    39.4&    12.9&    76.9&     4.6&    1.86&  0.89&       &  &Optical area excluded\\*
RXJ1413.5+4411&	14 13 30.94& +44 11 44.2&    32.6&    12.1&    81.1&     5.5&    1.51&  0.61& 	    &  &Optical area excluded\\

\cutinhead{RP700122} 	
RXJ1415.2+1119&	14 15 15.77& +11 19 32.6&    75.2&    12.5&    75.7&     4.6&    3.18&  0.63& 	&?OC1 1414+1123&z=0.3,within 7'\\

\cutinhead{RP800401A01}	
RXJ1415.8+2316&	14 15 50.54& +23 16 10.9&    19.0&     9.1&    64.2&     4.2&    2.01&  1.55&     &OC2 1415+2317&z=0.4,within 2.5'\\*
RXJ1415.9+2307&	14 15 57.16& +23 07 37.4&    99.1&     0.4&    67.9&     5.9&   10.20&  0.78&  	  &OC3 1415+2307&z=0.3,within 30"\\*
RXJ1416.3+2309&	14 16 22.51& +23 09 59.6&    28.1&     6.4&    66.5&     5.2&    2.90&  0.69&  	  &   &\\*
RXJ1416.4+2315&	14 16 26.88& +23 15 37.2&   499.0&    10.8&   109.6&     9.8&   51.34&  1.53&  	  &   &Optical area excluded\\*
RXJ1416.4+2302&	14 16 28.28& +23 02 31.6&    15.8&     8.5&    55.7&     3.3&    1.67&  1.16&     &   &\\

\cutinhead{RP700257N00}		
\nodata	&\nodata&\nodata&\nodata&\nodata&\nodata	&\nodata&\nodata&\nodata&\nodata&\nodata&\nodata \\

\cutinhead{RP701373N00}		
RXJ1548.8+2126&	15 48 52.57& +21 26 06.1&    61.9&    13.4&   110.8&     8.9&    5.08&  2.42&   d   &OC2 1548+2127& z=0.4,within 1.5'\\

\cutinhead{RP800239N00}		
RXJ1603.6+4316&	16 03 39.01& +43 16 19.1&    53.7&     9.6&    56.1&     3.0&    2.17&  0.36&  	    &   &\\

\cutinhead{RP300021N00}		
RXJ1605.0+2552&	16 05 04.66& +25 52 47.4&    32.9&     9.3&    52.8&     2.7&    1.76&  0.68&    d &?OC2 1604+2554&z=0.6,within 2.5',Cluster center excluded\\*
RXJ1605.5+2553&	16 05 30.99& +25 53 16.7&    40.0&     3.8&    74.8&     6.5&    2.06&  1.10&    d &OC4 1605+2553&z=0.2,within 1.5'\\*
RXJ1605.6+2548&	16 05 41.71& +25 48 27.7&    39.6&     3.2&    75.7&     6.8&    2.04&  1.05&  	   &   &\\*
RXJ1606.1+2558&	16 06 10.83& +25 58 42.8&    22.0&     9.1&    55.8&     3.1&    1.17&  0.69&      &   &\\

\cutinhead{WP170154}	
RXJ1626.3+7816&	16 26 23.25& +78 16 59.6&   303.9&    15.5&    87.0&     4.8&   10.63&  0.62&(0.580)&OC1 1626+7817& z=0.5,within 15"\\*
RXJ1627.1+7756&	16 27 10.77& +77 56 06.8&   102.7&    11.4&    74.1&     4.8&    3.47&  0.49&   d &?OC2 1626+7759&z=0.3,within 3'\\*
RXJ1629.7+7757&	16 29 45.43& +77 57 59.5&   122.2&     6.9&    78.6&     6.7&    4.12&  0.64&   d &   &\\*
RXJ1629.9+7819&	16 29 56.33& +78 19 18.6&    92.2&    14.5&    98.6&     6.8&    3.11&  0.75&     &   &Off frame\\*
RXJ1630.2+7815&	16 30 12.72& +78 15 35.3&    44.1&    10.9&    59.9&     3.1&    1.54&  0.57&     &   &Optical area excluded\\

\cutinhead{RP800395N00}
\nodata	&\nodata&\nodata&\nodata&\nodata&\nodata&\nodata	&\nodata&\nodata&\nodata&\nodata&\nodata \\

\cutinhead{WP701457N00}		
RXJ1700.6+6413&	17 00 40.58& +64 13 04.5&   551.7&    2.3&    44.6&     2.8&   24.54&  0.44&(0.22)&OC1 1700+6413& z=0.3\\*
RXJ1701.3+6414&	17 01 21.31& +64 14 16.5&   428.3&    3.2&    53.0&     3.8&   19.05&  0.44& (0.45) &OC2 1701+6412&z=0.7\\*
RXJ1702.0+6407&	17 02 00.94& +64 07 39.0&   35.0&     8.0&    58.2&     3.7&    1.56&  0.25&      &   &\\*
RXJ1701.6+6421&	17 01 42.96& +64 21 20.2&   42.7&    10.4&    83.5&     6.4&    1.89&  0.75&     &OC3 1701+6421& z=0.2\\*
RXJ1702.2+6420&	17 02 10.14& +64 20 00.3&    36.1&   11.1&    60.7&     3.2&    1.65&  0.37&(0.23) & OC4 1702+6420&z=0.3\\

%\tableline
\enddata
\tablenotetext{1}{Flux units are $10^{-14} \flux$.}
\tablenotetext{2}{The extended X-ray 
source RXJ1256.9+4720 is likely to be associated with the radio galaxy 3C280, the target
of the original observation. The X-rays could be from the galaxy or a cluster surrounding
the galaxy.}
\tablenotetext{3}{This source (Abell 1455) was obscured by a nearby F3 star in the
I-band data, but 3 optical
candidates were identified nearby. The matched filter algorithm may have picked up the outskirts of this cluster.}

\end{deluxetable}
% X-ray catalog ends here.

\begin{table}
\caption[]{Spectroscopic vs. Matched Filter Estimated Redshifts \label{z} }
\begin{tabular}{lllll} \tableline
Xray ID & Optical ID & Spec z & Est z & Comments \\ \tableline
RXJ0907.4+3342 & OC3 0907+3343 & 0.470 & 0.5 &  \\
RXJ1119.4+2106 & OC4 1119+2107 & 0.176 & 0.2 &  \\
RXJ1119.7+2126 & OC6 1119+2127 & 0.061 & 0.4 & foreground galaxy \\
RXJ1204.1+2807 & OC1 1203+2758 & 0.167 & 0.5 & MS1201+2823/A1455\tablenotemark{1} \\
               & OC4 1204+2805 & 0.167 & 1.0 & MS1201+2823/A1455\tablenotemark{1} \\
               & OC5 1204+2801 & 0.167 & 0.2 & MS1201+2823/A1455\tablenotemark{1} \\
RXJ1204.3-0350 & OC5 1204-0351 & 0.261  & 0.2 & \\
RXJ1256.9+4720 & OC1 1257+4719 & 0.997  & 1.0 & Optical $2.0\sigma$; 3C280 \\
RXJ1309.9+3222 & OC2 1310+3221 & 0.245  & 0.3 & EMSS1308.8+3244 \\
RXJ1626.3+7816 & OC1 1626+7817 & 0.580 & 0.5 & \\
RXJ1700.7+6413 & OC1 1700+6413 & 0.220 & 0.3 &  \\
RXJ1701.4+6414 & OC2 1701+6412 & 0.45 & 0.7 & \\
RXJ1702.2+6420 & OC4 1702+6420 & 0.23 & 0.3 & \\ \tableline
\end{tabular}
\tablenotetext{1}{The X-ray cluster was obscured by the diffraction spike of an
F3 star; the matched filter algorithm found 3 candidates in the vicinity, none
close enough for an official match.}
\end{table}

\end{document}